# The 2nd Workshop on Hacking and Making at Time-Bounded Events

**Ei Pa Pa Pe-Than and Alexander Nolte**

*(Editors)*

Institute for Software Research
School of Computer Science
Carnegie Mellon University
5000 Forbes Avenue
Pittsburgh PA 15213

This work was funded by the Alfred P. Sloan Foundation under grant number G-2015-13989, James Herbsleb, PI.

**Keywords:** time-bounded collaborative events; workshop; hackathons; sprints; codefests; radical collocation; community building; innovation; collaborative learning, cscw.

# Abstract

In hackathons, small teams work together over a specified period of time to complete a project of interest. Such time-bounded hackathon-style events have become increasingly popular across different domains in recent years. Collegiate hackathons, just one of the many variants of hackathons, that are supported by the largest hackathon league (https://mlh.io/)[1] alone attract over 65,000 participants among more than 200 events each year. Variously known as data dives, codefests, hack-days, sprints, edit-a-thons, mapathons, and so on, such events vary depending on different audiences and with divergent aims: for example, whether teams know each other beforehand, whether the event is structured as a competition with prizes, whether the event is open or requires membership or invitations, and whether the desired outcome is primarily a product innovation, learning a new skill, forming a community around a cause, solving a technical problem that requires intensive focus by a group, or just having fun. Taken together, hackathons offer new opportunities and challenges for collaboration by affording explicit, predictable, time-bounded spaces for collaborative work and engaging with new audiences. With the goal of discussing opportunities and challenges surrounding hackathons of different kinds, this one-day workshop brought together researchers, experienced event organizers, and practitioners to share and discuss their practical experiences. Empirical insights from studying these events may help position the CHI community to better study, plan and design hackathon-style events as socio-technical systems that support new modes of production and collaboration.

[1] Major League Hacking, https://mlh.io/

# Workshop Organizing Committee

| | |
|---|---|
| Ei Pa Pa Pe-Than | Carnegie Mellon University |
| James D. Herbsleb | Carnegie Mellon University |
| Alexander Nolte | University of Tartu, Carnegie Mellon University |
| Brad Chapman | Harvard T.H. Chan School of Public Health |
| Brittany Fiore-Gartland | University of Washington |
| Elizabeth Gerber | Northwestern University |
| Aurelia Moser | Mozilla Science Lab |
| Nancy Wilkins-Diehr | San Diego Supercomputer Center, Science Gateways Community |

# Workshop Participants

| Name and Personal website | Affiliation | Title | Email |
|---|---|---|---|
| Sian Brooke | Oxford Internet Institute, University of Oxford | PhD Student | sian.brooke@oii.ox.ac.uk |
| Karen Bhavnani | MCI Group | Innovation and Implementation Manager | karen.bhavnani@mci-group.com |
| John M. Carroll | Pennsylvania State University | Professor | jmcarroll@psu.edu |
| Brad Chapman | Harvard Chan School | Senior Research Scientist | bchapman@hsph.harvard.edu |
| Margaret Drouhard | University of Washington | PhD Student | mdrouhard@acm.org |
| Rosta Farzan | University of Pittsburgh | Assistant Professor | rfarzan@pitt.edu |
| Elizabeth Gerber | Northwestern University | Professor | egerber@northwestern.edu |
| Eureka Chen Yew Foong | Northwestern University | PhD Student | eureka@u.northwestern.edu |
| Robert Gradeck | Western Pennsylvania Regional Data Center, University of Pittsburgh - UCSUR | Project Manager | rmg44@pitt.edu |
| Sourobh Ghosh | Harvard Business School | PhD Student | sourobhghosh@gmail.com |
| James D. Herbsleb | Carnegie Mellon University | Professor | jherbsleb@acm.org |
| Erin Hoffman | University of Washington | PhD Student | erinrhof@uw.edu |
| Bonnie E. John | Bloomberg LP | Senior Interaction Designer | bjohn11@bloomberg.net |
| Greg Kiar | McGill University | PhD Student | gregory.kiar@mail.mcgill.ca |
| Peter Müller | Technical University Munich | PhD Student | pet.mueller@tum.de |
| Alexander Nolte | Carnegie Mellon University, University of Tartu | Assistant Professor | alexander.nolte@ut.ee |
| Je'aime Powell | The Texas Advanced Computing Center | Senior Systems Administrator | jpowell@tacc.utexas.edu |
| Ei Pa Pa Pe-Than | Carnegie Mellon University | Post-Doctoral Researcher | eipapapt@cs.cmu.edu |
| Anthony Viviano | Bloomberg L.P. | Senior Interaction Designer | aviviano@bloomberg.net |
| Shahtab Wahid | Bloomberg LP | Interaction Designer | swahid2@bloomberg.net |
| Mona Wong | SDSC @ UCSD | Software Engineer | mona@sdsc.edu |

# Introduction

In recent years, there has been a surge in popularity of time-bounded intensive events. These events - which are generally known as hackathons - typically engage enthusiasts in small ad-hoc teams to create artifacts - most commonly software prototypes - over a period of 1 or 2 days, motivating them with competitive awards such as prizes and job offers [3, 17]. Hackathons or similar collaborative events are often termed as data dives, codefests, hack-days, sprints, edit-a-thons, mapathons, etc. The popularity of hackathons has increased dramatically and that, for example, collegiate hackathons, just one of the many variants of hackathons, are able to attract over 65,000 participants across 200 events each year[1].

The hackathon model is applied across fields. Examples include informal and collaborative learning [10, 15, 20], creating startups [6], arts and culture [3], civic open innovation [1], corporate innovation [16, 17], computational biology [14, 19], and social issues [18]. It has also been used in academic conferences through workshops exploring alternative models of creation such as OCData Hackathon at CSCW [11], and CHI4 Good Day of Service [18] and Crowdcamp [2] at CHI. These hackathons may differ on the interaction style (competition or collaboration), the mode of collaboration (face-to-face or remote), the extent to which communication tools are used[2], whether newly formed or existing communities working on new or existing projects [13, 14, 16, 17], and goals and orientation (community building or advancement of existing projects) [5, 7].

Hackathon-style events introduce new and interesting opportunities as well as challenges for the study of collaborative work. For example, these events may provide unique opportunities for cooperation by affording explicit and time-bounded spaces for individuals to work more interdependently; access to new collaborators with needed background and experience, and existing collaborators who are otherwise difficult to reach [20]; predictable interactions that can serve to strengthen existing social ties and develop new ones [14]. At the same time, working on projects that are outside of one's normal workflow may provide challenges for continuity after the brief cooperative stint is over [20]. For example, continuing projects in a virtual setting may require carrying over social and work artifacts that are not in easily editable formats and highly context dependent [15], and keeping momentum and enthusiasm for completing projects presents a further challenge [13]. These events may also provide different pressures on team dynamics process. For example, team formation and common understanding need to happen relatively quickly.

[2] Mozilla Science Lab Global Sprint 2016. https://science.mozilla.org/programs/events/global-sprint-2016

Despite the plethora of research and public attention, little is known about how to design a hackathon to achieve intended outcomes, what benefits hackathons offers, what the immediate and longer-term impacts of hackathons are, and what the larger impacts of hackathons on CHI community and on the society as a whole are. To fill these gaps, we conducted our first workshop[3] on this topic at the 2017 ACM conference on Computer Supported Cooperative Work (CSCW 2017) [8] which the current workshop is built upon. There we brought together researchers and practitioners to share their hackathon-related experiences. The outcomes of our previous workshop were reported in a technical report, and distributed to all workshop participants. Building on our successful first workshop, we conducted the hackathon workshop for the second time at the 2018 ACM CHI conference on Human Factors in Computing Systems (CHI 2018) in order to continue with the discussions about the open issues that were identified in our first workshop and build the community of researchers and practitioners with an interest in hackathons.

This report presents an account of the one-day workshop[4] at the CHI 2018 which brought a diverse set of researchers and practitioners including past event organizers and individuals interested in organizing events in the future. In the remainder of this report, we describe the format of the workshop, including preparation and post-workshop activities, and summarize poster presentations and results of discussion sessions which took place during the workshop.

# Workshop

On Sunday, April 22, 2018 the "2nd Workshop on Hacking and Making at Time-Bounded Events: Current Trends and Next Steps in Research and Event Design" took place at the Palais des Congrès de Montréal in Montreal, Québec, Canada. The workshop was held in conjunction with the 2018 ACM CHI Conference on Human Factors in Computing Systems (CHI 2018).

[3] The CSCW 2017 workshop on hacking and making at time-bounded events. https://hackathon-workshop.github.io
[4] The CHI 2018 workshop on hacking and making at time-bounded events. http://hackathon-workshop-2018.com

This workshop had four main objectives:

- Facilitate networking between CHI and CSCW scholars and practitioners (both those who have experience organizing events and those who are curious about doing so),
- Develop an understanding of how to situate time-bounded events in the broader context of CHI and CSCW methods and theory,
- Identify and compile recommendations for organizers of events, as well as important tradeoffs, and
- Explore future directions for research in this area, including publication venues.

# Workshop Format

## Pre-Workshop Activities

This workshop was led by an 8-person organizing committee comprising both researchers working in the fields of CHI and CSCW, and practitioners with experience organizing events. We also informed and invited organizers and participants from our previous workshop as well as other potentially interested individuals to help us organize this second workshop. We specifically aimed to have both researchers and practitioners in this the organizing committee in order to bring multiple perspectives to bear on event organization and advertisement, and participant recruitment and selection.

CHI and CSCW researchers on the committee came from the Institute for Software Research at Carnegie Mellon University and Northwestern University. Practitioners on the committee came from the Harvard T.H. Chan School of Public Health at Harvard University, the Mozilla Science Lab at the Mozilla Foundation, University of Washington, and the Science Gateways Community Institute at the San Diego Supercomputer Center.

We launched our workshop website at http://hackathon-workshop-2018.com. The workshop was also advertised through several communication channels, such as CHI 2018 conference's website, the "Researchers of the Socio-Technical" Facebook Group, the CSCW and chi-announcements mailing lists, and invitations to our previous workshop participants. All workshop applicants were asked to submit a 2-4 page paper describing their interest in one or more of the workshop themes, presented as a research idea or a story that drew from their own event experience. After the paper submission deadline, members of the organizing committee and authors of submitted position papers were randomly assigned to submissions, and rated them on how well they represented the themes and their potential for discussion at the workshop. Each paper received at least two reviews. All accepted position papers can be found in **Appendix A** of this technical report.

## Workshop Activities

A total of 21 participants representing 15 institutions attended the workshop. 48% of participants were female, and 52% were male. The workshop started with a brief introduction by Jim Herbsleb, followed by a keynote presentation by Elizabeth Gerber. In her talk, she discussed the importance of accessing diverse knowledge and perspectives for open and collective innovation, and how hackathons could be a platform to foster knowledge exchange between people from diverse backgrounds. Moreover, she highlighted the importance of bringing together perspectives from human computer interaction, management science, design, manufacturing, and product development to drive innovation.

After this introduction, we conducted a general introduction session in which each participant introduced themselves and their interest in hackathons in 10 words. The introduction was followed by poster presentation session which aimed to facilitate a focused and interactive discussion among participants and encourage the development of new collaborations. During this poster session, participants had the opportunity to discuss mutual interest in more detail. Prior to the workshop, we had advised all participants to bring a poster containing a summarized description of their hackathon-related work or their submitted position papers. There was a total of 11 posters presented at the workshop and the poster session was divided into two 30-minutes sub-sessions. In the first session, the authors of the first 6 posters presented their work, giving the other workshop participants opportunities to discuss with authors of posters they were interested in. In the second session, the authors of the remaining 5 posters presented their work. All posters can be found in **Appendix B** of this technical report.

During the introduction and poster sessions, we collected ideas from participants which were clustered into three themes. These themes were then used as discussion topics for breakout groups. The following three themes emerged from this process: 1) organizing hackathons, 2) diversity and inclusion, and 3) measuring hackathon outcomes.

After a short lunch break, participants formed three groups around the previously-identified themes and each group explored their chosen theme in detail. This discussion session lasted 2 hours with a 30-minutes break in between. Each group was advised to use a Google Docs to record ideas resulted from their group discussion. At the end of this session, one person from each group presented a summary of their group work discussion. Each group presentation lasted about 10-15 minutes.

The breakout-group discussion session was followed by a 60-minute plenary session in which all participants proposed and discussed ideas and issues which could be considered to move hackathon research and practice forward. Part of this discussion were considerations of how to integrate hackathon-related research into the broader CHI and CSCW community including potential future research directions. In addition, participants also proposed materials which might be useful for practitioners, e.g., a tree-structured hackathon planning kit which guides hackathon organizers towards a suitable event design based on a set of questions (e.g., about their event goals). The workshop ended with a session during which all participants were asked to state one aspect of the workshop that they liked and one aspect that they wished to be different. During this session all participants sat in a circle together and took turns to share what they liked and what they wished to be different for a future workshop.

# Results

In this section, we present the results of the discussion by each aforementioned breakout group.

## Organizing Hackathons

This workgroup consisted of seven members, of which two were experienced event organizers, two were interested in putting on a scientific hackathon in the following summer, and the rest were PhD students studying hackathons. The purpose of this group was to compile resources that could be helpful for future event organizers and discuss challenges related to the organization of hackathons.

The group started off with identifying and listing available **online guides for planning and organizing hackathons**. Examples of identified planning guides include the organizer guide[5] by Major League Hacking (MLH), an organization that organizes and supports hackathons with college students, and the hackathon planning guide by Gartner based on NASA's international annual hackathons[6]. Other hackathon organizing guides identified by this group can be found in **Appendix C** of this technical report.

In addition to planning guides, this group also documented several **online websites that list hackathon events and projects** including the names and other online profiles of team members. Examples include Devpost[7] in which organizers can post their events and participants use this space to submit their projects, and MLH[8] which lists all hackathons organized by MLH. Other online hackathon listing websites can be found in Appendix C of this technical report.

This group also explored a number of questions: what is a hackathon? what are the toolsets for organizing a hackathon? what does the period before an event look like for you? With regard to the question about what a hackathon is, one member asked "what are the design dimensions of a hackathon?" The group proposed that a hackathon consists of five major components: collective innovation, communication or documentation, education (training component), incentives, and skill application to projects. In addition, they regarded that hackathons are time-bounded and can lead to something new being built.

Regarding the second question, Je'aime specifically wondered if there were any tools or templates that event organizers could readily utilize. Other group members pointed out CHI for good events and a paper about "typology of hackathons" from our previous workshop which categorized hackathons into three types. Robert and Erin also provided some tips for how to get participants motivated with the event. These include giving more awards than the participants would expect, having a "science fair" style project presentation at the event, and letting participants tell their stories as part of their hackathon outputs. Further, Robert mentioned that, based on the keynote presentation, the 6-week window period before and after the event is important for organizers to monitors event related activities including all preparatory work and follow up activities.

---

[5] MLH organizer guide. https://guide.mlh.io
[6] Gartner hackathon planning guide. https://www.gartner.com/smarterwithgartner/plan-a-successful-hackathon/
[7] Devpost hackathon listing. https://devpost.com/hackathons
[8] MLH hackathon listing. https://mlh.io/seasons/na-2018/events

## Diversity and Inclusion

The second workgroup consisted of seven members and the group discussed the challenges around diversity and inclusion in the context of hackathon. The group started off discussing the **boundary of hackathons**, and proposed that hackathons were time-bounded (but the meaning of "bounded" needed to be defined clearly), fast and scrappy (fidelity-based definition), intensive, face-to-face (in many cases) events, and can either be a long term or short term project, but they do not have to be coding and could also be editing, documentation, data, and work production. The group also noted that teams involved in hackathons are ad hoc in nature which creates particular diversity/inclusion challenges.

After defining the boundary of hackathons, the group brainstormed the **meaning of diversity and inclusion**. The identified diversity criteria include intellectual, ethic, gender, expertise, domain, family status (e.g., childcare availability at the event), and languages and skills, while the group defined accessibility, cost, family, and culture as criteria for inclusion. The group also noted that diversity is context specific depending on the purpose and expected participants of the hackathons. Sian added that hackathons are ritualistic and have been built around defined ideological tenets, so a simplification of diversity may be needed so that a person's identity isn't used to devalue their contributions.

In addition, the group also discussed using GenderMag[9] approach to encourage gender-inclusiveness in hackathons. In addition to a common way of assessing the level of diversity and inclusivity through self-reported measures, this workgroup also discussed how to measure the awareness of whose perspectives are not presented, especially when they have been systematically excluded. One proposed solution is to apply the critical mass theory, suggesting that a certain percentage of representation of minorities is needed in order for them to be able to influence and involve in the engagement. For example, in the context of political science, 30% has been regarded as the crucial cut-off point for getting women involved in politics but this percentage can be reduced by more prominent women figureheads in the group [4].

[9] GenderMag. http://gendermag.org

The group then examined **how to promote leadership and positive inclusive role models** in hackathons. Their proposal includes having a woman leader and a representative leadership team who are valued for their diverse sets of experiences and skills. The group further noted that it is important to have a system in place to support broader recruitment and support for people of diverse identities. Regarding the need for a broader recruitment effort, the group made a few suggestions to achieve this. One is related to personal invitations and recruitment in pairs, enabling participants to feel psychologically safe by being able to come to the event with a person they trust. The following statement could be included in the recruitment message: "I'd love to invite you personally and I think you'd have a lot of fun if you bring a friend." Jim suggested an alternative model of "She Innovates"[10] hackathon hosted by the University of Pittsburgh. This is a women-only hackathon which serves as a means of preparing women to get involved in other events with more diverse people. This group also noted that a better connection between event organizers would be helpful to promote more cross-pollination between events.

The group also recognized the importance of **branding hackathons to be inclusive** and explored how to achieve this objective. Jim, Brad, and Liz provided useful guidelines to make the events more inclusive. These include having less emphasis on competition (e.g., following the event structure used by Codefest or Collaboration Fest [14] and Crowdcamp [2]), making explicit valuation of other skill sets in addition to coding, and assigning roles (i.e., role-based coordination) to facilitate greater team participation.

MLH suggests that another important way that non-coder women contribute to code-centric hackathons is utilizing their subject matter expertise. Figure 1 represents the participation of women in hackathon from MLH "Hackathon Data" talk at Hackcon IV[11]. The red bars represent the percentage of the US population in each group, the yellow bars represent the percentage of EECS majors in each group, and the blue bars represent the percentage of MLH hackathon participants in each group. Notably, MLH hackathons had a higher percentage of female participants on average than EECS degree programs did. One possible explanation for this might be that among EECS majors, MLH hackathons were more likely to attract women, but MLH attributed it to the participation of non-EECS majors. In 2016, 20% of MLH hackathon participants had majors outside of EECS, and both women and African-Americans were disproportionately likely to have non-EECS majors.

---

[10] She Innovates Hackathon. https://www.innovation.pitt.edu/events-competitions/she-innovates/

[11] MLH "Hackathon Data" talk at Hackcon IV, https://youtu.be/NJCfKG4tt5M

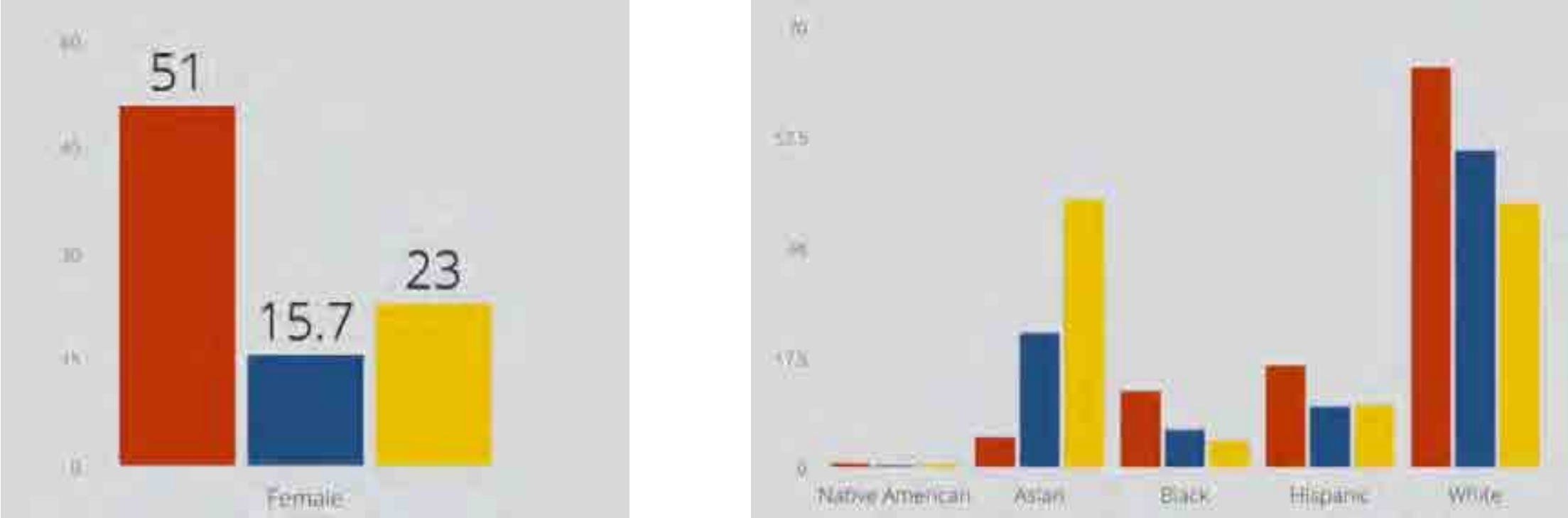

**Figure 1.** Women participation in hackathons from MLH "Hackathon Data"[11]

This group then discussed how and whether to include inclusiveness dimension as affirmative action plan in order to ensure that the event goal to attract diverse participants, rather than filling quotas. Related to this aspect, Jim pointed to the prior research work of Filippova et al. (2017) [9] in which the authors proposed brainstorming as a way of encouraging the participation of self-identified minorities in hackathons.

This group concluded their discussion with a provocative statement, "given the nature of hackathons, is it possible to have diverse hackathons? If the contention is, diversity occurs when people trust each other and are familiar with each other," then building up to the hackathon could lay the groundwork for diversity. The group also noted the need for design guidelines about how to make the hackathon event inclusive. The questions raised also include whether the duration of the event as well as funding for attending and event organizing are barriers to inclusivity.

## Measuring Hackathon Outcomes

The third workgroup consisted of eight members and their discussion focused on **how to define the success from hackathons, how to measure outcomes, and how to sustain them**. Similar to the two groups mentioned above, this group also started their discussions by identifying key characteristics of hackathons. This group identified the following aspects to be relevant: in order for an event to be considered a hackathon it has to be time-bounded, collaborative, multidisciplinary, product/development-driven, and supports participants to leave their comfort zone. Moreover, this group categorized hackathons into three different types:

- 1-off (come, hack, complete, done),
- recurring (come, join existing groups, hack, repeat), and
- progress-driven (propose desired outcome, hack, facilitate process, repeat until mature).

The group then proposed and discussed a number of possible quantitative and qualitative measures of hackathons. For each measure, the group identified suitable points of measuring specific outcomes: at the start of an event (SOE), at end of an event (EOE), and after the event has ended (EOE + T). The hackathon measures and associated points of measurement are listed in Table 1 below.

**Table 1. Possible quantitative and qualitative measures of hackathons and suitable points of measuring**

| **Measures** | **Suitable points of measurement** |
|---|---|
| **Quantitative measures** | |
| Number of participants | SOE |
| Resumes/job interviews | EOE and EOE + T |
| Corporate promotions | EOE and EOE + T |
| Multidisciplinary collaborations/ "gaps bridged" | EOE |
| Human-hours of dedicated (sprint) work towards specific topic | SOE, EOE and EOE + T |
| Companies founded | EOE and EOE + T |
| Commits/repositories/organizations created on GitHub | SOE, EOE and EOE + T |
| Community members | EOE + T |
| Network evolution (i.e., weak ties, strengthen relationships, new positions) | SOE, EOE and EOE + T |
| Initial projects merged/combined | SOE and EOE |
| Publications/white papers/case studies contributed to | EOE + T |
| GitHub issues opened (i.e., bug reports/feature requests) | EOE |
| Lines of code produced and/or deleted, “newly exposed” products/tools (i.e., link clicks etc.) | EOE |
| Intentions formed/goal set and subsequently achieved | EOE and EOE + T |
| **Qualitative measures** | |
| The number of ideas | SOE and EOE |
| “Innovative” projects (i.e., unexpected and beyond scope of the task) | EOE and EOE + T |
| People who worked on a new discipline/platform/project | EOE |
| Applications of newly acquired skill within event | EOE |
| “Value” of things accomplished | EOE + T |
| Participation equity | EOE |
| Diversity of a community | EOE + T |
| Strength and confidence in abilities/efficacy in area of interest | SOE, EOE and EOE + T |
| The type/quality of group interaction | SOE, EOE and EOE + T |
| Street cred/change in social barriers | SOE and EOE |
| Percentage of satisfied/happy participants | EOE |
| Percentage of “spin-offs” or translated ideas | EOE and EOE + T |

# Discussion and Conclusion

Based on three breakout group discussion sessions and plenary session, several important things that are worthwhile to consider for both research and practice were identified. The majority of participants agreed that it would be helpful to have a decision tree that allows organizers to search for event design guidelines based on a set of minimal questions depicting their event goals or objectives. Participants also wished to have a collection of resources about lessons learned from previous events and training materials which would lower the barriers for organizers to run future events.

With regard to diversity and inclusion, there are some open questions regarding how to promote the aspect of diversity and inclusion in hackathons. How can we design effective inclusive hackathons according to different outcomes? What evidence exists for different strategies? Which strategy is most effective? How do we measure the outcome of inclusive strategies such as feeling of inclusiveness? Could brainstorming be a model of inclusive hackathon? What motivates leaders to design for inclusion? How could the language of hackathons influence inclusion? What are the factors that detract from inclusivity and diversity at hackathons?

Other questions that are worthwhile to explore include how hackathons can be positioned in a broader theoretical space. Given the unique nature of hackathons, one might consider positioning hackathons as a new form of collocated work or the future of work [12]. Exploring the similarities and differences in event goals, participant's motivations and expectations, and outcomes across various types of hackathons such as civic, corporate, science, and collegiate. Further, hackathons could be positioned as one element of a collection of events, and their real goals are often to integrate with larger efforts (e.g., as a means to onboarding people to the process), especially in the case of hackathons for scientific community building.

Based on the feedback of "like and wish" session, we found that the participants were very positive about the experience provided by the workshop. The majority of participants expressed that they liked keynote presentation which laid some foundation of hackathons, as well as poster session which allowed them to meet and talk with people of similar interest in a more interactive manner, and less-formal breakout group discussion session which some participants wished to do it again next year. Participants also wished to extend the community beyond the workshop and produce a more concrete output such as an academic paper and technical report. A couple of participants wished that it would be good to run a pre-workshop survey to elicit participant's research interest beforehand. Interestingly, one participant wondered what would happen if this workshop is structured as a hackathon, i.e., applying the hackathon model to the workshop. Most participants found it useful to have both researchers and practitioners in the workshop and breakout groups. Taken together, our workshop was able to fulfil its purpose of facilitating the cross-fertilization of ideas by bringing in people with diverse perspectives, as well as future collaboration in this hackathon space.

# Appendix A

## Position Papers

# Organization for Human Brain Mapping Open Science Hackathons: Accessible and Inclusive Neuroinformatics

**Elizabeth DuPre**
McGill University
Montréal, QC Canada
elizabethm.dupre@gmail.com

**Gregory Kiar**
McGill University
Montréal, QC Canada
greg.kiar@mcgill.ca

**R. Cameron Craddock**
University of Texas at Austin
Austin, TX USA
cameron.craddock@gmail.com

**Chris Gorgolewski**
Stanford University
Stanford, CA USA
krzysztof.gorgolewski@gmail.com

**Felix Hoffstaedter**
Forschungszentrum Jülich
Jülich, Germany
f.hoffstaedter@fz-juelich.de

**Anisha Keshavan**
UW eScience Institute
Seattle, WA USA
anishakeshavan@gmail.com

**Jean-Baptise Poline**
McGill University
Montréal, QC Canada
jbpoline@gmail.com

**Matteo Visconti di Oleggio Castello**
Dartmouth College
Dartmouth, NH USA
matteo.visconti@gmail.com

**Kirstie Whitaker**
Alan Turing Institute
London, UK
kw401@cam.ac.uk

**Pierre Bellec**
Université de Montréal
Montréal, QC Canada
pierre.bellec@criugm.qc.ca

## Abstract

As leaders of the Organization for Human Brain Mapping's (OHBM) Open Science Special Interest Group, we host an annual hackathon to teach and promote open science among neuroimaging researchers. We have endeavored to design our hackathon events such that existing open-source software projects receive attention and support from experienced users while new adopters can gain familiarity with these tools. However, we have received community feedback that our events are still perceived as exclusionary to code-nervous researchers, and we have had difficulty quantifying overall outcomes to assess how to make improvements. We are applying to the CHI workshop with the hope of learning how to better design and organize our events to be more inclusive, and provide concrete evidence upon which we can advertise the successes of OHBM hackathon events and continuously improve.

## Author Keywords

neuroscience; informatics; software development; education

## Background

The Organization for Human Brain Mapping (OHBM) is an annual, international meeting of neuroimaging researchers dedicated to understanding the structure and function of the brain and their pathologies in health and disease. Since 2013, the OHBM Open Science Special Interest Group has hosted a hackathon adjacent to the annual meeting to promote open science for replicability, collaboration, and innovation in neuroscience research.

The hackathon and its adopted structure have grown out of the Brainhack initiative [2]. Brainhack has three core aims: 1) educating new community members on open scientific best practices, 2) encouraging the development and maintenance of open-source software, and 3) promoting the free exchange of ideas to encourage future collaborations. These ideas are operationalized as short educational courses such as *Brainhack 101*, hackathons, and brief, informal presentations to share research ideas (see Figure 1).

*Motivations for joining the workshop*

This format has seen significant success within the OHBM community, providing many neuroscientists with an initial exposure to hackathons. As we continue to grow in attendance, however, the authors—as leaders of the Open Science Special Interest Group—hope to refine our workshops to better engage the OHBM membership and encourage the adoption of open science within our community.

This direction is largely driven by feedback that our hackathons are perceived to be oriented towards "power users" who are already strong coders, rather than all members of the community. We therefore hope to explicitly target less code-experienced and junior neuroscientists in future OHBM hackathons. A concern this introduces is how best to measure event success if not with direct outcomes (such as lines coded or papers written), since these measures are likely to miss the collaborative and educational aspects we hope to emphasize. For this reason, we are applying to the CHI 2018 Workshop: *Hacking and Making at Time-Bounded Events* in order to better design for and measure outcomes of future OHBM hackathons.

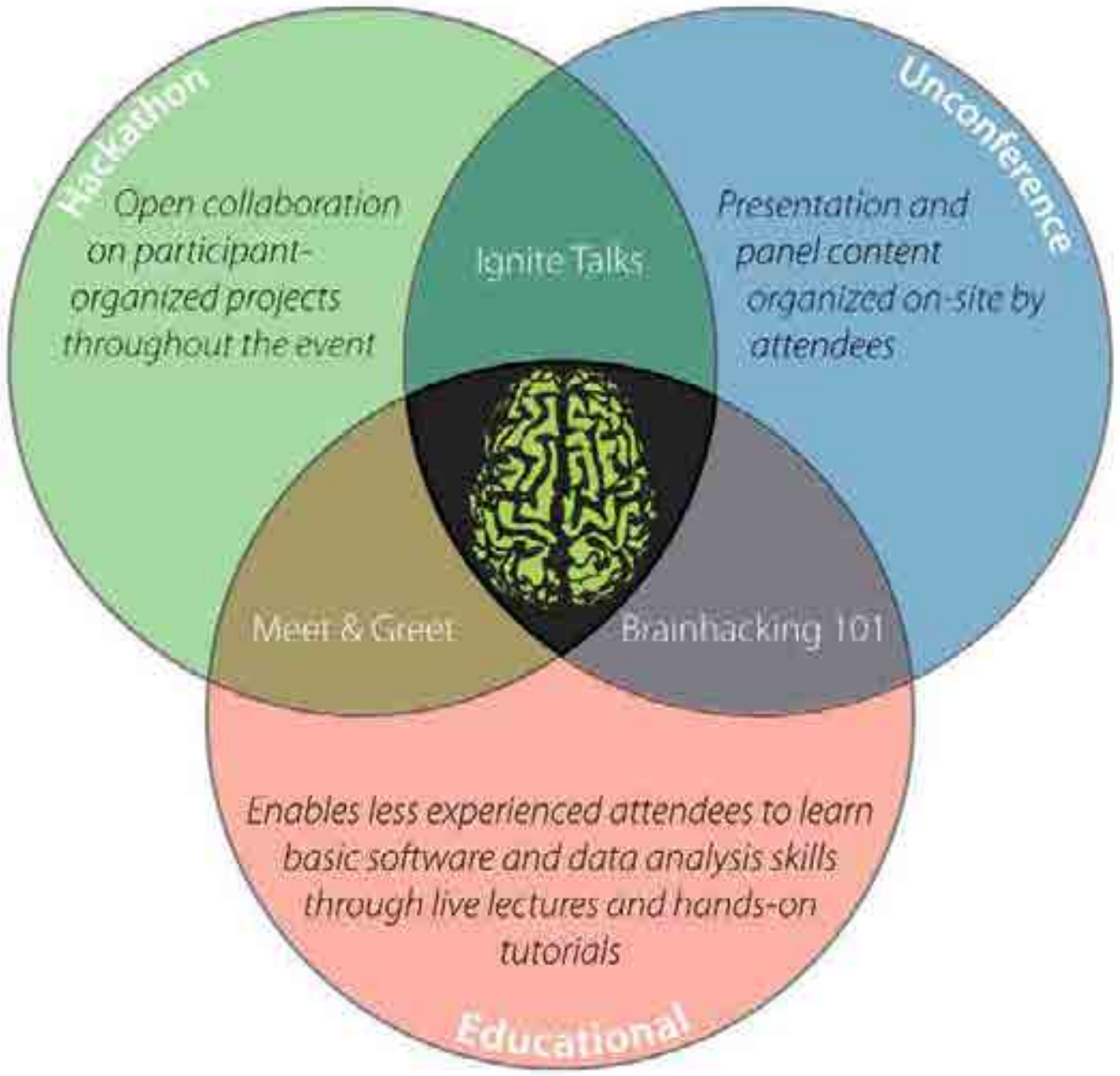

**Figure 1:** The Brainhack recipe combines education, hacking, and informal presentations into short-format events. Figure adapted from [2].

## Themes of Interest

There are two themes in particular which the authors believe will be especially valuable towards accomplishing our goals: *Design Variations* and *Outcome Measurement*.

While we benefit from a well-defined application area and a wealth of experience running both local and distributed workshops, we've been limited by our ability to engage a broad community of scientists and create/evaluate measurable event outcomes.

*Design variations*

The OHBM community consists of scientists from a wide range of backgrounds, including (but not limited to): physicians, physicists, statisticians, psychologists, and informaticians. Although these diverse skills enable interdisciplinary efforts, they also create ideological silos that the OHBM Brainhack aims to break down. An additional challenge in creating Brainhack events is that these scientists arrive with varying levels of comfort in creating and using code.

We have considered variations in the design of our event, such as designing around a theme and on-boarding attendees to that topic, ensuring that all participants have a minimal background in the Brainhack focus. However, it is difficult to find a topic with which at least one scientific background is not already overly familiar, and this approach still does not address the concern of varying technical experience.

We therefore wish to learn how variations in the design or presentation of future OHBM Brainhacks can be made more accessible, increasing the attraction of these events to additional members of our community, without detracting from the experience of established participants.

*Outcome measurement*

Although previous OHBM hackathons have resulted in published collaborations (e.g., [1]), it is unclear that publications are the best outcome by which to measure hackathon success. Related outcomes, such as lines of code written or git commits generated, are not sensitive to attendees' varying levels of experience.

We have considered alternative short-term outcome measures such as participants' self-reported satisfaction after the event or relative increase in comfort with new programming languages and/or tools; however, it is unclear if these provide an appropriate index of success. Long-term outcomes, such as building an inclusive, open-science-focused neuroimaging community, are similarly in need of quantification. One option that has been proposed would be to aim to increase the number of institutions represented at each year's hackathon; however, due to space limitations, at some point this metric would cease to be meaningful. We are therefore interested in exploring alternative methods for quanit the impact of future OHBM hackathons.

## Conclusions

In designing for and evaluating future OHBM Brainhacks, we believe that the feedback and learning opportunities provided at the CHI 2018 workshop will be invaluable for our success. We would therefore be thrilled to participate in this year's event, and thank you for your consideration.

# SIGCHI Extended Abstract File: Hacking Creativity

**Peter Müller**
Technical University Munich
Munich, 80333, Germany
Pet.mueller@tum.de

## Abstract
This abstract introduces my academic and personal background and briefly describes my sociological PhD project on technical creativity and inventiveness, featuring ethnographical field studies of hackathons. Although primarily an empirical research project, I try to conceptualize my findings with abstract theoretical framework concerning the (social production of) technical creativity and inventiveness. I present my preliminary findings on how hackathons, conceived as particular social situations, constitute a fruitful environment for creative, inventive (social) practices. Namely by the openness of communication, the diversity of participants, and their specific, short-period temporality.

## Author Keywords
Hackathons; Sociology; Creativity; Collaboration

## Introduction – Personal and Project Background

After finishing my M.A. in sociology, I have joined the Munich Center for Technology in Society (MCTS), the first German "Science & Technology Studies" (STS) institute, as a graduate student. As a technology enthusiast, I am intrigued by phenomena of technical invention. Hence, I combined this personal interest with my academic profession and am investigating how invention happens and is done, from a sociological perspective. This is a complex question, taking into account the phenomena of inventing and thinking 'new' ideas in the absence of a concrete, plainly given problem. For the contrary scenario is rather a commonplace: people who think of solutions when facing a particular problem, defined by a present problem situation. My empirical access to the social dimension of technical creativity lies in the observation and participation at hackathons and similar events. For they claim to foster creative and inventive ideas.

## Hacking Creativity

As indicated, my research project is empirically focused on hackathons concerning the investigation and understanding of meaning and setup of social situations around technical creativity and inventiveness. Thereby, I follow two different research tracks: one that tries to sociologically understand practices and structures within and around events like hackathons, and how creativity is articulated and used as an end, mean or (symbolic) resource. And one that tries to rethink sociology as an 'applied science' that can contribute to a technical understanding of creativity by providing insights in the mechanics and requirements of 'social creativity'. Creativity is, in general, a very fuzzy term. This can be very confusing for one who tries to use creativity as an empirical, scientific concept. For creativity, even defined as technical inventiveness, still cannot be properly operationalized. Nevertheless, 'technical creativity' is, epistemologically, more disposable for it features the differentiation of 'works/works not'. Furthermore, 'creativity' here is identified with all outcomes that, somehow, surprise, i.e. results that had not been expected. This is also a qualitative operationalization that takes into account not only coded results but also new ways of reinterpreting outcomes. Although a tautology, it is a viable sociological approach for, functionally, it does not matter in how far something might 'actually' be creative but how and whether creativity can be successfully attributed as a (socio-economic) quality. This common conceptualization thus also integrates my two research tracks.

However, this abstract highlights the latter, 'applied' track of my research. I ethnographically studied seven hackathons, applying methods of hidden, participating observation (i.e. taking part without revealing my actual research intentions). During those hackathons, I have learned a lot about the explicit and implicit diversities of hackathons, how they integrate different types of participants, e.g. designers, coders, citizens, different experts, professionals, enthusiasts and stakeholders) and topics (open data, AI, IoT, public issues, media and even music); but also how they differ regarding the setup of issues: giving defined tasks and problems to the participants, offering mere thematic frameworks or assigning the invention/discovery of new (possibly) problems and issues. Although hackathons cover all kinds of open/defined problems/solutions and tinkering, my research is focused on the 'creative' aspect of open, non-defined (not ill-defined!) problems.

The more 'present' and (however)-defined problems are, the more they are accessible to systematic, logical structuring. Hence I suggest that the specific, peculiar, and 'abductive' perspective of sociology can shed some light on processes that do not happen on an explicit level and therefore are incommensurable to classical, deductive(-nomological) research approaches.

I have also been able to analyze some of the social mechanics of hackathonian collaboration that render those events 'creative' and to identify first requirements of creative hackathons that produce something 'new'. I have conceptualized three of those hackathonian creativity features (or requirements) that go beyond the ergonomically informed organization of hackathons like starting with knowledge assessment units (e.g. keynotes): Ideational and communicative openness, instant diversity of participants, and short-period temporality.

Ideational and communicative openness refers to the particular capacity of hackathons to set up a realm of low-threshold compatibility of ideas and communication. The presented, communicated and offered ideas by each participant are likely to be accepted as a (proto-)productive contribution. It is yet hard to explain this particular hackathonian feature but the hackathon-typical emphasis on amusement and fun (I call this: 'funnification') is probably one main reason of this open, casual atmosphere. Although it is no imperative, 'funnification' seems to work like an informal hackathon code of conduct. I have, with all studied cases, observed that even virtually rejected ideas were discussed in a friendly and appreciating way. This results, however, in more than mere social convenience; it is often a vital requirement for unexpected resumptions of ideas, either by trying to somehow integrate such ideas into one's own thinking or by looking for productive aspects of the idea which often differ from the interpretation of the original contributor. Furthermore, original contributors tend to accept and follow those reinterpretations. Those highly irritating and perturbative interaction-structure then leads to project developments and drafts that were not expected by the participating individuals and thus lead them on tracks which each of them alone would not have followed.

However, the utility or integration of such results is another issue. Often (external) hackathonian projects fail to transfer their results into established organization contexts of e.g. greater companies. Vice versa, hackathonian projects work well and even long-term in private or start-up contexts. Hence, this is less a genuine difficulty of scaling time from the event to a continual elaboration but an issue of discrepancies between instant conceptions and established organizations which operate under certain standards.. E.g. one hackathon was won by my team and it was part of our prize that we were given the opportunity to present our project idea to the sponsoring company. However, there was no proper format to integrate our project and work into the static structures of that organization. Although there was interest in our ideas and suggestions, they were almost complete incompatible to this company's technical and organizational infrastructures. In another case, a winning team was invited to cooperate with the hackathon's hosting company. But that cooperation failed as well; this time because the participants could not be motivated to engage themselves in a long-term project that suddenly appeared to be plain work. The company's mostly monetary incentives just did not apply to the interests and expectations of the hackathon team which was rather looking for technical challenges. However, those problems might rather concern 'external' hackathons; unfortunately, I have almost no experience with internal ones.

On an abstract conceptual level, this disadvantage derives systematically from the very creative features of such hackathons: to think off the beaten track. In order to explain this feature of social creativity at hackathons, neuro- or cognition science theories can be used. Those disciplines identify technical creativity with the ability to solve inherently difficult problems (e.g. nine dots problem) which, from their point of view, actually requires to omit certain heuristics or pattern recognition[1]. Heuristics can force individuals to interpret situations in a specific manner that renders a needed solution inconceivable (like drawings lines beyond the assumed bounds of a dot square). While individuals often struggle to omit their common, internalized heuristics, the perturbations produced by the communicative irritations described above resemble the absence of heuristics on a social, inter-individual level, breaching with common plausibilities not within but between the participants by means of openness. However, cognition science conceptualizes creativity contrary to disciplines like management studies or ergonomics which identify creativity with holistic capabilities of overviewing the whole situation, having in mind both the little, subtle details and the larger frameworks[2]. Contrary to the concept of technical creativity I have introduced, ergonomic creativity is focused on large-scale innovation projects instead of micro-events of invention. While innovation projects rather require tact and foresight for enterprise and thus have to understand the social meaning of situations in order to respond appropriately, invention seems to need quite the opposite. With hackathons fostering and focusing on this latter type of creativity, the former, project-management orientated perspective rather falls aside and can thus be missing in greater innovation scenarios. Thus, hackathons underlie an inherent trade-off between inventive, technical and innovative entrepreneurial creativity. However, this effect might be regulated, to some extent, by providing participants with concrete, given assignments so that given frameworks can be defined and taken into account. There is a continuum between open hackathons without defined problems and those that feature specific technical challenges. However, finding the right equilibrium of inventiveness and integratability can be tough. The instant diversity of participants also amplifies the perturbative quality of interactions. Diversity not in terms of social inclusion; actually, especially for civic hackathons, like open data day events, inclusion and (self-)selectivity is a notable issue. Here diversity means a micro-level heterogeneity in terms of experiences, backgrounds, skills, mindsets and things taken for granted. Intriguingly, this feature corresponds also with ergonomic concepts of technical creativity [3]. The plain diversity of participants increases the likelihood of irritations, of facing unexpected and (personally) unstandardized styles, contents and logics. In sociological terms, stressing Luhmannian system theory [4], they lack of 'moral', i.e. binding precedencies and established interaction orders. The thus constituted heterogeneity within interactions, again, resembles the demanded heuristic breach. It furthermore contributes to the aforesaid qualities of openness: Participants often do not know each other. Hence, they are also unable to assess each other’s capacities. As a result, hackathons provide a practical application of the philosophical ‘principle of charity’ [5]: Every statement is interpreted in the most useful and reasonable way. However, since there is no common interaction routine, discrepancies of communication are often reproduced by attempts of benevolent interpretation. This continual bridging of communicative differences results in unexpected, thus creative, interactions. The diversity and initial anonymity of participants also amplifies the funnification since the success of hackathon events, not in terms of productivity but in terms of an ‘awesome happening’, cannot be realized by mere means of organization but highly depends on the participants themselves, who thus tend to comply in the performance of casual unconventionality. Also, the said ‘lack of moral’ renders explicit declarations in terms of ‘funnification’ more influential concerning the actual conduct of hackathons.

The short-period temporality of hackathons means more than general time-boundedness, but their very own rhythm. All observed hackathons that featured schedules with multiple intermissions like keynotes, presentations, lunch, joint events, etc., produced many ‘creative’ outcomes, and vice versa. Those event schedules function as ‘tacit project schedules’: hackathon teams tend to use these schedules to temporally structure their own project work. They apply them as binding deadlines for work steps, e.g. having a concrete idea at lunchtime, finding a technical solution until the afternoon presentation, etc. This is not only crucial for a proper project execution but often forces hackathon teams to deliver premature results which impels them to deal with unforeseen situations . Eventually, this can result in serendipity, when the actual significant project outcome is the solution of such a sudden problem instead of the initially appointed objective – e.g. my team wanted to create a passenger counting system for streetcars but suddenly had to find a way of determining directions of movements using only one ultra sound sensor (because more would have interfered with each other) which happened to be our main achievement. To understand this social creativity feature, imagine a person who is assigned to do a radical new art work within infinite time. This person will most certainly end up with a product that is exactly like one’s imagination of ‘radical new art’ (and probably never finish). Give that person temporal bounds and the result might inevitably appear as ‘radically new’ because there was no time to adapt it to given concepts of novelty. [6] This virtual reproduction of creative cognition is a particular important feature because it even works for rather homogenous and routinized groups.

# Breaking Gender Code: Hackathons, Gender, and the Social Dynamics of Competitive Creation

**Siân JM Brooke**
Oxford Internet Institute
University of Oxford
1 St Giles, Oxford, OX1 3JS, UK
sian.brooke@oii.ox.ac.uk

## Abstract

The hacker is 18-25-year-old, college educated, white, and male. This is a stereotype that not only accurately depicts much of computing professionals but creates active barriers for anyone who deviates more than slightly from this expected, and legitimate, identity. Hackathons are spaces in which the dominant identity is most visible. This project will account for the apparent absence of women in creative coding – e.g. hackers - by examining the forums of hacker culture; gaining skills, sharing knowledge, and collaborating to become technically proficient in programming. I will conduct an ethnography of gender-based collaboration in the rituals of hackathons. I hypothesise that hostile masculinity both masks and deters women's participation. It will be proposed that collaboration and mixed gender spaces are more productive, and that the domination of hackathons by 18-25-year-old, college educated, white, males deprives hacking and hackathons of substantial technical talent. The proposed project contributes to debates on gender performance and representation in hackathons and associated computational cultures.

## Author Keywords

Gender; Masculinity; Visibility; Representation; Programming; Hacking

## ACM Classification Keywords

H.5.3 Group and Organization Interfaces; Collaborative computing, Computer-supported cooperative work.

## Background

My doctoral research at the Oxford Internet Institute (Oii) focuses on the gendered disparity of women's visibility in hacking culture in online and offline spaces. The spaces of identity politics in computational culture is a field I have researched beyond subversive and creative hacking prior to my doctorate. My undergraduate degree thesis focused on the power relations of sexuality and gender in geographical dating apps, such as Tinder and Grindr. The project examined the radicalisation of the self as an overt object and subject of consumption and disposability, looking at how desire in the dating market is repackaged in the refinement of distaste, such as "no fats, no femmes" or "over 6ft only". My Master's thesis focused on the performances of gender in anonymity on the social networking site Reddit.com. It surveyed: Advice Animals meme genre (r/AdviceAnimals); anonymous self-portraiture for the purposes of crowdsourced abuse (r/Roastme); and collective action by volunteer content moderators (the Reddit Revolt). My background focuses on gender performances and representation in computational culture.

## Workshop Motivation & Themes

The CHI hacking and time-bounded events workshop will give me the chance to critically examine methods for investigating spaces of hacking and creative coding. The project is divided into three ways in which gender in hacking and creative coding might be understood. They are directed at differing levels of technical competencies and context: Knowledge sharing; anonymous participation; and the lived experience of coding. Firstly, knowledge sharing will be evaluated through those seeking and providing educational materials and advice on Python programming. Secondly, the invisibility of gender, and thus assumed maleness, of accessible and subversive technical spaces shall be examined through natural language processing on dark web forums and expert-exchanges such as StackOverflow.com. Finally, competition, cooperation, and visibility as aspects of the creative coding identity will be examined through an ethnography of hackathons.

The participant observation made possible by hackathons is a crucial aspect of this workshop. The opportunity for research exchange, sharing of practical experiences, and exploration of potential applications of the research is unique. The workshop provides an exciting opportunity to explore the methodological approach and implication of my research. As such, the primary themes of interest are the examination of the theoretical spaces of hackathons and the mediation of interactions. In looking to the gendered etymology of these spaces, I advocate an approach founded in critically assessing the telling of history, that is, explanation of women's absence beyond origins. The historically based approach shall be built on to form an understanding of the ecology of, and interaction within, these spaces in which the anonymity of online participation is revoked, in favour of an embodied experience. The project will focus on gender in the ritualism of hackathons, and how performances of gender are reshaped in a culture that relies on immateriality and disembodiment.

## Brogrammers

Overt sexism and hyper-masculinity has emerged in hackathons and similar social computational events in recent years; the brogrammer. Reiterating internet culture's fondness for portmanteaus, a 'brogrammer' is a combination of the term 'bro' (as in brother) and programmer. In April 2012, the term brogrammer erupted into the public consciousness due to *Mother*

*Jones* article titled *'Gangbang Interviews' and 'Bikini Shots': Silicon Valley's Brogrammer Problem.'* The article pointed to how conceptions of geek-ness have been recast with "a competitive frat house flavour", exhibiting the hypermasculinity commonly associated with the culture of fraternities. There are many notable instances of such hetero-masculine behaviour, usually justified through employing a measure of humour. At the TechCrunch Disrupt 2013 conference the fictional mobile app 'TitStare' was pitched. The app was intended to be satirical, based on images of men staring at women's breasts. Such discourses are typically of the masculinity that surrounds the emergence of the brogrammer culture as an evolution of the uncool 'geek'. Such evolution points to how masculinised discourses and misogynistic actions are evidence of a replicating pattern in computational development, rather than the growing pains of a maturing field.

## The Oblivious Hacker

Hackers are radicals, new revolutionaries, and rebels with a cause (Coleman, 2015; Taylor & Jordan, 2004). They are on the frontier of computing, yet, despite claims to a meritocratic community there is less women in hacking than other computing cultures (Adam, 2004; Jordan, 2016). In exploring the depths of online anonymity, a challenge to the dominant masculine narrative has not been voiced. Scholarship often points to the dearth of women in creative technical labor, with minimal causal exploration of the absence (Jordan, 2016; Levy, 2010). However, Tanczer (2015) has shown that women do participate in the social spaces of hacking online, masked by the obliviousness of male discourse. My doctoral thesis addresses how such women negotiate the assumed gendered hostility of hacking's technical spaces, and how femininity may be rendered invisible in a bid for legitimacy. The following section outlines the roles and stereotypes attributed to women in computational culture than can account for a historical absence of a visible critical mass, despite prominent individuals.

### *The Social Justice Warrior*

Eric S. Raymond (2015) dismisses those who critique the male dominance of programming as Social Justice Warriors (SJWs) and the "enemy" of the hacker's "cult of meritocracy". Whilst SJWs can be taken to refer to any individual with socially progressive ideas surrounding identity politics, in computational culture it is a term more commonly used pejoratively. Its negative use gained prominence during the Gamergate[1] controversy, with Massanari (2015b) highlighting it as an epitome of toxicity in technology culture. A SJW is motivated by personal validation rather than political conviction. In providing a voice counter to the painting of hacking as meritocratic, feminist voices calling for more women in computing can be perceived as a moralising project or imposition (Ahmed, 2010). Such dominant perspectives leave little room for critical analysis of women's absence, and examination of the benefits of their inclusion in technical spaces and events.

[1] Concerns issues of sexism and progressivism in video game culture, stemming from a harassment campaign of women in the industry beginning in August 2014, nicknamed GamerGate.

*The Hacker's Wife*

Adam (2003) proposes that scholarly works have done much to marginalise the role of women in hacking, whilst pointing to their absence. Levy's (2010) work *Hackers: Heroes of the Computer Revolution* begins with a "Who's Who" of hacker culture. The list references fifty-two men, ten computers, and three women. All the women are discussed in terms of being a wife, permitted into the physical spaces of hackers by marriage, despite their own hacker credentials. In reflecting on the history of hacking, Adam (2003) asks how the rhetoric of equality and open meritocracy arose in the first place. Himanen's (2001) *The Hacker Ethic* is proposed as a source of such a narrative, placing little emphasis on actants (Coleman, 2015). Adam (2003) holds that material barriers to women's entry into hacking are rarely noted, such as the long hours and late nights of hackathons which are juxtaposed to narratives of domesticity and childcare in femininity. In seeing hacking as being normatively masculine in the retelling of history, the justifications for women's absence default to precedence: if there were never women in technical spaces, there must be an explanation – usually based on pseudo-psychology or 'personality differences'.

## Breaking Gender Code

Women have founded projects that reimagine the physical spaces of hacking. Stemming from the ideology of and discussion on *The Geek Feminism Wiki* (2008), hackerspaces have physically emerged as spaces in which femininity in hacking is normative and necessary. A hackerspace is an arena in which individuals with a common interest in computers or technology can socialise and collaborate, operating like an informal hackathon. Since 2010, hackerspaces have spread quickly across the USA, influenced by the dominant German based model in their understandings of openness based on interest, rather than having a pre-proven ability, unlike *The Hacker Ethic*. In recent years, several feminist hackerspaces have emerged, such as the 'Hacker Gals' in Michigan (est. 2014). Whilst scholarly contributions address the closed off nature of the hacking community and privilege of meritocracy, little work addresses the progressive and feminist politics of alternative spaces (Toupin, 2014).

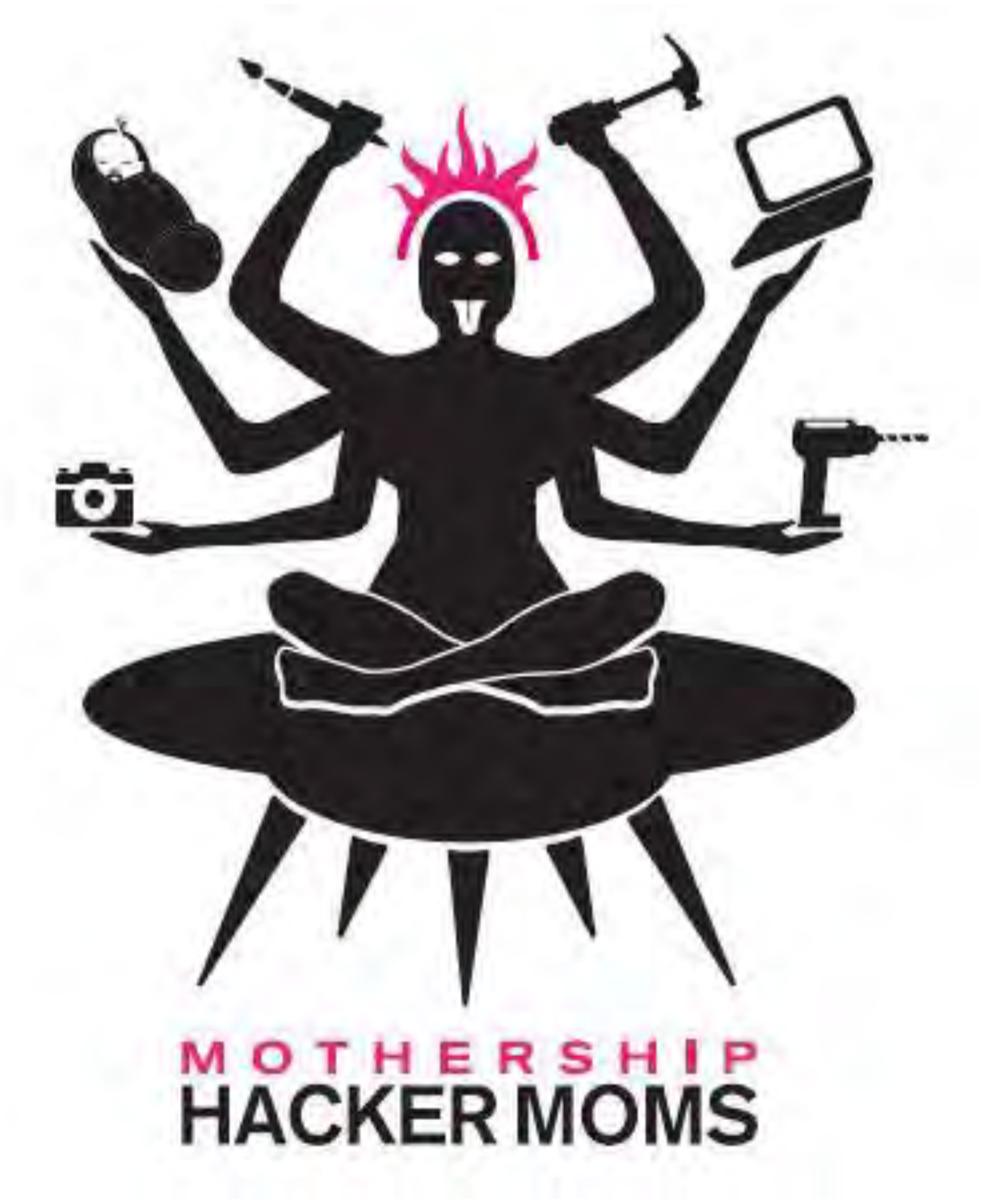

**Figure 1**. Mothership HackerMoms was the first-ever women's hackerspace in the world founded in April 2012.

*Organised Resistance: Hackathons & hackerspaces*
Hackathons are fundamental to the culture of inclusive creative coding in hacking. Coleman (2010, p. 49) argues that online interaction or "networked hacking" should not be seen to displace physical interaction, rather, the "two modes powerfully reinforce each other". Hackathon's are ritualistic and "ingrained in the ethos of coding" (Leckart, 2012, p.109 cited Jones et al , 2015, p. 325). Jones (2015) proposes that they are not merely occasions of technical work, but rather they express ideological tenants such as individual competencies. For example, 'Hackermoms' is a women's only space which promotes a 'DIY ethic' with on-site childcare encompassing domesticity into hacking. In a more traditional sense of time-bounded technical events, hackathons are emotionally charged, as interpersonal relationships manifest and the prosaic nature of hacker's online social world is ritually embodied. In these transient physical spaces, a hacker's identity as a woman is made salient no longer digitalised and anonymous. Women also organise ideologically in these spaces, often developing informal procedures to encourage female participation (Coleman, 2010). Thus, spaces of resistance, and the code they create, are fundamental to the analysis of gendered practise in creative coding.

## Deconstructing Resistance

It could be speculated that visibility in hackerspaces continues in the defining of femininity corporeally; one may only be accepted as a woman hacker in the presentation of a female identifying body[2]. Thus, it would be problematic to see feminist hackerspaces and women-only events as cleanly progressive spaces of gender politics; the absence of masculinity defines women-only spaces. As such, masculinity's cultural privilege and normativity in relation to the feminine other is reinforced, and the social hierarchy of hacking as a male space is recreated.

## Conclusion: Feminine made visible

In opposition to equality through gendered segregation, what could be considered as meritocratic in hacking spaces would be androgynous collaborative spaces. Research points to how gender balanced teams are more productive on a collaborative level (Song et al, 2015). Female participation in group assignments raises the performance levels of other members of the group, even when gender is not disclosed. Song et al. (2015) speculate that the reason for such improvement is that women act more cooperatively. Terrell et al.'s (2017) study into gender in proposed changes to a software project's code, documentation, or other resources found that women's contributions are accepted more often than men's if their identity is unknown. However, when gender is made salient, women's contributions are 15% less likely to be accepted (Terrell et al., 2017).

Work on gender in computational culture is necessary as women's participation in computing culture and professions has plummeted in recent years. For instance, in 2013, only 26% of computing professionals were female, down from 35% in 1990 (Corbett & Hill, 2015). This decline in participation is unique to computing, with other STEM fields seeing a moderate increase in the same period (Corbett & Hill, 2015). Assessing and increasing the gender balance in hacking can thus serve to benefit creative coding cultures, as even fewer women are perceived to be involved in hacking than other computing fields (Jordan, 2016). Therefore, whilst women-only spaces do provide an arena for women's engagement in creative coding, the gendered division and emphasis on physical visibility reaffirms the narrative of hacking as a masculine practise.

[2] The conception that an individual can only be believed to be female upon showing a feminine body is a well-known 'rule' of online anonymous forums, often expressed in the adage "Tits or GTFO" (get the fuck out).

# Hackathon Team Leadership: Supporting Innovation through Teaming at Time-bounded Events

**Eureka Foong**
Northwestern University
Evanston, IL 60208, USA
eureka@u.northwestern.edu

**Elizabeth Gerber**
Northwestern University
Evanston, IL 60208, USA
egerber@northwestern.edu

## Abstract
Organizers of time-bounded events bring together professionals to form interdisciplinary teams and generate breakthrough technological solutions. While such teams need support to innovate, event organizers invest more effort managing logistics rather than supporting individual team processes critical to innovation. Through a six-week participant observation of a weekly civic hackathon in Chicago, we analyzed the extent to which individual team leaders follow Edmondson's Framework for Effective Teaming. While team leaders had opportunities to define meaningful goals, they faced challenges encouraging diverse perspectives, seeking rapid feedback, and organizing knowledge gained. Event organizers can support individual teaming by modeling behaviors that foster psychological safety, coordinating community partnerships, providing alternatives for recording team progress, and setting success criteria around project iteration. Our work fills an important gap in extant research that describes the goals, outcomes, and operations of time-bounded events, rather than individual team leadership processes that lead to innovation.

## Author Keywords
Hackathons, civic, leadership, teams, collaboration, feedback, community, innovation

## Introduction and Background
Corporate, non-profit, and government organizations increasingly turn to time-bounded events to solve challenging, ill-defined business and societal problems [10]. Each year, Microsoft hosts one of the largest private hackathons, inviting employees across 75 countries to develop novel technology [12]. Internet of Elephants, a social enterprise based out of Kenya and the U.S., hosts hackathons to support the conservation of endangered animals [16]. In 2016, more than 3000 people participated in GovHack, a hackathon using open government data sponsored by the Australian government [17]. While the goals [2], duration, and frequency [6] of these events vary, event leaders share the belief that collocating talented professionals with diverse ideas and skills will produce creative products, solutions, and business ideas.

Although exposure to diverse ideas and skills is certainly essential to individual team innovation, alone it is insufficient [7]. In a study of innovation teams, leaders actively encouraged capturing and testing diverse ideas to learn more deeply about team members' skills [7]. Based on 30 years of empirical research, teams researcher Amy Edmondson [4] finds that individual teams that produce innovative solutions have leaders and teams that prioritize the following five processes:

- **Aiming high:** Leaders set challenging, meaningful goals that spur and sustain action, such as building an energy efficient, symbolic stadium for the Beijing Olympics in 2008.

- **Teaming up:** Leaders prevent communication breakdowns in interdisciplinary groups by being genuinely curious about others and creating a space of psychological safety [5]. For example, web developers working with designers should feel comfortable clarifying website functionality instead of assuming the designer's intent.

- **Failing well:** Innovative teams maximize the number of "intelligent" failures (p. 87) by continuously experimenting, such as testing the demand for an upcoming product with a small group of users.

- **Learning fast:** Leaders learn fast by fostering psychological safety so that teams can easily diagnose problems, generate and test solutions, and reflect on knowledge gained from testing. For example, product design teams at IDEO used multiple observations of people sleeping to generate quick prototypes for an innovative mattress [3].

- **Repeating the process:** Innovative teams form habits around failing and learning to improve on their ideas, such as by applauding intelligent failures and holding brief meetings to reflect on what the team has learned.

While these leadership behaviors are critical in teaming for innovation, hackathon event organizers currently pay little attention to these individual team processes because they are busy addressing other aspects of the event. In our personal experience leading five hackathons in the last 10 years, we spent the majority of time identifying venues, purchasing food, recruiting participants and partners, and scoping projects, instead of managing effective team leadership behaviors and team interactions. The lack of attention to these activities is echoed in the accounts of other event organizers (i.e., [15]). Time constraints that limit the ability to form social ties [13] make it even more crucial for event organizers to facilitate effective leadership. To date, scholars and practitioners have described the goals and outcomes [1,8,10,11] and knowledge-sharing processes [9] of time-bounded events without critically examining the individual team processes that lead to innovation.

Our goal is to bring awareness to individual team processes that influence innovation and offer ways that event organizers can plan for more effective individual teaming. We use a six-week participant observation of a recurring civic hackathon to understand the extent to which a team and its leader follow Edmondson's Framework for Effective teaming. While the event provides opportunities to create personal and challenging goals, an individual team leader can still face difficulties encouraging participation from new members, seeking rapid feedback, and organizing the knowledge gained across meetings. To produce more innovative solutions within individual teams, we recommend that event organizers model behaviors that promote psychological safety, help individual team leaders connect with community partners, offer alternative solutions for recording team progress, and incentivize teams to iterate on their solutions.

## Description of the Field Site

Our analysis is based on the first author's six-week participant observation of a weekly civic hackathon in Chicago from January to April 2016. The event, which is held from 6:00 p.m. to 10:00 p.m., is part of a network of hackathons in the U.S. focused on engaging citizens to help solve civic issues such as improving government transparency and preventing bacterial outbreaks. In this paper, we define event organizer as a person who plans the larger hackathon event and team leader as a person who leads a project team within the hackathon. All names used in this paper are pseudonyms for actual participants in our study.

Every week, 40-150 working professionals, freelancers, and students in the technology industry attend the event, at least half of whom are return visitors. The event begins with an hour-long lecture from a guest speaker who has worked with government data. Next, individual team leaders from previous weeks invite attendees to join project "breakout" groups. Any attendee of the hackathon can propose a project and become a team leader; the event organizers, Peter and John, do not instruct teams to follow a work process or set specific goals. The first author joined Justice in the City, a project team led by a nonprofit attorney named Kelly. Each week, the team worked to reduce the rate of recidivism by providing information that would prevent parolees from violating parole. Because attendees are free to join and leave teams at any time during the hackathon, the size of the team varied between three to seven people each week. At the end of the six weeks, team members had developed a deep understanding of the design problem but had just started creating content for one solution.

After each hackathon, the first author wrote field notes and memos, drawing themes through continuous reflection on the data as well as academic literature [14]. The first author attended project meetings for the same Justice in the City team and observed chats in the team's online communication channel, *Slack*, with permission. The first author also attended one leadership council meeting that included John, an event co-organizer, and four men who had previously attended the hackathon. All attendees were welcome to join the meeting, which occurred during the hackathon. Neither the co-organizers nor the attendees of the hackathon were paid; however, Peter, an event co-organizer with formal training in public administration, was paid for his work managing the larger network of civic hackathons. As part of his role, Peter prepared leadership guidelines for other hackathon event organizers. The author also interviewed Peter about his motivations for joining the civic hacking movement.

## Results

While we found evidence of activities that supported processes Edmondson's framework [4], we also found significant challenges for team leaders to balance multiple roles while producing a testable solution. In spite of this, participants at the leadership council meeting spent little time discussing the progress of individual teams, discussing instead issues of event funding, creating promotional materials to attract sponsors, and connecting with local political officials. Below we discuss the extent to which one team's experience followed Edmondson's [4] framework.

**Aiming high** refers to setting challenging and meaningful goals that unite and drive teams to action. The event organizers excelled in showing individual team leaders examples of challenging goals presented by charismatic guest speakers. For example, one guest speaker rallied attendees to help the government update its technology, making "a 45-year journey in four years." The structure of the hackathon, which allowed any attendee to pitch or join a project team, gave individual team leaders the freedom to create personally meaningful goals. For example, one team leader's goal was to "stop people from getting sick" by helping predict levels of bacteria in the water. The structure of the event provided many examples of and opportunities to present challenging goals, yet the goals themselves were largely dependent on the presentation skills of individual team leaders.

**Teaming up** refers to managing team interactions so that members feel safe to share ideas and failures and prevent miscommunication. Kelly modeled leadership behaviors that promote psychological safety, such as acknowledging when she lacked understanding about the parolee community and asking follow-up questions when team members shared ideas. Nevertheless, new members contributed little to the discussion because they were unfamiliar with the team's work history. At the first meeting, the author felt useless because returning members referenced tasks for a different project. Only when she asked how she could help did the leader ask about her interest in typing notes.

**Failing well** refers to incrementally improving team processes by testing the team's assumptions. Kelly tested the team's assumptions about challenges parolees face by inviting people who have worked with parolees to meetings. During one meeting, team members learned that finding employment with a criminal history is one of the most difficult components of reentering society. This prompted the team to design a more complete guide on reentry, instead of an application that merely alerts users of possible parole violations. However, because Kelly coordinated these meetings, the team did not test their assumptions about parolees' needs when guests could not attend.

**Learning fast** refers to rapidly creating and testing prototypes to better understand problem constraints. Because it was important to generate buy-in from local organizations, Kelly focused the team's initial efforts on need finding rather than on the creation of rapid prototypes. For example, the team learned that a hotline for employment assistance was unreliable after a team member called the hotline during the meeting. Only in the last two weeks did team members begin writing content for a guidebook for recent parolees. The team may have had more opportunities to develop their prototype with more frequent meetings and additional support from event organizers to gather information about the parolee community.

**Repeating the process** refers to the team forming habits around the four previous activities. Because the hackathon was held weekly, team leaders had opportunities to repeat this learning process. Yet, it was difficult for Kelly to document the work and research done *outside* of meetings. Because the team was focused on need finding and creating written content for a guidebook, they could not rely on GitHub code repositories to document their progress as other teams did. The first author introduced the team to *Trello*, a program for creating visual task boards; however, few members had updated the *Trello* task board during the week and valuable time was spent updating online documents at the hackathon.

## Recommendations for Supporting Individual Team Leaders at Hackathons

Given the range of responsibilities of individual team leaders at civic hackathons, we highly encourage event organizers to support them in conducting effective teaming. Within teams, leaders can choose to prioritize support in areas they are most lacking; for example, if a team often receives new members, the team leader may need to be reminded more frequently to invite these members to participate. Our study suggests the following as opportunities for improvement:

- **Encourage psychological safety in teams, particularly for new members:** Team leaders may not be aware of the need to invite new members to participate in discussion. Hackathon event organizers should consider modeling this behavior from the beginning, for example by presenting a brief video of team leaders inviting new members to share ideas.

- **Help team leaders form and maintain connections with community partners:** To generate solutions with a high chance of being adopted by the community, civic projects must consider the needs of existing community organizations. Hackathon event organizers can help individual team leaders connect with community partners and coordinate transport of partners to the event.

- **Provide alternatives for documenting team progress beyond code repositories:** To iterate on solutions, teams that gather knowledge of a user community must be able to document their progress outside of code repositories. Event organizers can support team leaders by providing alternative tools for organizing their knowledge.

- **Set criteria for success around team processes:** To incentivize teams to seek feedback and iterate, event organizers should ask teams to describe how their solution has evolved in light of user feedback.

## Limitations

Edmondson's framework assumes that teams can develop over time. Although we observed a recurring hackathon, the team may not have had the time to practice these leadership processes because team members were free to join and leave during the hackathon. Another limitation of this research is the focus on one team's experience at a recurring hackathon, which may not generalize to other hackathon models. For example, at a one-time hackathon, team leaders may not face as many difficulties integrating new members as all members will have the same work history.

## Future Work

In future work, we will test the effectiveness of these recommendations by conducting additional interviews with event organizers. Together with organizers, we will develop materials and activities that address these recommendations. We will use surveys and participant observation to assess how these activities influence the frequency of participation of new members, the involvement of community partners, a team's ability to iterate using previously acquired knowledge, and the effectiveness of solutions produced by teams.

## Conclusion

Exposure to diverse perspectives is necessary but not sufficient for innovation. By analyzing one hackathon team's experience using Edmondson's Framework for Effective Teaming, we find several opportunities for event organizers to support individual teaming: helping team leaders encourage participation from new members, seek rapid feedback, and organize their team's knowledge. Even if event organizers recognize the importance of team processes to innovation, producing appropriate interventions may be challenging due to time and attention constraints at hackathons. We hope to gain valuable feedback about the feasibility of our recommendations in the context of time-bounded hackathons so that we may support innovation.

## About the Authors

Eureka Foong is a PhD student and a Segal Design Cluster Senior Fellow at Northwestern University studying crowdsourcing applications that support design education. She has spoken at TEDx about user research and problem solving at civic hackathons.

Elizabeth Gerber, PhD, is the Charles McCormick Professor of Design at Northwestern University and Faculty Founder of Design for America, a nationwide network of student design teams. Her current research focuses on understanding the work of social innovators in online and offline communities.

## Acknowledgements

We would like to thank the organizers and participants in our field study, Professor Gary Fine in the Department of Sociology at Northwestern University, as well as our colleagues in the Delta Lab for their helpful feedback and NSF Awards IIS 1217225 and 1530837.

# Collaborative community coding events in open source biological research

**Brad Chapman**
Harvard Chan School Bioinformatics Core
http://bioinformatics.sph.harvard.edu/
bchapman@hsph.harvard.edu

## Abstract

The Open Bioinformatics Foundation Codefest is a multiple day collaborative working session. Codefest provides a venue for real time collaboration between researchers who have established relationships through decentralized open source work, as well as a place for new developers to integrate with a welcoming community. We'll describe the unique collaborative structure of Codefest and discuss approaches to help improve the event to sustain long term collaborations while training a diverse set of attendees.

## Author Keywords

bioinformatics, collaboration, training, diversity, open source, biology

## Background and motivation

The Open Bioinformatics Foundation (https://www.open-bio.org) is a community of scientists creating open source code to solve biological problems. A yearly conference, started in 2000, provides the opportunity for in person discussion and presentation on technical work about code development and biological analyses.

In 2010, we recognized a need for a more practical hands on working session in addition to the conference and developed a two day coding event called the OpenBio Codefest (https://www.open-bio.org/wiki/Codefest). This event contin-ued the past 8 years in a wide diversity of locations, with the most recent taking place at a non-profit, community-run hackerspace in Prague (https://www.open-bio.org/wiki/Codefest_2017).

This summer, we've combined with another open source community to create a full bioinformatics community conference including dedicated training, traditional conference talks and four days of collaboration (https://gccbosc2018.sched.com/).

Our goals at the CHI 2018 Hackathon Workshop are to describe the unique collaborative structure of Codefest, connect with other organizers building long term community relationships through collaborative events, and learn about how we can improve at training a diverse set of attendees.

## Collaborative event design

Codefest initially started as a space for community members who were already collaborating remotely to sit together and work. Over time, it expanded to better incorporate new members into the community by serving as a fun and open environment for sharing work and meeting like-minded researchers.

We plan to share some unique design elements we've learned in organizing Codefest:

- The value of collaboration over competition. Codefest has no prizes or competitive structure, and instead focuses on producing useful practical code that we can share at the associated conference and more widely through blog posts and scientific papers.
- The power of self-organizing groups. We do not predefine the agenda for Codefest and let the attendees suggest areas of focus and then provide introductions so working groups can form. This allows newer community members to work alongside more experienced developers in areas they'd like to learn, and to allow the community to shift focus with new technologies and approaches.
- The advantage of in person discussion for developing interoperability standards. One successful outcome of Codefest have been the development of tool communication standards which allow different communities to share development resources. Like other projects at Codefest, standards creation happened organically due to the need for larger projects to be able to better to re-use analyses. These standards have been essential for forming new long term collaborations for building necessary research infrastructure.

## Training and community building

The biological problems we work on at Codefest require collaboration across a diverse set of research areas. We're continually focused on strengthening and improving our community and are hoping to learn from other organizers at the Workshop:

- How to attract a more diverse set of community members. Like many programming and bioinformatics conferences, we struggle to attract a diverse crowd of attendees. As a result, Codefest can feel intimidating or unwelcoming to those outside the community. We've received universal praise that we're welcoming once overcoming that initial hurdle, but would like ways to project this welcoming attitude so under-represented researchers feel comfortable investing their time and expertise at Codefest.

- Incorporating teaching and training into the content of Codefest. As we've increasingly tried to attract new community members, we've developed the need to help integrate them into the community. In many cases, new members will be experts in some areas but not in the projects or languages under active development at Codefest. We need to develop methods to quickly get them comfortable and productive so they can contribute within a reasonably short time frame. At this year's upcoming Codefest we plan to evaluate the impact of having dedicated training prior to the collaborative hands on event.

- Scaling events to incorporate new members and approaches. As we actively recruit new attendees we're running into the issue of figuring out how to support them at larger scale. Our approach of having a few mentors who make connections and provide orientation on projects will need improvement if we're successful in recruiting new, diverse attendees.

Attending the Hackathon Workshop is a chance to share areas where we've been successful and to learn how to be better organizers. We hope to continue to expand and improve Codefest and related events for the open bioinformatics community.

# CHI 2018 Hackathon Workshop Submission: Initial Findings from Hackathon Trace Data

**Erin Hoffman**
3rd-Year PhD Student
University of Washington
Seattle, WA 98195, USA
erinrhof@uw.edu

## Abstract

Competitive overnight coding and prototyping events known as “hackathons” represent a large and growing phenomenon, with tens of thousands of participants and millions of dollars spent each year. The social computing community has made progress in understanding certain hackathons through ethnographic and research-through-design methods, but has not yet answered a number of important questions about hackathons as a whole. This paper analyzes hackathon-related trace data from the websites Devpost.com and GitHub.com to address some of these questions about the geographical distribution of hackathons, the distribution of projects across hackathons, long-term hackathon outcomes, the content of hackathon projects, and hackathon participant networks. This analysis has generated four data visualizations and seven insights from those data visualizations. These visualizations and insights form the first step in a larger mixed-methods investigation.

## Introduction

I have been a hackathon attendee and organizer since 2012 and am in the process of conducting a mixed-methods investigation of hackathon trace data. I am thus both a hackathon practitioner and a hackathon researcher. I hope to contribute to the 2018 CHI "Hacking and Making at Time-Bounded Events" workshop in the contexts of *design variations, short-term and long-term outcomes, practical support for hackathon organizers,* and *theoretical space of hackathons*.

*Experience as a Hackathon Attendee and Organizer*

My first hackathon was a Windows Phone hackathon in 2012, which was hosted by Microsoft representatives and took place in Michigan State University's Computer Science and Engineering department conference room. It was a small event with Jimmy John's catering where I first heard someone use the term "hack" to mean "a clever but brittle quick fix" rather than "a data breach". My teammate and I were the only women in attendance, and we won the people's choice award (and two Windows Phones and an Xbox) for our prototype (which tried to improve Bing Maps on campus). I left feeling like I hadn't wasted my Saturday, but didn't anticipate going to many more hackathons.

When a close friend persuaded me to go to HackIllinois in 2014, I changed my mind. Instead of 30 people and a handful of projects in a conference room, HackIllinois had hundreds of attendees and dozens of projects spread out over multiple campus buildings at the University of Illinois Urbana-Champaign. We were alternating between hours of coding and debugging and sleeping on the hard floor of a brightly-lit classroom, but the warm community and collective effervescence [1] were delightful. I also felt that I had "leveled up" as an engineer by learning how to use tools we had never used in class (like REST APIs and GitHub), and by networking (and playing cards) with employees of elite tech companies. This hackathon experience made me realize that I really enjoyed this type of large hackathon, which is also known as a "collegiate hackathon" and is often a member event of an organization called Major League Hacking.

Since HackIllinois in 2014, I have competed and mentored at dozens of collegiate and other hackathons across the United States, and in the past year also mentored at a hackathon in Abu Dhabi, UAE. I also helped to found SpartaHack, Michigan State University's collegiate hackathon, which welcomed 300 participants in 2015 and successfully completed its 600-person fourth edition in 2018. More recently I helped organize DubHacks, the University of Washington's collegiate hackathon. As of February 2018 I have experienced thousands of hours as a hackathon attendee, volunteer, mentor, and organizer.

*Research Interest in Hackathons*

When I began my PhD, I didn't intend to study hackathons. I planned to build tools for moderators on sites like Wikipedia, Reddit, and Facebook – tools that would use natural language processing to provide a quantified basis for removing harmful content. After spending more than a year investigating possibilities, I concluded that accuracy standards in computational linguistics are too different from accuracy standards in online content moderation for this kind of tool to be viable with today's computational linguistics tools [2]. Maybe I will return to this project in a few years when hackathons are better understood by the research community!

I first turned to hackathons as a research topic when I took a qualitative methods class. I conducted a partial Grounded Theory [3] participant observation study of my fellow hackathon organizers at DubHacks, and found tensions and contradictions around how organizers conceptualized the purpose of hackathons [4].

While performing the literature review for that project, I found that relatively few papers had been written about hackathons thus far, and that almost all of them have been qualitative or mixed-methods studies with small numbers of participants [e.g. 5, 6, 7] and/or research-through-design studies where the researchers participated in the organization of a hackathon geared towards some particular design outcome [e.g. 8, 9, 10]. Most of these studies investigated only one hackathon, and all (that I could find) investigated fewer than ten hackathons. While these kinds of studies are able to answer interesting questions, their narrow scoping prevents them from painting a broader picture of the hackathon phenomenon. I want to be able to answer larger questions about hackathons like:

- What do people build at hackathons?
- Why do people build what they build at hackathons?
- Are hackathon projects discarded after the hackathon ends, or do they become long-term projects?
- Who participates in hackathons? How often do they participate more than once, and what factors predict this?
- Are hackathon participants likely to learn new skills at hackathons that they will apply in future projects?
- How do prizes impact all of the above?

Many or most of these questions would be practically impossible to answer in a generalizable way with the research methods that have been applied to hackathons thus far. Almost all of the previous work has relied heavily on interview data, which provides rich insight but cannot reasonably scale to dozens or hundreds or thousands of hackathons.

Fortunately, there is another source of data that researchers have not yet utilized: online trace data. As a hackathon participant, I learned how participants use the websites Devpost.com and GitHub.com to showcase and store their work. As a researcher, I have begun to utilize the traces left on these websites as a data source to help address these big questions.

## Methods

Under the instruction of Kate Starbird, I am working on a mixed-methods research project with a similar methodology to her paper titled “(How) Will the Revolution be Retweeted?” [11]. This methodological process involves multiple iterations of quantitative data/network analysis and qualitative grounded theory analysis in order to construct a broad and deep understanding of a set of online trace data. This process is currently a work in progress at the initial quantitative stage, and I will present my current findings in this paper.

*Data Collection*

In order to begin to address my questions about hackathons more broadly, I collected data from the Major League Hacking (MLH) Fall 2017 season.[1] Major League Hacking is an organization that supports hundreds of college and high school hackathons, and I used its event listing as a starting point. Eventually, I plan to collect data from a significantly larger number of hackathons.

Based on the listings on the MLH website, I identified 79 hackathons that took place in North America between August and December 2017. For each of the 79 hackathons, I collected (or attempted to collect) the name, location, link to the Devpost.com submissions page, and the timestamp of when winners were announced.

Of these 79 hackathons, 73 had submissions on Devpost.com. Devpost is a software portfolio website that also allows hackathon participants to submit summaries of their projects for hackathon judging. Each project submission on Devpost has information about the project, such as a name, tagline, and description, and some Devpost project submissions also have links to GitHub.com repositories. GitHub is an online source control tool that allows users to post their code as well as a history of changes to their code. From the starting point of 73 MLH hackathons, I found 4,637 project submissions on Devpost. Of those 4,637 project submissions, 2,742 had links to GitHub, although only 2,506 of these links led to public GitHub repositories.

For each of the 4,637 project submissions, I collected (or attempted to collect) the name, tagline, description, "built with" tags, Devpost link, GitHub link, hackathon name, and links to up to 6 author profiles on Devpost (plus a count of total authors).

For each of the 2,506 public GitHub repositories, I collected the number of commits (a term describing a set of changes made to a repository) both before winners were announced for the hackathon and after winners were announced for the hackathon.

Data collection was completed in January 2018, so some data may be stale. For example, some Devpost users might have edited their submission descriptions, or some GitHub users might have made additional commits.

*Analysis*

Following my mixed-methods trace data methodology, I used data visualization tools to get a broad overview of the data I collected. This is the first step in the process, and my next step will be to begin a grounded analysis of the content of the project submission descriptions on Devpost and the GitHub commits.

## Results

In order to get a high-level view of my data, I created four different data visualizations. The first is a map view with each hackathon represented by a circle, showing the geographical distribution of these events. The second is a bar graph with each hackathon along the x-axis and the number of projects and number of long-term projects represented on the y-axis. The third is a word cloud showing the most common tags listed in the "built-with" field. The fourth is a network graph, showing collaboration links between Devpost users.

[1] https://mlh.io/seasons/na-2018/events (Fall 2017 events are part of the "2018 season")

**Insight 1:** The majority of hackathons and hackathon project submissions in this dataset are in the eastern half of the US, not in or near Silicon Valley

**Insight 2:** The largest hackathons by number of submissions are clustered at some prestigious universities with strong engineering programs:

- University of Washington (DubHacks)
- University of California, Berkeley (Cal Hacks 4.0)
- University of California, San Diego (SD Hacks)
- University of Michigan (MHacks X)
- Georgia Tech (HackGT)
- University of Waterloo (Hack the North)
- University of Pennsylvania (PennApps)
- Princeton University (HackPrinceton)
- Yale University (YHack)
- Harvard University (HackHarvard)

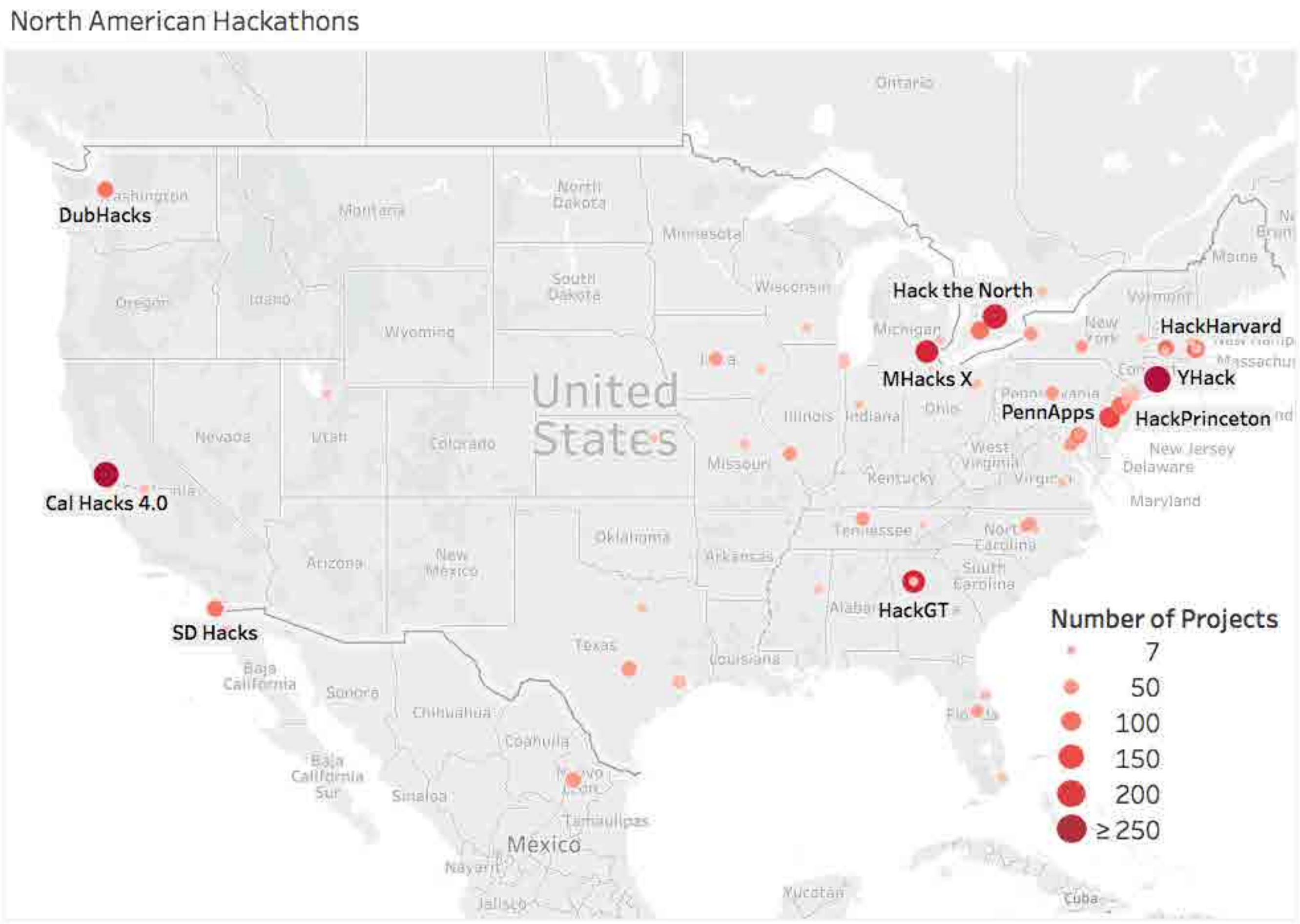

Figure 1: This visualization shows the distribution of hackathons across North America (USA, Canada, and Mexico) for the Fall 2017 season. Each circle representing a hackathon is larger and darker depending on the number of projects submitted at that hackathon. This visualization was created using Tableau.

**Insight 3:** The majority of hackathons have fewer than 100 submissions. The minimum is 7, maximum is 271, mean is 63.52, and median is 44.00.

**Insight 4:** A fairly small proportion of projects have any commits on GitHub from after their respective hackathons ended.

- Overall, 11.5% of projects had GitHub commits after their hackathons ended
- Out of the 2,506 projects with public GitHub repository links, 21.2% had GitHub commits after their hackathons ended
- Purdue University's hackathon (BoilerMake) had the highest percentage of projects continued (32.4%)
- 8 hackathons out of 73 had 0 projects continued
- Earlier hackathons (further left on the graph) seem to be slightly more likely to have projects be continued

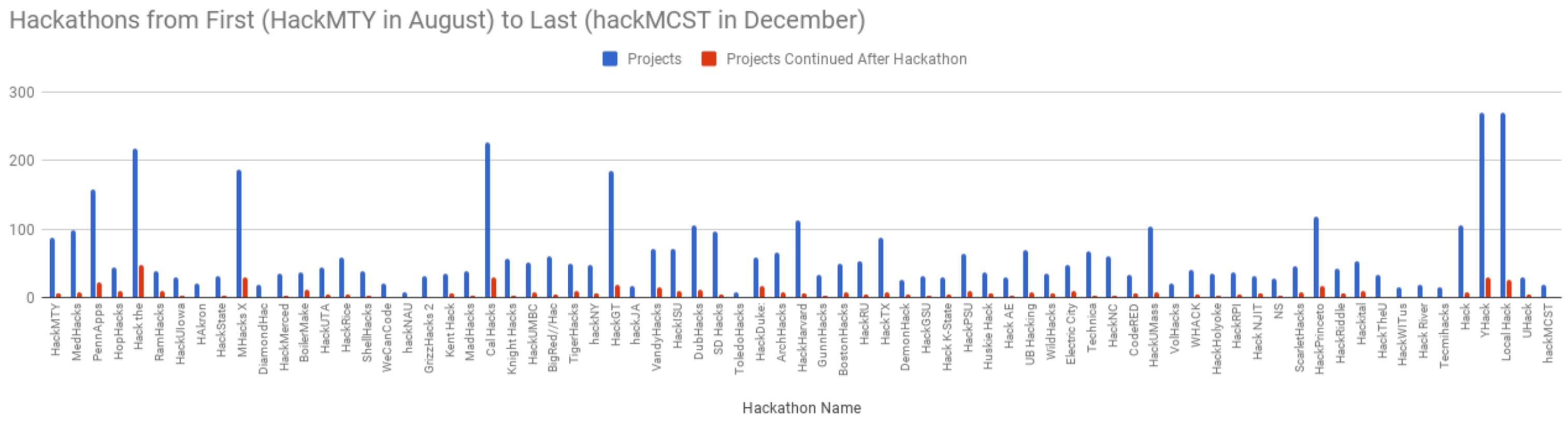

Figure 2: This visualization shows the distribution of projects and continued projects for each hackathon. Hackathons are arranged horizontally in chronological order, starting with HackMTY in August and ending with hackMCST in December. Blue bars represent the number of projects submitted on Devpost, and red bars represent the number of projects with at least one GitHub commit (set of code edits) after the hackathon's winners were announced. This is likely to be an underestimate of how many projects have actually been continued, because it only includes projects with publicly-available GitHub repositories. This visualization was created using Google Sheets.

**Insight 5:** The "built with" tags are clustered around 3 tags (N.B.: these are percentages of the projects whose authors filled in the "built with" field and did not leave it blank):

- "javascript" was used in 1,580 out of 4,480 projects with tags (35.3%)
- "html" was used in 1,474 out of 4,480 projects with tags (32.9%)
- "python" was used in 1,458 out of 4,480 projects with tags (32.5%)

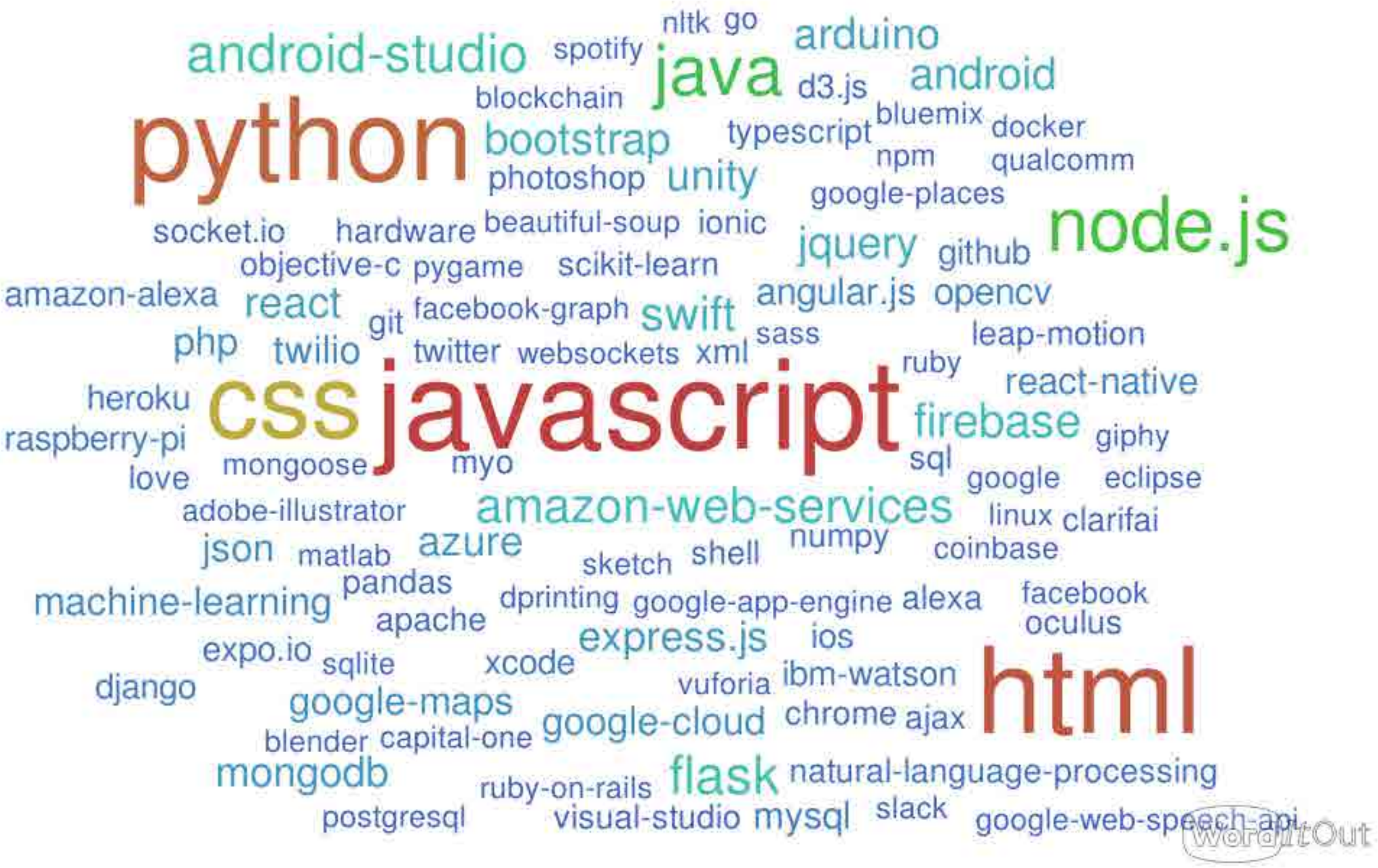

Figure 3: This visualization shows the distribution of tags used in the Devpost projects' "built with" field. Larger font size indicates the tag was used more times. (N.B.: only the most popular tags are displayed out of 520 distinct tags.) This visualization was created using WordItOut.com.

**Insight 6:** The large clusters near the center represent highly active hackathon participants who frequently form teams with others. The largest, darkest circle represents a user who collaborated with 19 different people over the course of 7 different hackathons.

**Insight 7:** Most people had collaborated with 2-3 other people (N.B.: this graph only includes people who worked with at least one person)

- 84.3% worked with 2 or more people
- 57.6% worked with 3 or more people
- 15.4% worked with 4 or more people
- 5.5% worked with 5 or more people
- 2.8% worked with 6 or more people

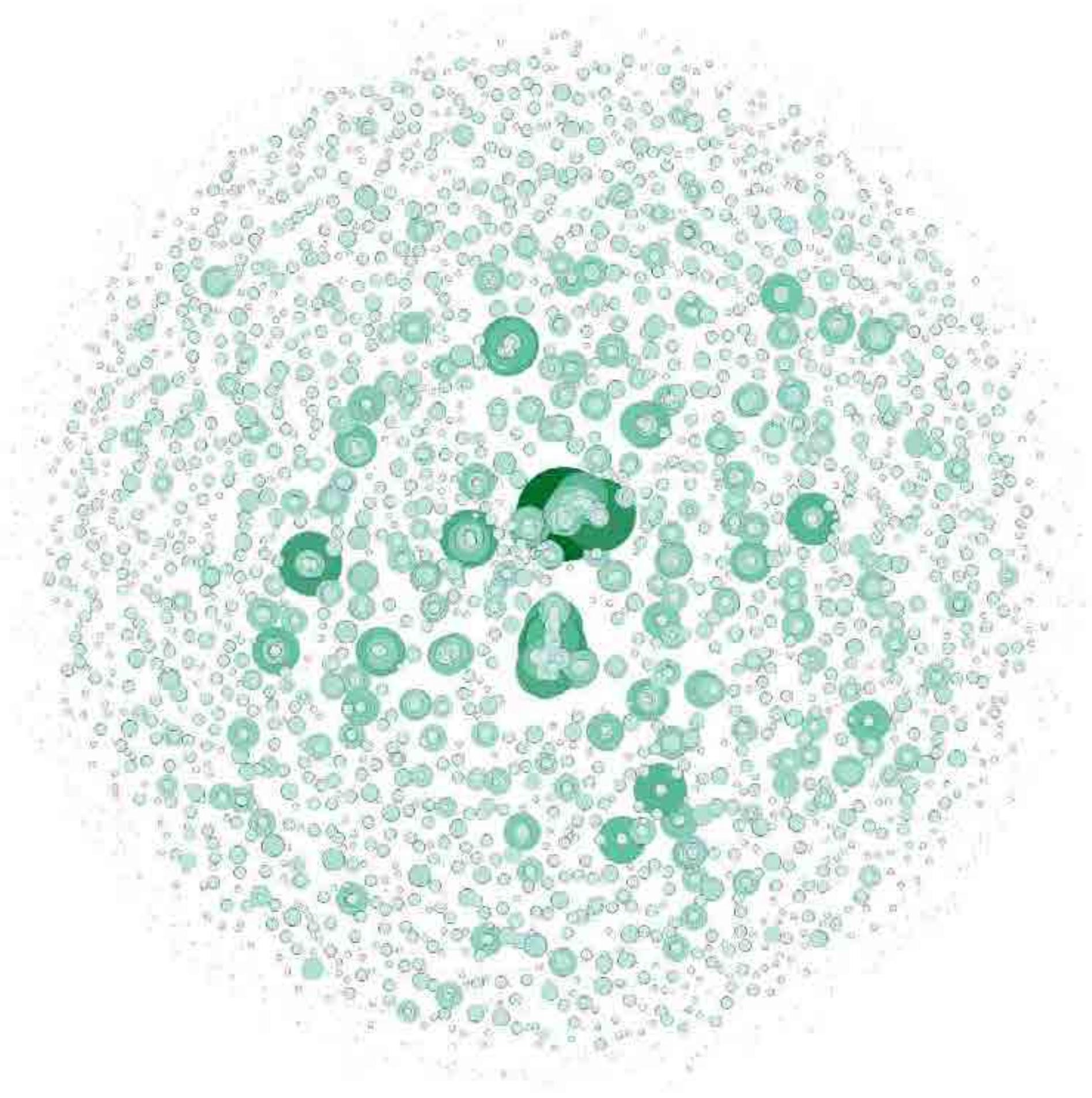

Figure 4: This visualization shows collaborations between participants at hackathons based on Devpost submissions. Each circle represents a participant, and the circles are larger and darker when the participants have collaborated with more people. Circles are spatially located nearer to the circles of the people they have collaborated with and further from the circles of the people they have not collaborated with. This visualization was created using Gephi.

## Conclusion

Overall, the four visualizations have led me to seven insights. (I use the term “insight” rather than “finding” or “conclusion” because this is only a preliminary step.)

First, MLH hackathons in Fall 2017 were geographically clustered around the eastern half of the United States, with a particularly dense cluster going from Washington, DC to Cambridge, MA.

Second, the largest of these events (by number of project submissions) took place at prestigious universities with strong engineering programs. Despite this, some prestigious universities with strong engineering programs had relatively small hackathons (e.g. Carnegie Mellon University, Northwestern University).

Third, these hackathons each had between 7 and 271 project submissions, with a median of 44 project submissions. This was smaller than my original guess, which was a median of 60 or 70 project submissions.

Fourth, a fairly small percentage of hackathon projects appear to be long-term projects. When “long-term” is defined as “having at least one commit on GitHub after hackathon winners are announced”, only 21.2% of projects with publicly-available GitHub commit data were long-term.

Fifth, out of 520 distinct “built with” tags, “javascript”, “html”, and “python” were the most popular. Each was present in about 1/3 of all Devpost submissions where the “built with” field was filled out.

Sixth, a few participants attended many hackathons and formed large numbers of connections with different collaborators, the highest being 19 collaborators. The participant with 19 different collaborators attended 7 of the 73 hackathons in the dataset.

Seventh, and finally, most participants who collaborated with at least one other person collaborated with at least 3 other people (57.6%). A very small proportion of participants collaborated with at least 6 other people (2.8%).

The next step of this project will be a qualitative grounded theory analysis of the actual content behind these numbers. Following the advice of Howard [12], I will use insights from these quantitative visualizations to guide me in sampling the qualitative data. For example, I might investigate the kinds of projects created at BoilerMake in order to better understand why that particular hackathon had the highest rate of long-term projects. I also might look into the differences between the small minority of people who have had more than five collaborators and the large majority of people who have had three or fewer. (In particular, I am interested in finding out whether this is tied to geography – the people with the highest numbers of collaborators seem to live on the East Coast where the geographic density of hackathons is high.) Finally, I might investigate what kinds of projects include the most popular “built with” tags (“javascript”, “html” and “python”) and what kinds of projects include more-obscure tags like “photoshop” or “love”.

At this CHI workshop, I hope to share these initial findings with the group and to get your feedback on what to investigate next.

# Engaging Time-Poor Designers in Philanthropic Activities through Time-Bounded Events

**Bonnie E. John**
**Paddy O'Flaherty**
**Shahtab Wahid**
**Anthony Viviano**
Bloomberg L.P.
New York, NY, USA
bjohn11@bloomberg.net
poflaherty2@bloomberg.net
swahid2@bloomberg.net
aviviano@bloomberg.net

## Abstract
As a company, Bloomberg L.P. is committed to encouraging and facilitating the philanthropic activities of its employees [1]. With initiatives in arts and culture, education, the environment, public health, and community, among others, Bloomberg employees donated almost 150,000 hours to philanthropy in 2017 [2]. Most of these hours are divorced from the skills used in their work; anyone can plant a tree or assemble meals for seniors. However, volunteering in areas where your unique skills are brought to bear can be have a great impact on the causes you support and can inspire young people in career choices and life-long philanthropy. As UX designers at Bloomberg, we want to explore the opportunities to do "design for good" through time-bound events.

## Introduction

In keeping with Bloomberg L.P.'s commitment to encouraging and facilitating its employees' philanthropic activities, UX designers look for opportunities to bring their unique skills to "design for good". Such activities can benefit the causes with which we work and also inspire young people in career choices and life-long engagement with philanthropy.

While these goals are lofty, as professionals in a competitive field, we are "time-poor", with little leisure time to spend long hours on personal philanthropy. However, we believe that these philanthropic goals can be coincident with company goals, like increasing Bloomberg's reputation and supporting recruiting of talented employees.

Bloomberg has developed extensive experience in hosting events centered on programming, e.g., the global CodeCon hackathons [3] and we are considering extending a similar framework to UX design. We wish to participate in the Time-Bounded Events Workshop at CHI 2018 to (1) understand how time-bounded events might help us meet our diverse goals and (2) provide an industry perspective on academic research about time-bounded events.

## Our Goals and Constraints

Our goals for a time-bound event are threefold:

- Do impactful design for good as a philanthropic activity.
- Provide a meaningful experience for young people to help in a UX design career choice and
- Start them on a path to lifelong engagement in philanthropy.
- Increase the reputation of Bloomberg L.P. among UX academics, professionals and students for the purpose of recruiting.

The constraints under which we would like to achieve these goals are:

- A time-bound event of no more than one full day of professional designers' or students' time
- Follow a user-centered design process

In creating an event under these constraints, to achieve these goals, there are a myriad of design decisions. We would like the input from workshop participants' research and experience to help make these decisions.

## Questions to Explore

Some of the questions we would like to explore during the workshop include the following.

*How can a user-centered design process fit into a 1-day event?*

As so often happens in UX design, the constraints fight against each other, making it hard to achieve the goals. A user-centered design process takes time; how best to use the limited time in a day and still follow the spirit of that process? Some possibilities, each with their own pros and cons, include the following.

- Perform interviews of appropriate users ahead of the event and make videos and transcripts available during the event. Participants will not exercise their data-collection skills, but will have material from which to do user-centered design.

- Have the philanthropic partner provide appropriate users to be interviewed by the event participants in the early part of the day. Have participants share their findings before starting design.
- Pick a philanthropic domain where the event participants themselves are appropriate users and have them interview each other.

*How best to engage time-poor students in this event?* Not only are UX professionals time-poor, but students have class, and sometimes work, commitments to balance against opportunities like this event. Some ideas include the following.

- Have the event at or near a university with the cooperation of faculty in a User-Centered Design class. Perhaps participation could somehow be part of the curriculum to help free up some time for the students.
- Have the event at or near a UX conference with traditionally high student attendance. This arrangement would likely get participation by students from many universities, offering the opportunity for experience more broadly, as well as spreading Bloomberg's reputation more widely.

*What is the role for UX professionals in the event?* Are we observers, swooping in if we see a group get stuck or go down a rabbit hole? Are we the experienced half of a master/apprentice relationship? Are we team members, co-designing along with the student participants? What role(s) would best balance success toward the producing goal, the education/inspiration goal, and the company branding goal?

*What format of deliverable can best achieve the goals?* Are "pitch decks" a good deliverable to for allowing the philanthropic partners to make use of the results and provide portfolio content for the participants? Do we need to provide sufficient support so running prototypes can be delivered?

*Should the procedures be piloted before the event?* An academic would never run and experiment without one or more pilot runs. Should we try out the format internally with our employees to establish procedures, flow, timing, tools, and deliverables? Do we risk our reputation if we don't have such a run-through?

*Finally, how do we measure the success of the event?* Since there are multiple goals, both short and long-term, how can they be measured?

- Has there been a positive impact on the philanthropic partner in the event? How can the output be something they can use to further their work after the event?
- Did the students have a meaningful experience that helps them in their life?
- Did this event start the students on a path of philanthropy? Do they continue on that path?
- Has Bloomberg's reputation been enhanced, especially with respect to future recruiting?
- Might partnering with researchers interested in studying the effects of time-bounded events benefit the participants, sponsors, and researchers alike?

## Applicants' Backgrounds

Bonnie E. John is a user experience designer and researcher working on discoverability tools and tools for finanical quants at Bloomberg. An academic for Carnegie Mellon for 25 years, and head of the Masters in HCI there for over a decade, she is interested in helping students realize happy and successful lifework.

Paddy O'Flaherty is a user experience designer working on tools for financial quants at Bloomberg. He has organized several one day conferences for designs and is interested in combining this with his philanthropic interests. He has over 25 years of industry experience.

Shahtab Wahid is a user experience designer working on collaboration tools at Bloomberg. He has mentored research and design interns from various schools and looks forward to continuing relationships with academic institutions through events such as design-a-thons. He earned his doctorate degree studying Human-Computer Interaction at Virginia Tech.

Anthony Viviano is a user experience designer, currenlty at Bloomberg, producing both client-facing mobile applications and tools for engineers to manage systemuptime. He will bring his experience informally mentoring young people with an interest in design to the design of effective time-bound experiences.

# Hacking Academia: How Academic Communities are Evolving a New Breed of Hackathon

**Margaret Drouhard**
eScience Institute
Department of Human Centered Design & Engineering
University of Washington
Seattle, Washington, USA
mdrouhard@acm.org

**Brittany Fiore-Gartland**
eScience Institute
Department of Human Centered Design & Engineering
University of Washington
Seattle, Washington, USA

**Anissa Tanweer**
eScience Institute
Department of Communication
University of Washington
Seattle, Washington, USA

## Abstract

Adapted from their roots in open source and tech industry cultures, hackathons that facilitate intensive collaboration or co-working are increasingly prevalent in academic communities. Many features of hacking culture challenge entrenched norms for pedagogical models and research, so expectations and norms around this new breed of hackathon are still emerging. Our ethnographic study follows multiple iterations of academic hackathons, including organizers' sensemaking processes toward developing best practices. Based on extensive participant observation, interviews with organizers, and participant survey responses, we analyze some of the factors that motivate participation in academic hackathons, as well as how their evolution relates to broader challenges in academic institutions.

## Introduction

As ethnographers embedded within academic data science environments, numerous hacking events have intersected our study of the emerging culture and practice of data-intensive scientific discovery. We conducted more than 175 hours of participant observation at multiple types of hackathon events, including multiple iterations of academic hack weeks. We began characterizing the sociotechnical di-mensions in a preliminary typology, which we workshopped with hackathon researchers at CSCW [1]. In this work, we focus our analysis on academic hack weeks, an example of the communal hackathons we described in our initial typology. Like other communal hacking events, hack weeks are designed to build out infrastructure, practices, and culture for a particular community, in this case members of an academic discipline or sub-field. Our access to longitudinal data from multiple iterations of various hack weeks, as well as our ability to triangulate analysis using survey responses from participants, have allowed us to explore some of the motivations that draw participants to hack weeks and how these events are perceived to fit into the broader ecology of academia and scientific exploration.

## Motivations for Participation

Given that we initially described hack weeks as an example of *communal* hackathons, it seems intuitive that finding community emerged as a core motivator for participating in these events. Perhaps less intuitively, the community of practice that comes into being at these hack weeks draws participants largely from academic and research institutions with a particular disciplinary focus, which might be considered fairly niche communities in and of themselves. The particular community that participants and organizers describe seeking at hack week events is even more specialized, and perhaps it is the unique and uniquely bounded "space" of hack weeks that afford the evolution of practices and culture for this community.

*The Hack Week Community*

Many participants referenced getting to know "like-minded" peers as a central component of the hack weeks. One participant described the event as, "An opportunity to meet like-minded scientists in the field and learn about tools and methods to benefit our research." After this phrase cropped up repeatedly in surveys and chats, we heard from one participant in more detail what it meant to be in a community of "like-minded peers." Responding to our inquiries about what makes the hack week community distinct within academic communities, the participant explained,

> I guess I would say it is an "openness to new tools and techniques." So in both academia and industry where I've worked before, there is this sense that people are trying to build new knowledge in the field, but the ways of doing that are set. So everyone has their standard places to go for data, and their standard analysis routines, and then the data hopefully signals some new insight. But here, people are much more interested in working reproducibly and recognizing the value that new techniques and tools can bring to the field in terms of the questions we're able to answer. [Paraphrased except where text is in quotes.]

Hack week organizers have also commented on the opportunity they see for hack weeks to offer the sort of community that allows "computationally-minded" researchers to "break from the isolation of their academic departments" [3]. Both participants and organizers have expressed enthusiasm around the possibility that these events may spark new connections and collaborations both within and across disciplines, and build infrastructure that could support the opening of new lines of scientific inquiry.

*"Space" for a Different Kind of Work*

A related theme that we have seen emerge from multiple iterations of academic hack weeks is an interest in taking advantage of the "space" these events create to spend time on work and learning that are difficult to prioritize in their normal work settings. In a survey about one of the hack weeks, a participant described the event as: “A workshop for learning new computational techniques and to experiment with new projects that you might not otherwise have time for.” Our reflections about the “space” or environment that hackathons create for a particular mode of engagement align with prior work focused on the “publics” created by hackathons. According to Fiore-Gartland and Geiger, the publics that may be created by hackathon events represent a “respite from day-to-day research activities and provide a low-stress venue to learn new skills and attempt high-risk projects.” [2]. Often when we spoke with participants who lived locally, they emphasized the value of having this time dedicated to learning and practicing new skills, when they wouldn’t be interrupted with emails or normal day-to-day tasks. A unique “space” apart and community of “like-minded” peers surfaced from our thematic analysis as strong throughlines across multiple iterations and different types of hack week.

## Hackathons and Institutional Change

Academic hack weeks have grown in popularity throughout our observation, attracting larger applicant pools and spreading across disciplines and fields. It seems clear that participants and organizers are realizing meaningful outcomes through these events. One of the aims of our ethnographic work is to recognize work that falls outside conventional academic incentive structures (e.g., publications) and to observe how such work is valued by the communities involved. As such, we were interested in how hackathons relate to the ongoing changing of well-established institutions within academia.

One possible lens is to think of them as stopgap measures that are filling some need or needs that are not being met by established institutions. They might be a short-term fix until our institutions can evolve and catch up with the times and grow to fill that need. An academic hack week could indicate, for example, that we don’t have sufficient formal course offerings in advanced computational methods. We could consider which needs are being served through these time-bounded, relatively low-resourced hackathon events and what that says about shortcomings or gaps in our institutional configurations.

Another way we can try to make sense of both the diversity and commonalities we see across various hackathon events is through a sociomaterial lens that sees this phenomenon as being interwoven with the material exigencies of computing environments. One of the things that is distinctive about academic data science is that most projects evolve in a unique, configurable, and customizable software environment. It’s not a one-size-fits-all landscape, and the rapid evolution of computing capabilities and tools requires researchers to expand their knowledge and skills constantly. In part, hackathons are a response to these exigencies. Since you can’t just learn something once, academic data scientists need nimble and flexible pedagogical approaches and structures like hackathons to support the kind of ongoing learning that is required.

These lenses correspond to the various ways in which the term “hack” is often employed. A hack can be a quick fix, something that works well enough for now to patch up a problem, but should be addressed systematically later. And hacking is also often used almost synonymously with coding or programming, indicating a mainstreaming of these activities and skills, which seems in sync with viewing hackathons through a lens of emergent adaptations that are being normalized in response to material exigencies.

## Conclusion

Over multiple iterations, academic hack week organizers have assessed how well the events are serving the needs of their respective communities. They have adapted and redesigned various components to better align with their objectives for developing community infrastructure and to better support participants in acquiring and practicing new skills. One hack week is transitioning into a multi-week "hackademy," while others have changed and redeveloped tutorial modules over the years. Whereas some other types of hackathon may have a relatively stable structure over numerous iterations, the progression of hack week design reveals pathways for small, dedicated communities to ensure that their evolving needs are met. The success of academic hack weeks also highlights the opportunity for communal hacking events to be drivers of institutional change.

## Author Bios

**Margaret (Meg) Drouhard** is a PhD candidate in the Department of Human-Centered Design and Engineering at the University of Washington (UW). She works with Dr. Cecilia Aragon in the Human-Centered Data Science Lab. Along with other researchers on the ethnographic team at the eScience Institute at UW, Meg studies the emerging practice and culture of data-intensive scientific research. As part of her ethnographic work, she has observed and participated in several academic and community-led hackathons.

**Brittany Fiore-Gartland** is the Director of Data Science Ethnography at the eScience Institute and a Senior Research Scientist in the Department of Human Centered Design and Engineering at the University of Washington. Her research focuses on the emerging cultures and practices of data science and the social and ethical implications of technological change.

**Anissa Tanweer** is a PhD candidate in the Department of Communication at the University of Washington and a research assistant in the Human-Centered Data Science Lab. She is interested in the ways people organize with and around data, and how the use of increasingly large, heterogeneous datasets is transforming the way we construct knowledge and make decisions.

## Acknowledgements

The authors wish to thank the organizers and participants in the hackathons we observed for sharing their experiences and insights. This work was supported in part by Washington Research Foundation Fund for Innovation in Data-Intensive Discovery and by the Moore/Sloan Data Science Environments Project at the University of Washington.

# You Hacked and Now What? – Exploring Outcomes of a Corporate Hackathon

**Alexander Nolte**
Carnegie Mellon University
Pittsburgh, PA 15213, USA
anolte@cmu.edu

**Ei Pa Pa Pe Than**
**James Herbsleb**
Carnegie Mellon University
Pittsburgh, PA 15213, USA
eipapapt@cs.cmu.edu
jdh@cs.cmu.edu

**Anna Filippova**
GitHub Inc.
San Francisco, CA 94107, USA
annafil@gmail.com

**Christian Bird**
Microsoft Research
Redmond, WA 98052, USA
cbird@microsoft.com

**Steve Scallen**
The Microsoft Garage
Redmond, WA 98052, USA
sscallen@microsoft.com

## Abstract
Time bounded events such as hackathons, data dives, codefests, hack-days, sprints or edit-a-thons have increasingly gained attention from practitioners and researchers in recent years. Existing work around such events however has mainly focused on the event itself while potential outcomes of hackathons have received limited attention so far. In this paper we will present preliminary findings from a case study of the outcomes of a large scale corporate hackathon. Our findings provide insights into the continuation of projects, the sustainability of teams and the potential effects of hackathon participation on individuals.

## Author Keywords
Hackathons; project sustainability; learning

## ACM Classification Keywords
H.5.3. Group and Organization Interfaces: Collaborative-supported cooperative work

## Introduction
In recent years time-bounded events such as hackathons, data dives, codefests, hack-days, sprints or edit-a-thons have seen a steep increase in popularity. During such events people form – oftentimes ad-hoc – teams and engage in intense collaboration over a short period of time. Collegiate events alone attract over 65.000 participants worldwide among more than 200 events each year [14]. But it is not only collegiate events alone. Hackathons have spread into a variety of different contexts ranging from corporations to higher education and civic engagement [15]. Hackathons come in varying forms where participants might be present face-to-face or collaborate remotely [16]; they may involve newly formed as well as existing teams working on new project ideas or well-defined agendas [8,13]; they may involve prizes while others do not [17]; they may have different goals such as creating startups, support civic open innovation strengthening interaction in specific scientific domains and teaching specific skills.
While there is a growing body of research around hackathons, existing work mainly focuses on the event itself. This work contains descriptions of events [2] and covers themes such as how hackathons teams self-organize [13], how teams and organizers deal with diverse audiences [5] and how non-software hackathons can be conducted [8].

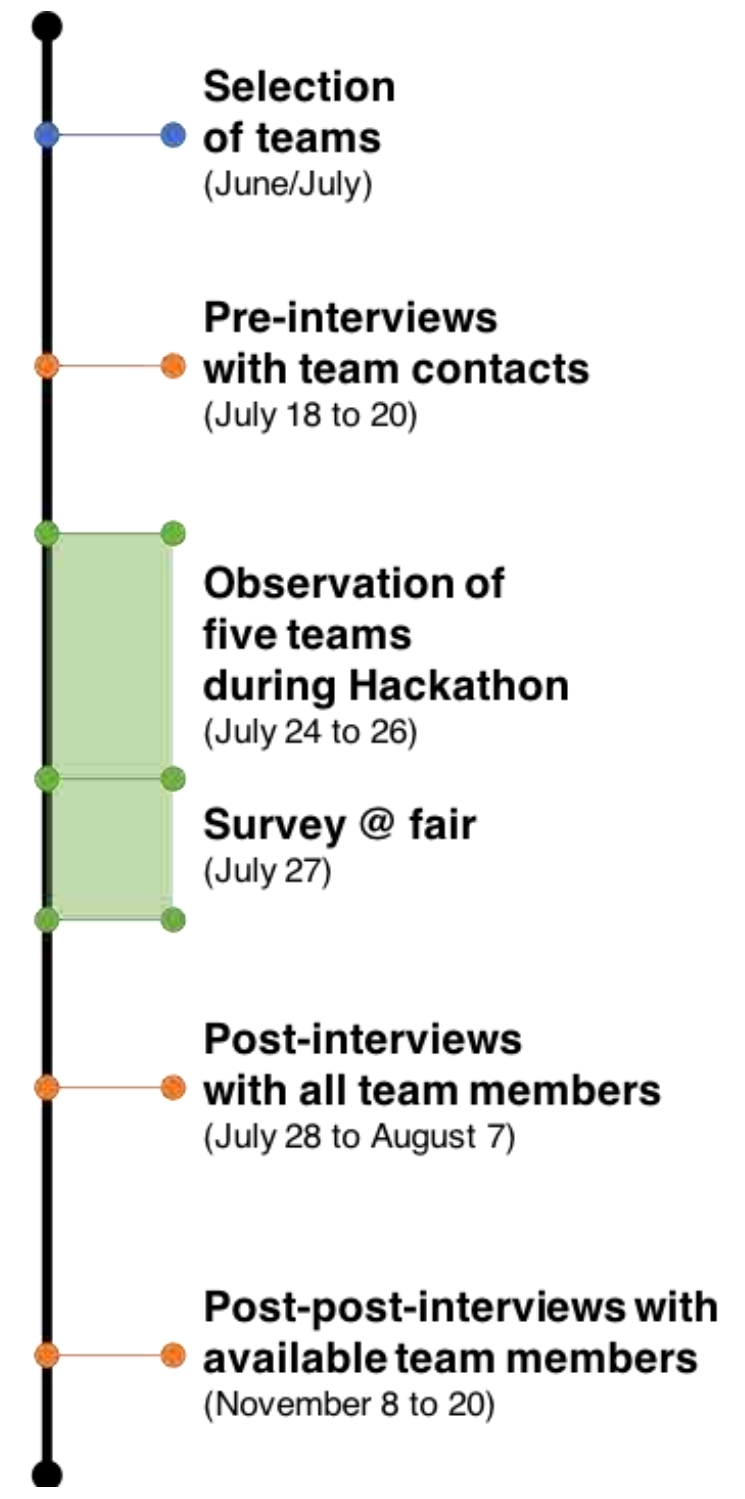

Figure 1: Data collection points before, during and after the hackathon.

While there is research around potential outcomes of hackathons, this work is still fragmented and scarce. Work around hackathon outcomes mainly focuses on identifying reasons for the lack of sustainability of projects [1,3,4] and on learning outcomes [9,11,12]. A general overview over potential effects of hackathon projects, project teams and participants so far is missing. Furthermore, it has to be noted that most studies are conducted in student or civic spaces while little attention has been paid to corporate hackathons. The lack of research around potential outcomes of hackathons is particular surprising in this domain appears surprising since corporations increasingly invest in hackathons to foster internal innovation [18]. This in turn means that corporations have a vested interest in conducting hackathons that focus on creating sustained outcomes in the form of projects that can become future products. Our focus however is not on the sustainability of projects alone. Corporations also aim to provide opportunities for their employees to expand their competencies [10], their network [6] and generally create a positive and motivating work environment [7]. Hackathons can support existing approaches in these areas since they require forming teams and acquiring new skills or expanding on existing ones. We are thus taking a wider perspective on potential outcomes of hackathons by focusing on the following research questions:

- **RQ1:** What happens to projects that were developed during a corporate hackathon?
- **RQ2:** What happens to teams that participated in a corporate hackathon?
- **RQ3:** What effect does participation in a corporate hackathon have on its participants?

In order to answer these questions, we conducted a study on a large corporate hackathon. We will describe the procedure of our study in the following before reporting on preliminary findings.

## Empirical study

Our study took place during Microsoft's One Week hackathon in summer 2017. One Week is an annual global 4-day event during which employees of Microsoft engage in intense collaboration to conduct any project that they are interested in. The last day is reserved for a presentation session which is organized as a fair.

During this fair teams can present their products to the wider Microsoft public. Participation in the hackathon is voluntary but encouraged by management. In order to participate, employees had to register in a web-based tool in advance of the hackathon. The tool allowed employees to join an existing project team, propose their own project, register as a team and search for additional project members. The teams had between three and four members on average.

We focused our study on five teams that collaborated at the largest hackathon site in Redmond which hosted around 5.000 participants in two large tents. The teams were carefully selected based on the dimensions of familiarity among team members and relationship between their hackathon project and their everyday work. Two of the teams consisted of employees that work together on a daily basis while three teams had been specifically formed for the hackathon. The teams had between three and seven members.

Our data collection includes semi-structured interviews which were conducted before, directly after and four months after the hackathon with the aforementioned teams (c.f. Figure 1 for an overview of the data collection procedure). We attempted to conduct both follow-up interviews with all team members but could not reach them all. We did however interview at least two participants of every team four months after the hackathon. In addition to interviews we also observed the teams during the entire duration of the hackathon and conducted a survey at the fair.

We focus our analysis on the interviews that were conducted four months after the hackathon, since they are most directly related to our research questions. These interviews lasted between 13 and 29 minutes each. We also include results from two survey questions which focused on intentions to continue with the project and with the team to help us understand the prevalence of individual continuation intentions. The interviews that followed immediately after the event were not formally analyzed for this paper, but they provide additional context.

## Preliminary findings and outlook

Using an open coding procedure, we focused on the continuation of projects (RQ1) and teams (RQ2) as well as potential effects on individual participants (RQ3).

**Projects:** About 70% of the members of the five teams we studied reported an intention to continue working on the project they started during the hackathon. An analysis of the follow-up interviews revealed that indeed two out of the five projects will be continued. Our analysis however also showed that none of the original team members will be involved in the continuation of the projects they started. One of the projects will be continued by a group that was already planning to develop a similar software before the hackathon (*"X told to Y: I think these guys have built what you are trying to build"*) while the other project will be continued by a group that perceives the project as a suitable addition to their existing product (*"they have a fairly similar app"*). It appears reasonable that projects are continued by groups with a fitting product since all hackathon projects we require additional resources to reach a shippable state (*"I would say that it needs a month to make it really usable"*).

This finding subsequently made us focus on identifying how those groups became aware of the project that they will take over. We found that the connection was mainly facilitated by presentations that took place after the hackathon *("we presented our project to multiple groups"*). Some of these presentations were a direct result of a group's participation in the fair (*"a manager came by and I showed her/him our project"*) while others were based on individual networks (*"our group leader has connections"*).

When investigating the three projects that had not been continued we found one project that will not be continued despite fitting to an existing product line. As potential reasons for the discontinuation of this project, study participants mentioned that there was *"no immediate demand"* and that *"the project will create a new business case that we are not ready for"*.

The continuation of a project in our case thus first required exposure to other groups and second a fit to existing projects as well as a suitable demand.

**Teams:** Similar to the aforementioned projects about 70% of our study participants stated their intention to continue working with the same people after the hackathon. However, despite those intentions only two teams continue working together which *"were actually all from the same work team"*. While this is hardly surprising given the way the company is organized we also uncovered follow-up activities of other teams. They *"met for lunch"*, *"chatted about next steps"* or even *"connected with some folks individually to tap into their skills for my current job"*. Teams thus continued based on the work situation as well as individual affordances.

**Individuals:** We found a large variety of different effects of hackathon participation on individuals. First, we found that participants acquired additional technical skills (*"I learned how the 3D stuff works"*). Some of those skills were directly applicable at work (*"I use some of the skills I learned"*) while others were not (*"I cannot do AR in my current job"*). Team leaders also reported that they gained skills related to project management *("I had the opportunity to organize something from start to finish"*). Second, participants also reported that the hackathon *"sparked an interest to develop other skills"*, instilled confidence in the ability to acquire them (*"I feel more equipped now that I have a background in those [technical] topics"*) and had a positive effect on individual's perception of their workplace *("that Microsoft does hackathons [...] has become something important to me"*). Third, we found that participation in the hackathon had direct as well as indirect impacts on the workplace of the participants. Three of our study participants got promoted after the hackathon. Participation in the hackathon was not the only reason for their promotion, but certainly played a role as evident by the following statement: *"Success in the hackathon shows creativity and capability"*. In addition to promotions one participant also mentioned that participation in the hackathon positively affects her/his manager's perception her/him *("participation in the hackathon is in my annual progress report […and I receive…] positive feedback by my manager"*). We thus observed direct effects on individual skills, interests and confidence as well as effects on the workplace for individual hackathon participants.

The findings presented in this work provide insights into potential outcomes of corporate hackathons. They also point towards future research directions such as identifying antecedents of project and team continuation and ways to influence them.

## Author background

**Alexander Nolte** is a Lecturer at the Institute of Computer Science at the University of Tartu in Estonia and an Adjunct Assistant Professor at the Institute for Software Research at Carnegie Mellon University in Pittsburgh, PA.

**Ei Pa Pa Pe Than** is a postdoctoral researcher at the Institute for Software Research at Carnegie Mellon University in Pittsburgh, PA.

**James Herbsleb** is a Professor of Computer Science at Carnegie Mellon University in Pittsburgh, PA.

**Anna Filippova** is a data scientist at GitHub Inc. in San Francisco, CA.

**Christian Bird** is a researcher at Microsoft Research in Redmond, WA.

**Steve Scallen** is a principal design researcher at The Microsoft Garage in Redmond, WA.

# Team Familiarity, Goal Setting, and Process: A Case Study of a Corporate Hackathon

**Ei Pa Pa Pe Than**
Carnegie Mellon University
Pittsburgh, PA 15213, USA
eipapapt@cs.cmu.edu

**Christian Bird**
Microsoft Research
Redmond, WA 98052, USA
cbird@microsoft.com

**Alexander Nolte**
**James D. Herbsleb**
Carnegie Mellon University
Pittsburgh, PA 15213, USA
{aun, jdh}@cs.cmu.edu

**Steve Scallen**
Microsoft Garage
Redmond, WA 98052, USA
sscallen@microsoft.com

**Anna Filippova**
GitHub Inc.
San Francisco, CA 94107, USA
annafil@gmail.com

## Abstract

Time-bounded events such as hackathons are increasingly popular, becoming a common feature of many large software companies including Google, Facebook, and Microsoft. With a widespread adoption of hackathons and a wide range of decisions about how to form teams and manage events, it is important to understand how characteristics of teams can impact desired outcomes. In particular, hackathons often include teams who are strangers before the event as well as teams who have regularly worked together. It is not clear how these different levels of familiarity impact choice of projects, coordination, and team dynamics. We collected interview data from members of five teams who participated at the 2017 Microsoft OneWeek Hackathon. We found that "pre-existing teams" (higher familiarity) used the hackathon space to get needed but non-routine work done, and chose projects that were riskier and long-term and set higher expectations on outcomes. In contrast, newly formed "flash teams" (lower familiarity) aligned their goals with official hackathon outcomes of lightly-engineered demos and videos, and had a substantial focus on personal goals. Flash teams experienced more conflict and misaligned expectations, yet were largely satisfied with the experience and intend to participate in the future.

## Author Keywords
Hackathons; time-bounded events; collaboration; familiarity; goal setting; team process; team dynamics; mixed methods study.

## Introduction
Time-bounded intensive events have become increasingly popular in recent years, variously called hackathons, data dives, codefests, hack days, sprints, edit-a-thons, map-a-thons, and so on. Their popularity is attested to by the fact that collegiate hackathons alone attracted more than 65,000 students from 16 different countries in 2017 [5]. Further, the scope of hackathons has broadened from the tech industry to other sectors and disciplines such as astronomy, arts and humanities, biology, social goods, and many more, taking on many different forms such as collaborative or competitive, and focused on innovation, community building, or learning [2, 7].

Regardless of design variations, all hackathons share a set of common features. People divide into small groups to innovate, improve, learn, and network within a specified timeframe, typically 2-5 days. These groups consist of people of often diverse backgrounds, experience, and expertise, and gather in one location. Due to their potential to leverage collective intelligence and foster innovation outside the usual constraints and processes of the workplace, hackathons have become a common feature of large software companies including Facebook, Google, and Microsoft. Yet little is known the ways different compositions of team harness, or fail to harness, the members' creative capabilities.

Prior research on team familiarity (e.g., [1, 3]) suggests that teams with higher familiarity of members have higher performance. In particular, prior studies have found that when members of a team work together over time, they have increased familiarity with the task domain and with each other, clearer expectations and communication, a common knowledge base, and better coordination. Another factor important for team coordination process is goal setting (e.g., [4]). As goals or conscious ideas regulate people actions, the choice of goals may determine team coordination process.

Hackathons where teams vary widely in familiarity provide an interesting context in which to study familiarity and the strategies that non-familiar teams develop to accomplish work in extremely compressed time scales where members may be required to deliver a substantial result 2-5 days after they first meet. In addition, hackathon teams are generally free to work on anything they want with very few constraints other than time and potentially a desire to appeal to a judge. Goal setting becomes a critical process for enabling the team to work together effectively. It is therefore interesting to examine how both familiar and unfamiliar teams set goals and approach their attainment. Thus, we aim to address the following research questions to advance our understanding of team process in time-bounded settings:

*RQ: In time-bounded settings, how do teams with higher familiarity differ from those with lower familiarity in 1) setting goals, and 2) coordinating their work?*

## Methods
We chose the 2017 Microsoft OneWeek Hackathon to study. This is a Microsoft's annual global event held with more than 16,000 participating employees in 2016. The data was collected using a mixed methods approach which included interviews, observation, and survey. In particular, our research group interviewed event organizers and each team leader prior to the event, one member of the research group shadowed one team during hackathon days, interviewed members of that team within a week after the hackathon and again three months after the hackathon. We administered a survey at the science fair which was held on the last day of the event where participants showcased or demoed their projects. Of these five teams we shadowed, two were pre-existing teams with members who regularly worked together before (P1-7), and three were newly-formed “flash” teams whose members had (mostly) not worked together before (F1-16).

All interviews were transcribed and analyzed following empirical grounded theory procedures describe by Strauss and Corbin (1998) [6], using Deedose, a web-based qualitative data analysis software. First, three authors conducted open coding on the interview data, in which familiarity, goal setting, and coordination were used as sensitizing concepts. In the second phase, we shared and wrote descriptive memos. We then discussed in a highly collaborative manner and combined codes that had similar meanings to yield second-ordered codes or themes. The resulting coding scheme was used for the remaining text.

## Preliminary Findings

In this paper, we report some of our preliminarily results based on the analysis of interview data collected within a week after the hackathon.

**Goal setting**: We found that pre-existing teams utilized the hackathon as a dedicated time and space to get the needed non-routine work done (P4, P5). Their chosen projects seemed to be riskier and have long-term potential compared to those of flash teams. In contrast, flash followed the goals that the hackathon set out for them, and aimed at producing lightly-engineered demos of their solutions (P4, P5, P6). In addition, majority of flash team members used the hackathon space to explore new skills and roles unrelated to their regular work, and network with and learn from people who are outside of their regular workgroup (F2, F4). In contrast, pre-existing team members’ participation was closely related to their regular job (P4, P5). In that regard, flash teams were more inclined toward innovation, suggesting that there is a trade-off between familiarity and innovation in terms of goal setting.

**Coordination:** Pre-existing teams’ members uniformly picked out tasks that they were familiar with or they could leverage their existing skills or knowledge (P1, E1, E3). In contrast, flash teams’ members adopted divisions of labor loosely based on standard team roles at Microsoft (e.g., developer, marketing, UX designer, program manager) and performed activities expected to be performed by these roles (F5, F6, F16). When organizing their processes, pre-existing teams were found to fall back on their regular work practices whereas flash teams’ members tended to coordinate based on their taken roles (P4, F16). Here, it is important for flash teams’ members to modify their taken roles considering the constraints of hackathon.

**Expectation mismatch**: We found that flash teams were prone to a problem of expectation mismatch,

especially when some members had no prior hackathon experience. These newcomers seemed to have set high or unrealistic expectations on outcomes of the hackathon, either by putting their personal goals first or holding to their professional engineering norms (F6, F15) rather than focusing just on a demo. Such expectation mismatches led to confusion or mild conflicts in flash teams.

**Conflict avoidance**: Both types of team seemed largely to avoid open conflict during hackathon. This may be due to the short duration of the hackathon which did not really allow time for conflict resolution, and that flash team members were unconcerned about allowing unspoken conflict to continue, since they were unlikely to continue to work as a team after the hackathon (P6, F11).

## Conclusion
Our results suggest that goals and expectations set out by hackathon teams are contingent on how familiar their members are with each other. Depending on the types of goal they pursued, teams adopt different mechanisms to organize themselves in such a way that would maximize the goal attainment. Having prior hackathon experience enabled them to realize the differences between hackathon and regular work and modify their hackathon roles accordingly.

## Author Background
**Ei Pa Pa Pe Than** is a postdoctoral researcher in the Carnegie Mellon University's Institute for Software Research. Her research focuses on understanding how technologies can be leveraged to create new forms of collaboration that improve engagement, productivity, and outcome quality.

**Alexander Nolte** is a postdoctoral researcher at the School of Information Sciences at the University of Pittsburgh and a research associate at the Institute for Software Research at Carnegie Mellon University.

**James Herbsleb** is a Professor of Computer Science at Carnegie Mellon University, where he serves as Director of the PhD program in Societal Computing. His research interests focus on global software development, open source, and more generally on collaboration and coordination in technical domains.

**Anna Filippova** is a data scientist at GitHub Inc.

**Christian Bird** is a researcher at Microsoft Research working on Empirical Software Engineering.

**Steve Scallen** is a principal design researcher at Microsoft Garage.

## Acknowledgements
We wish to thank Microsoft Garage team and Microsoft Research for collaborating with us on this study. We also thank members' of hackahton teams we shadowed for sharng their expreinces.

# Community Data Hackathons

**John M. Carroll**
**Jordan Beck**
Pennsylvania State University
University Park, PA 16803, USA
jmcarroll@psu.edu
jeb560@psu.edu

## Abstract
We are investigating community data: data gathered, analyzed, interpreted, and used by members of a local community. Community members are already engaged in community data practices. Our goal is to help make these more visible throughout the community, and to engage the community at large in deliberation and planning with respect to its data. We are hoping to organize a set of community-wide hackathon events as part of this effort.

## Introduction
As part of a long-term research project focused on community data, we are investigating the role of hyperlocal data in contemporary community. Data pertaining to a community and its locale, that is, data gathered, analyzed, interpreted, and used by members of a local community, is community data. There can be many different kinds of community data. Examples include but are not limited to: water and air quality, demographics, narratives about historically significant places, photographs of weather events, and so forth.

Through community data, the community describes the community to itself in order to understand its past, regulate its present, and plan its future. For example, several organizations in State College, PA, collect water quality data with the goal of supporting community leaders and citizens in making sustainable decisions about water use and land development.

However, it is a known issue within the open data and open government literature that availability of and access to data is insufficient to produce pervasive community participation (Janssen, Charalabidis, & Zuiderwijk, 2012). We are interested in the possibility of leveraging data hackathons as means to engage citizens with community data. Our goal is to identify needs and opportunities for citizens to better understand and contribute to data-driven civic participation as a focus for community innovation.

Citizens of the past needed to possess basic levels of textual literacy. More recently, it has been argued that students ought to develop data literacy (Koltay, 2015). In our view, data literacy is crucial for today's citizens. Citizens need to have facility in understanding and using data, and they need to be able to think critically and creatively about the many uses of data they might encounter. They will likely encounter a lot. From 2005-2010 the amount of data produced in the world increased tenfold. And, by 2020, it has been projected that the amount of data in the world might exceed 40,000 exebytes (Uhl & Gollenia, 2014).

In our view, data literacy will help citizens identify relevant data, analyze and interpret data, participate more actively as community members, and use data in everyday civic contexts, such as public hearings about rezoning and land development and casual conversations with colleagues and neighbors. The use of data and data-driven argumentation to shape decisions and policies at the local level provides one opportunity to strengthen local democracy and democratic practices in general.

Current information infrastructures, digital devices and sensors can empower citizens to initiate data-driven investigations of community concerns, such as: heritage protection, water quality management, energy regulation, and public by-way safety. Citizens can identify pertinent issues and research questions, coordinate with other citizens, gather and publish data sets online, and moderate community discussions and deliberations. Their role would be akin to those of citizen scientists with the exception that there would not necessarily be oversight from a subject-matter expert. Citizens would have greater autonomy and agency with regard to such projects. We want to help to make such initiatives easier to organize and carry out and more visible to the larger community.

These threads converge in the transformative possibility that data-enabled citizens could more constructively and effectively participate in and shape local governance that is itself data-oriented. Local governance has often failed through conflicts grounded in irreconcilable judgment and self-interest (Coleman, 1957). Since there are other factors driving decision-making, data may not be a panacea for human conflict. Moreover, it may not be possible to realize a discourse that is free from judgment and interests. However, in our view, data can create an opportunity for more reasoned, rational discourse to come into being. At the very least, data are a shared community resource that has been under leveraged.

What could public hearings look like if more citizens were able to identify data relevant to a pressing civic issue and accurately analyze and interpret it? We believe that their contribution would almost certainly be more substantive and convincing.

With this in mind, we want to facilitate a series of community data hackathons: workshops where multiple stakeholders and subject matter experts generate ideas, plans, designs, or prototypes in response to a design brief that we will provide (Morelli et al., 2017). Hackathons and other kinds of community workshops have been used to engage citizens around open data and pertinent civic issues. We are aware that the term "hackathons" is packed with meaning, and so we are interested to discuss the strengths and limitations of referring to events like the ones we envision as hackathons.

A key element of our approach would that participants (citizens) contribute ideas and proposals directed at possible future courses of action. They do not merely generate diverse ideas (as in brainstorming) or criticize existing approaches. For example, at an early-stage hackathon, it might be possible to articulate different levels of data literacy, identify learning objectives to achieve those levels, and plan a series of community workshops aimed at achieving those objectives. This means that our participants would not only be charged with advancing planned work but in planning the work as well, including (potentially) defining the metrics for its success. We see this level of citizen participation as a crucial step towards sustaining a long-term community data project.

We want to leverage our experience in community-scale participatory design (Carroll et al., 2000; Carroll, 2012; Carroll & Rosson, 2013) to engage a wide range of community stakeholders in large-scale action research. Participatory design is both an inclusive and equitable approach to design and a method for engaging with and learning about stakeholder values, practices and knowledge (Béguin, 2003; Simonsen & Hertzum, 2012). But we also want to fully leverage best practices in intensive, one-day hackathon events.

## Acknowledgements
This work is supported by a Faculty Fellowship from Student Enagegment Network of Pennsylvania State University.

# Appendix B

## Posters

# You Hacked and Now What? – Exploring Outcomes of a Corporate Hackathon

## Background

- Hackathons are becoming more and more popular across various domains.
- Most studies **focus on the event itself**
- Studies that focus on hackathon outcomes:
  - Are conducted in the space of collegiate or civic events
    *(e.g. Almirall, E., et al. (2014), Nandi, A., & Mandernach, M. (2016), Tandon, J., et al. (2017))*
  - Focus on singular outcomes
- Existing studies found:
  - That project continuation is an issue
    *(e.g. Carruthers, A. (2014), Guerrero, C., et al. (2016), Cobham, D., et al. (2017))*
  - Effects of participation on individual networks
    *(e.g. Lapp, H., et al. (2007), Leclair, P. (2015))*
  - Effects on individual competences
    *(e.g. Nandi, A., & Mandernach, M. (2016))*

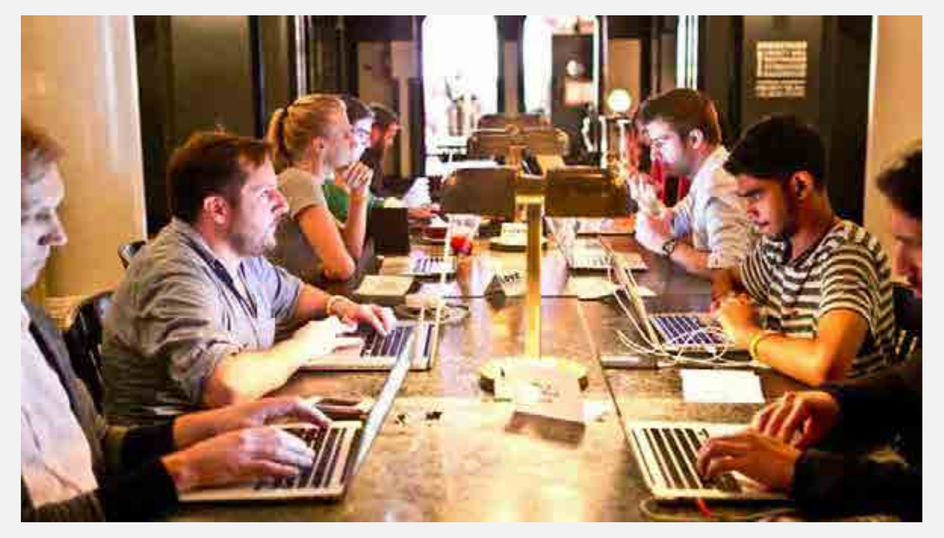

*Civic data meetings*

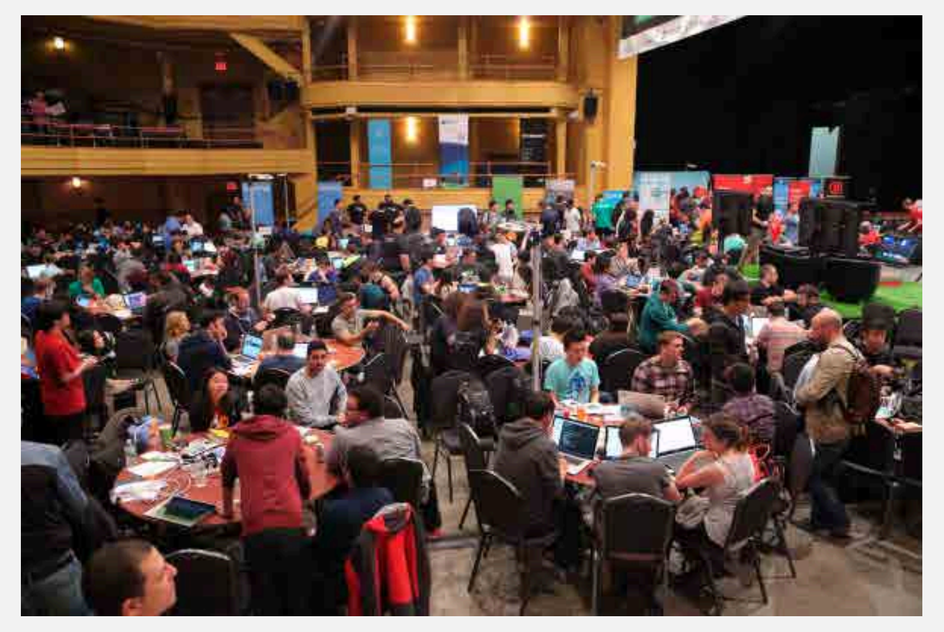

*Collegiate event*

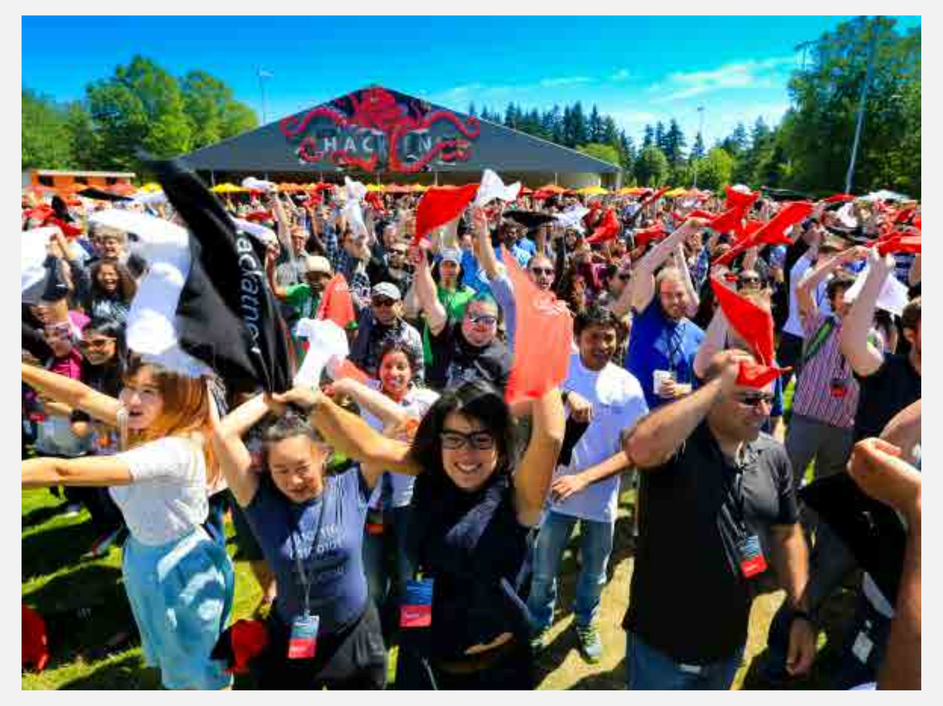

*Corporate hackathon*

**RQ:** What are potential outcomes of corporate hackathons related to projects, individuals and teams?

## Case study

### Microsoft One Week Hackathon 2017

- 4 days (3 days hacking and 1 day fair)
- 6.700 participants, 1.800 projects in Redmond alone
- Study population:
  - 5 teams (3 to 7 members each)
  - Selected based on relation of project to work and familiarity of members
- Mixed-method study consisting of:
  - Pre-interviews
  - Observations
  - Post- and post-post interviews
  - Surveys

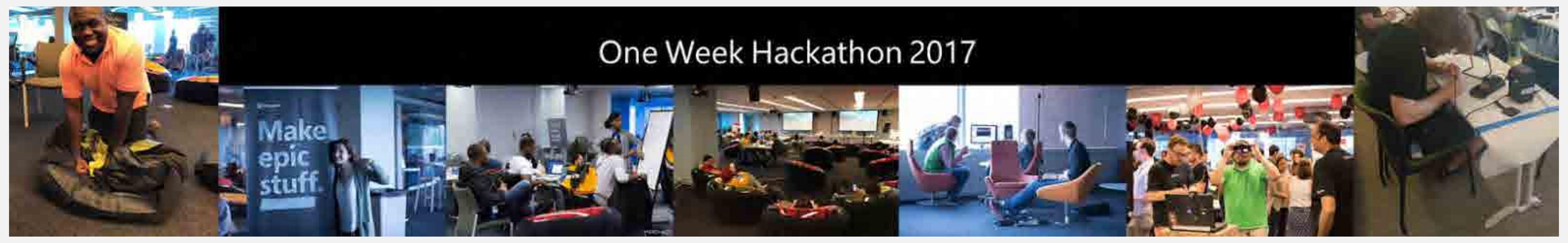

### Timeline

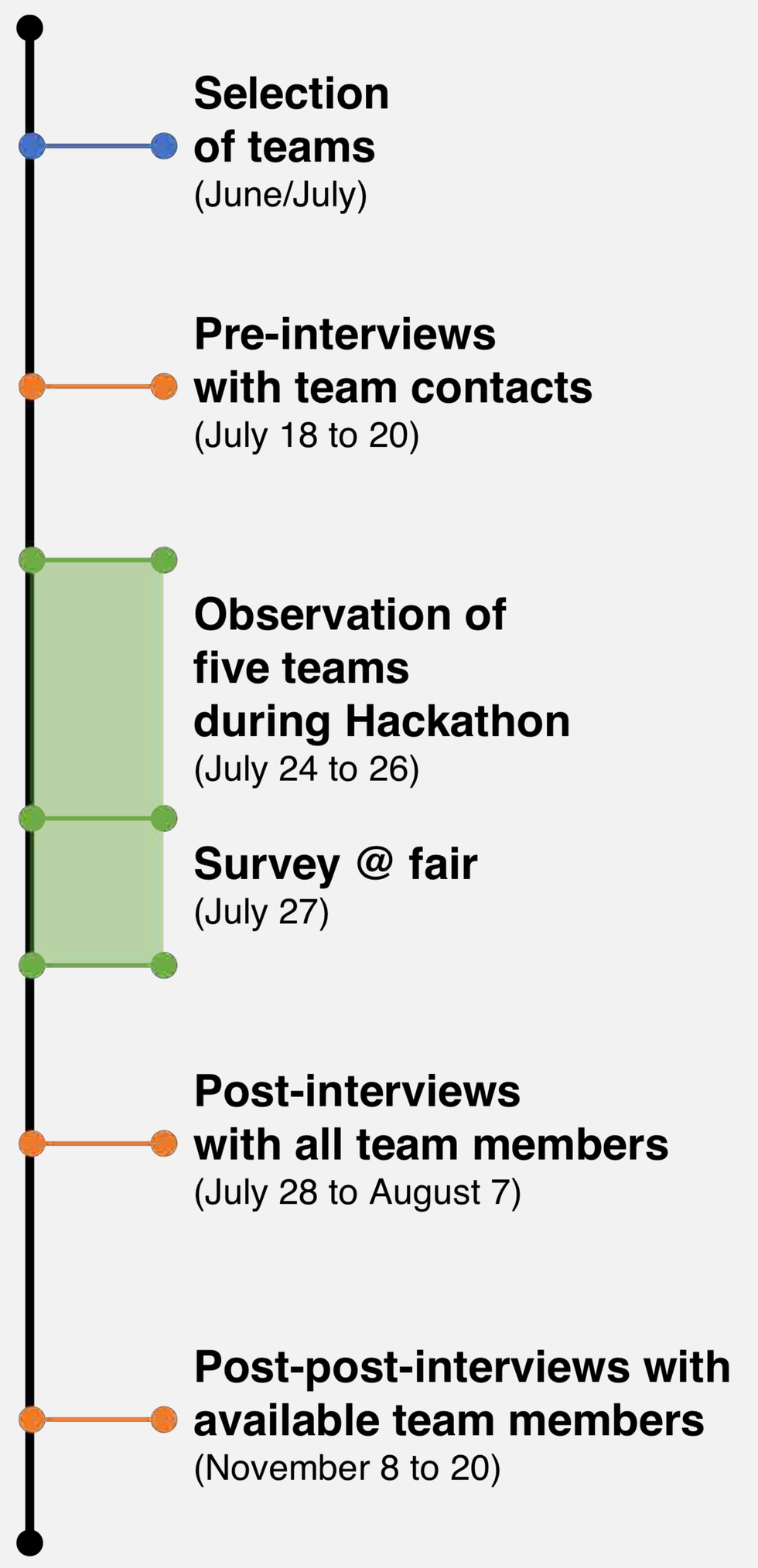

## Insights

### Projects

- Most participants **wanted to continue** their projects (70%) and wanted to continue working with the people in the team (70%)
- **Two out of five** projects will be continued
- **No original team member** will be part of the continuation
- One group leader attempts to stay involved
- Continuation required:
  - Individual intentions
  - Internal marketing (outside of fair)
  - Fit to existing products
  - Market need
  - Resources

### Individuals

- Effects on individual **skills**:
  - Acquired new technical skills or extended them
  - Project leaders improved project management skills
  - Sparked interest to acquire new skills and instill confidence in the ability to acquire them
- Effects on individual **careers**:
  - Promotions based on hackathon performance
  - Improved perception of participant by manager
  - Improved perception of participant by other employees
- Effects on individual **networks**:
  - Networks of newly formed teams increased through hackathon
  - Networks of project leaders improved through exposure

## Open questions

***How can we scaffold activities towards outcomes?***
Organizing hackathons in a certain way might lead to specific outcomes.

***Which outcomes are peripheral?***
Some outcomes might happen anyways while others can be influenced actively.

***How to integrate hackathons?***
They could become an integral part of innovation and software engineering practice.

***How much should we plan?***
Too much planning might turn hackathons into just another work task.

Carnegie Mellon University · Tartu Ülikool · University of Pittsburgh · Deutsche Forschungsgemeinschaft DFG

**Alexander Nolte – alexander.nolte@ue.ee – www.anolte.com**

partially funded by DFG under grant no NO 1302/1-1

# Organization for Human Brain Mapping Open Science Special Interest Group

## *Accessible and Inclusive Neuroinformatics*

### Motivation & Objective

- We wish to **educate community members** on best practices in open science and data sharing in neuroscience.
- We encourage the development, maintenance, and community engagement of **open-source software**.
- We promote the **free exchange of ideas** to encourage future collaborations and foster better science.

### Outcomes & Resources

- We promote **community-led demonstrations** for various neuroscientific tools through an **online web-series.**
- We create and curate learning resources such as **Brainhack 101**, which are all made publicly available.
- Brainhack is an **annual OHBM event since 2013**, and **Brainhack Global** has been held at over **40 sites in 16 countries**.

### The Brainhack Community

- The **Brainhack website**, https://brainhack.org, showcases projects, proceedings, and events from around the world.
- The **Brainhack Slack** is home to thousands of members and acts as a collaborative environment both in and out of workshops.
- Our Winnower post, "A step by step guide for organizing open collaborative 'Brainhack' events" provides **a guide to organizers**.

### Emphasis on Accessibility

The atmosphere we aim to create at our events is one of inclusivity and accessibility. Established participants engage with less experienced attendees, helping teach them skills for shared projects.

Of the group shown below, 5 of 6 members were new to Brainhack events, 3 of them new to hackathons of any kind, and 1 had never made a pull-request prior.

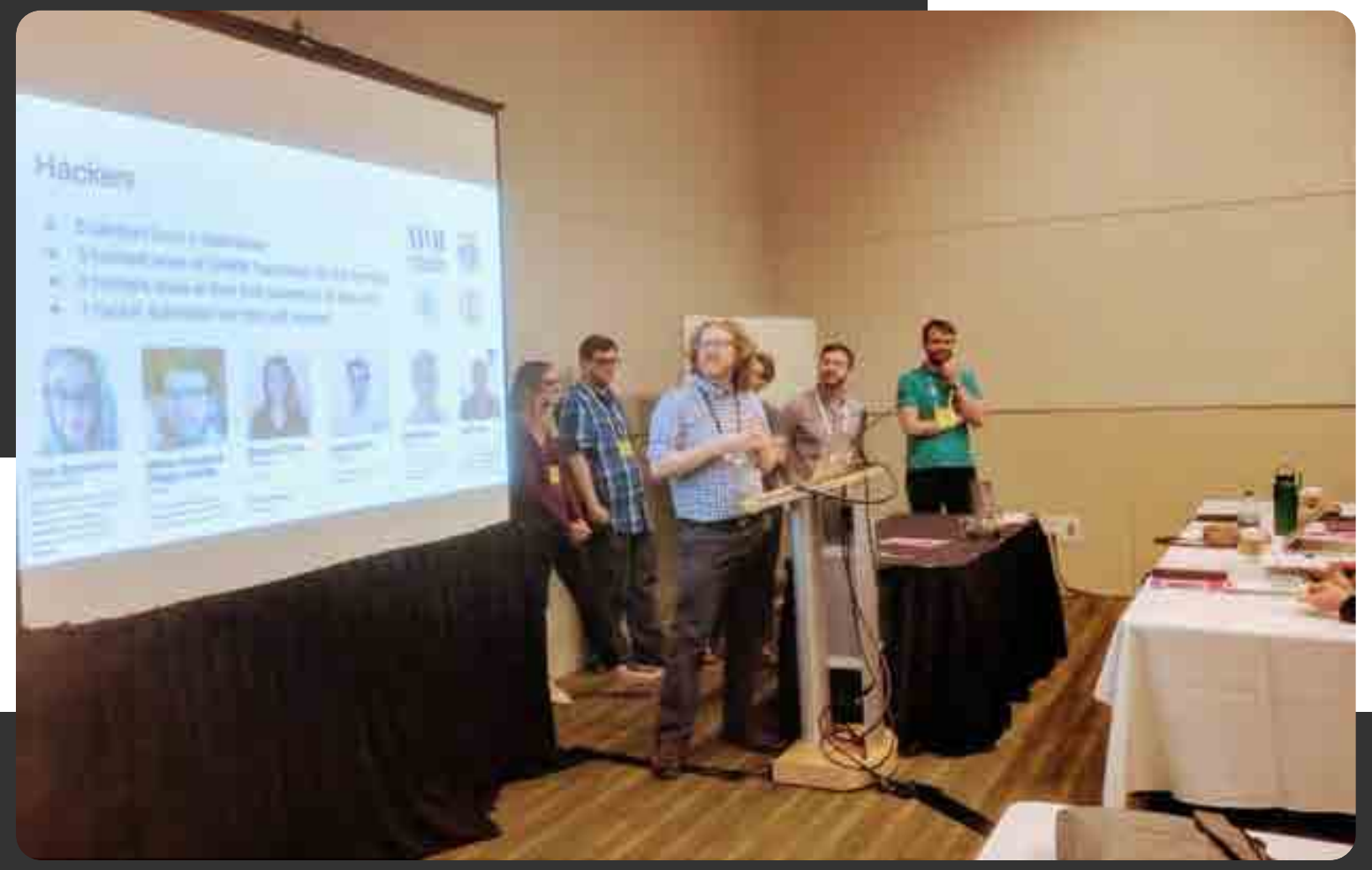

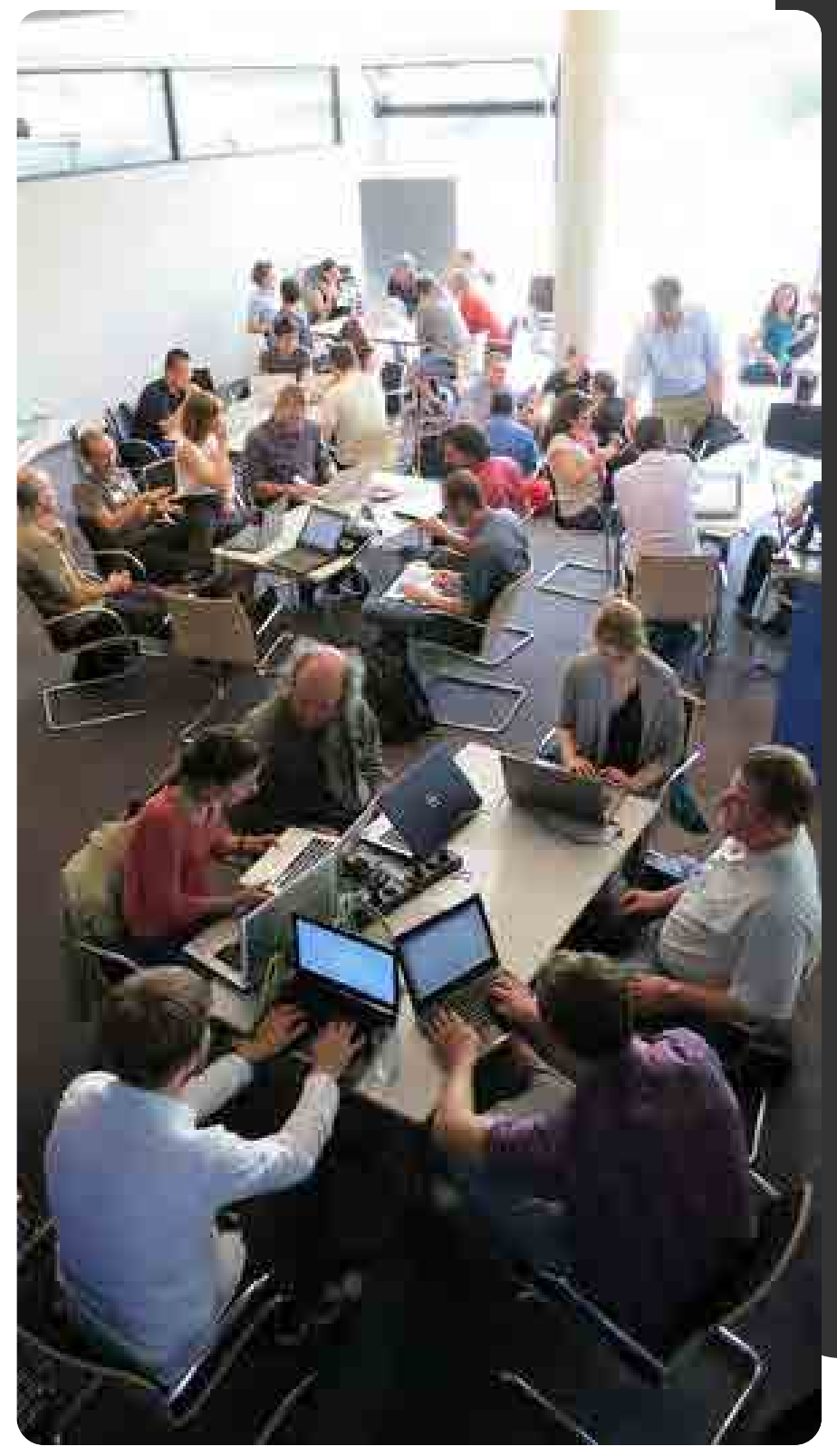

### Encouraging Collaboration

Throughout events, an emphasis is placed on collaborating across existing groups and boundaries.

Icebreakers and project pitches help identify overlapping interests between participants, and unconference sessions provide a structured outlet for discussion.

Regularly, event teams consist of members from different institutes. Reccurring projects such as Nipype, BIDS, NIDM provide accessible entrypoints for newer participants.

### The Brainhack Structure

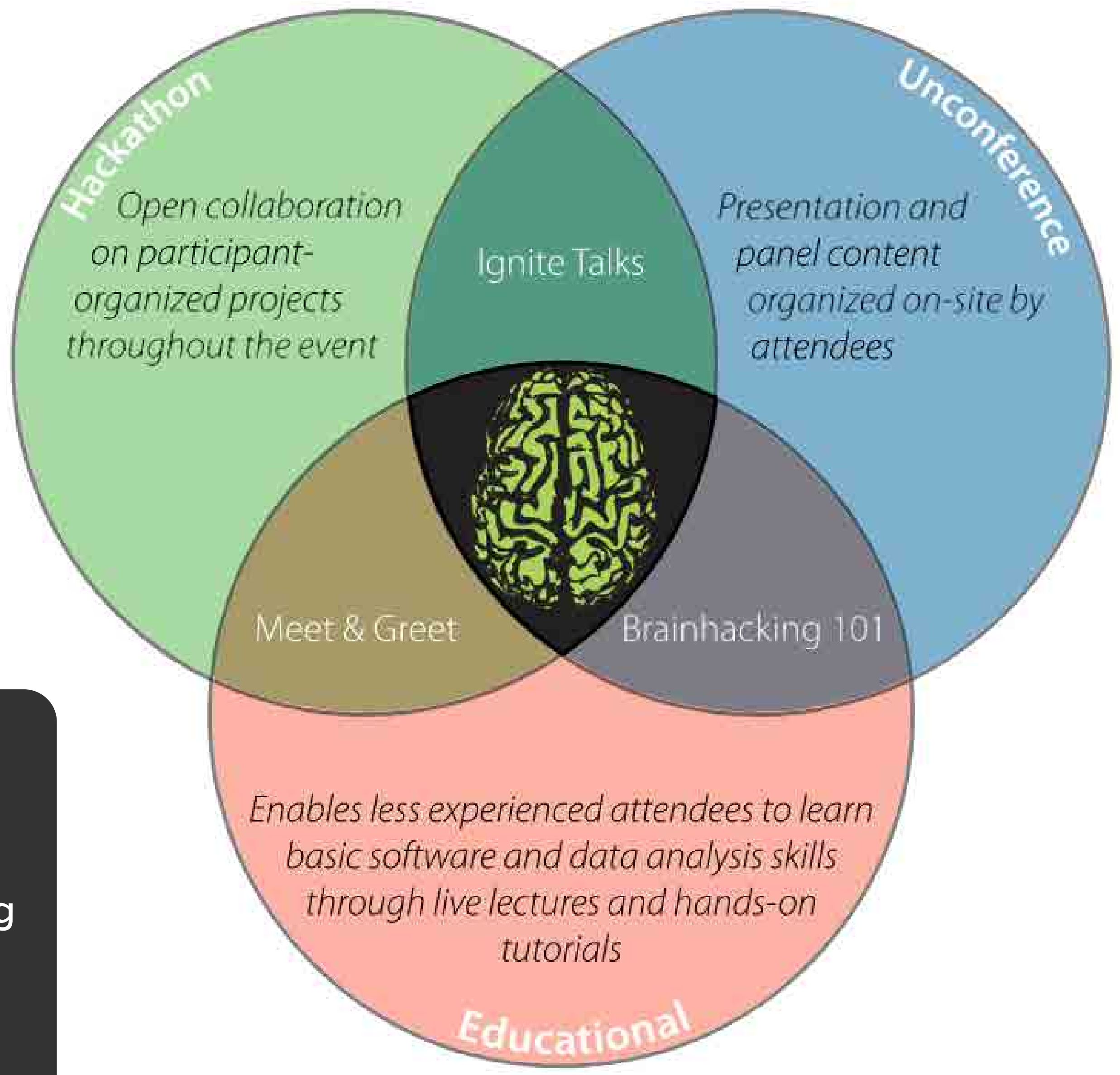

### Example Format

We strive to balance education, hands-on hacking, and built-in collaboration, providing a full schedule for novice and more experienced participants.

| Time | HackTrack | TrainTrack |
|---|---|---|
| Event Opening | Check-in | |
| | Welcome, Intro, and Icebreakers | |
| | Mingling and Project Pitches | |
| Repeated Throughout Event | Meal | |
| | Unconference Discussion | |
| | Hacking | 1-2 Hour Tutorial |
| | | 1-2 Hour Tutorial |
| | ... | ... |
| Event Closing | Project Presentations | |
| | Survey and Wrap-up | |

The schedule here is adapted from our 2018 OHBM event.

### Resources

**CHCI**

Center for Human-Computer Interaction

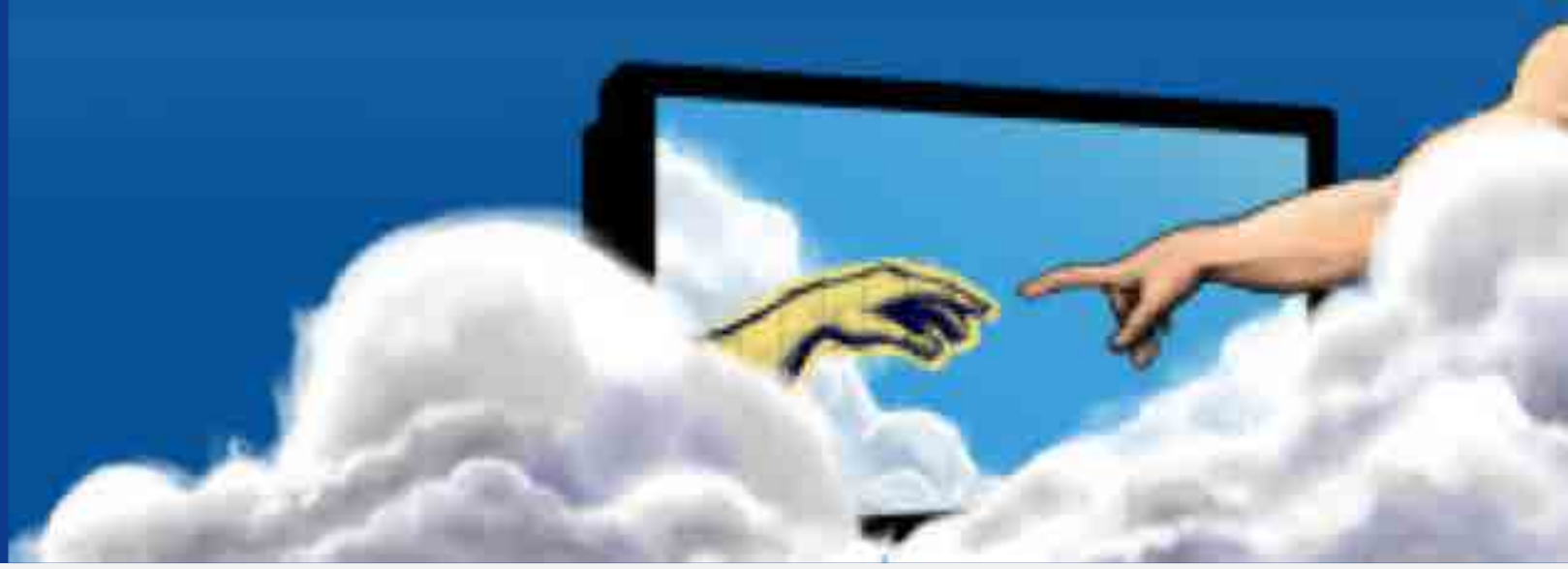

# Community Water Data Hackathons

## Awareness + Participation + Engagement

- PD with 100,000 people
- Leverage existing hyperlocal network of people, practices, data
- Informal learning about data, data-driven thinking
- More effective citizen deliberation, policy, governance
- Continuing projects in everyday innovation

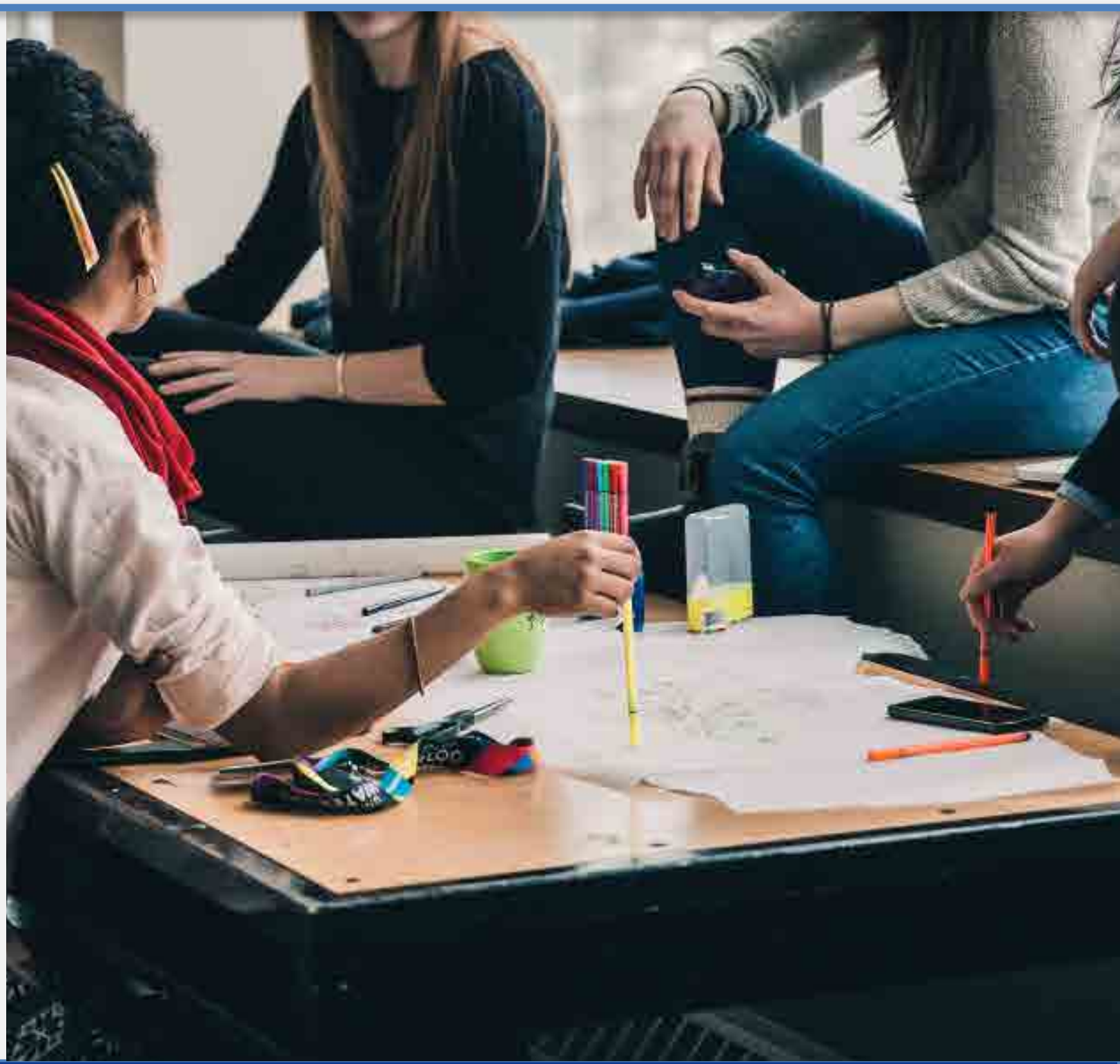

# The disappointing adventures of...

# HACKERGIRL_1852

By Siân JM Brooke

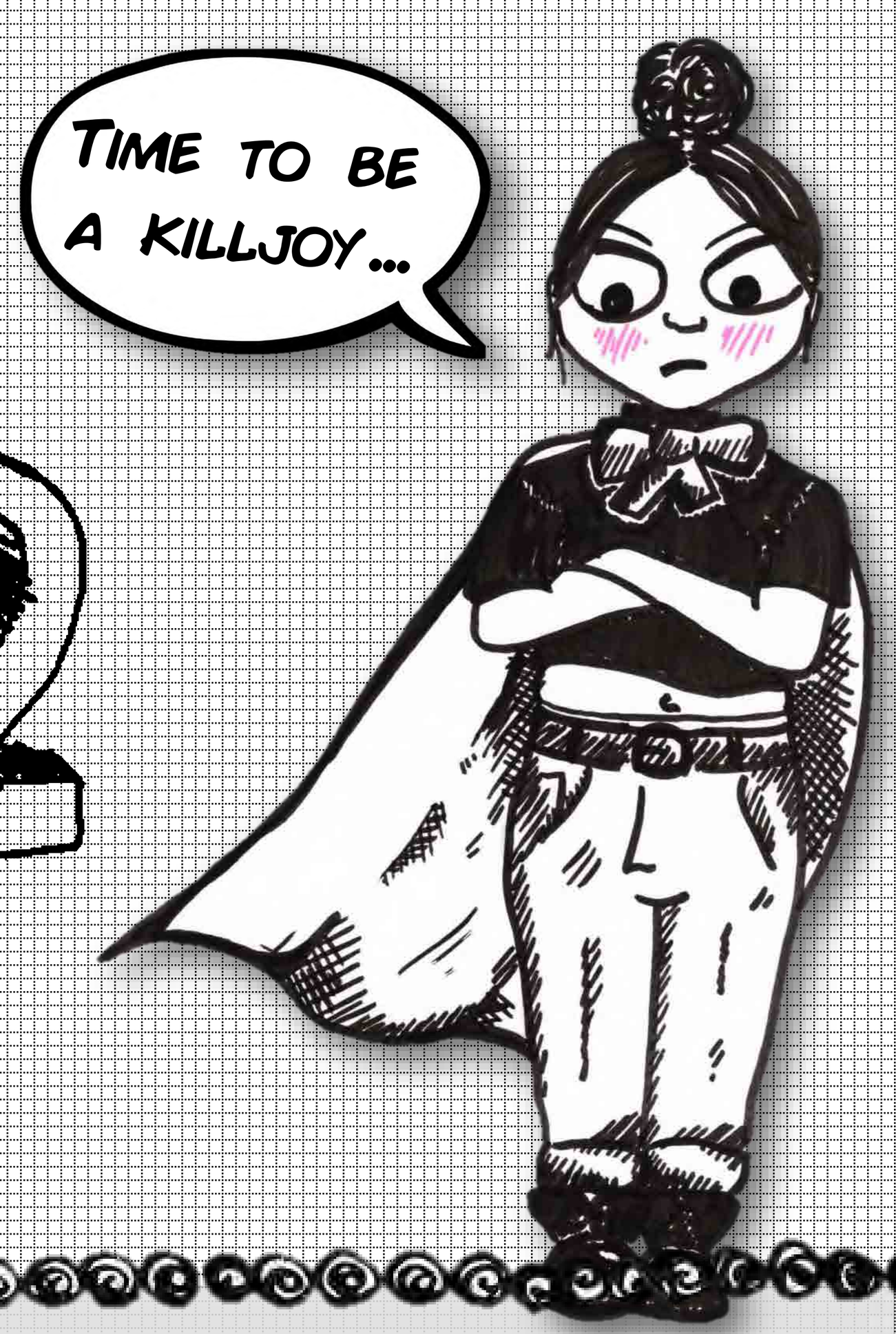

## The importance of telling stories

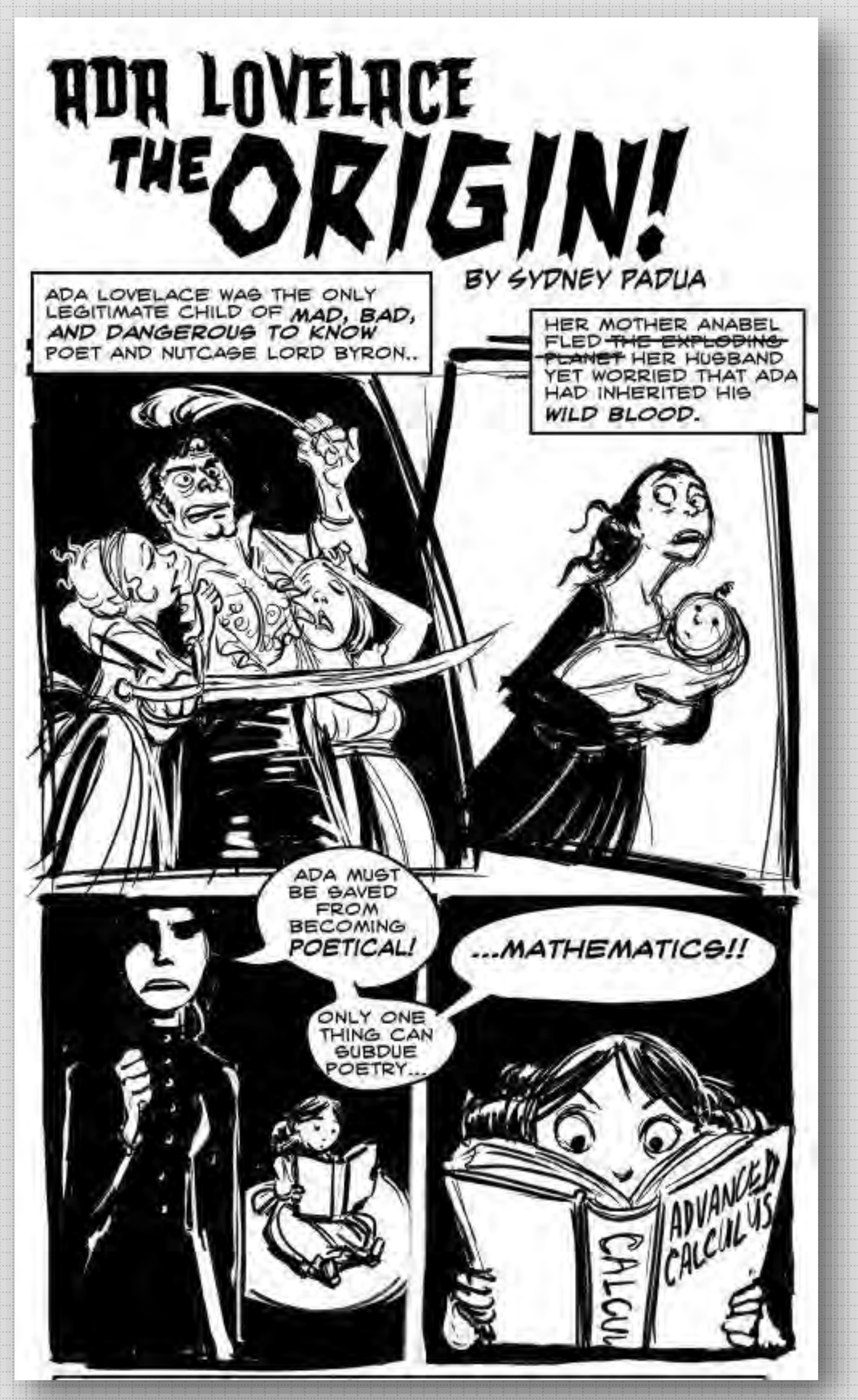

Cyberpunk fiction and comics have become a key element of how hackers and creative programmers perceive themselves since the 1980's. The genre often romanticises historical Victorian figures, giving new life to "forgotten" women of hacking.

Sydney Padua (2015) wrote the steampunk comic based on the life of Ada Lovelace, the inventor of the first computer programme. This poster is (loosely) based on this style to reflect the remembering of women 's history in technology and in spaces of hacking.

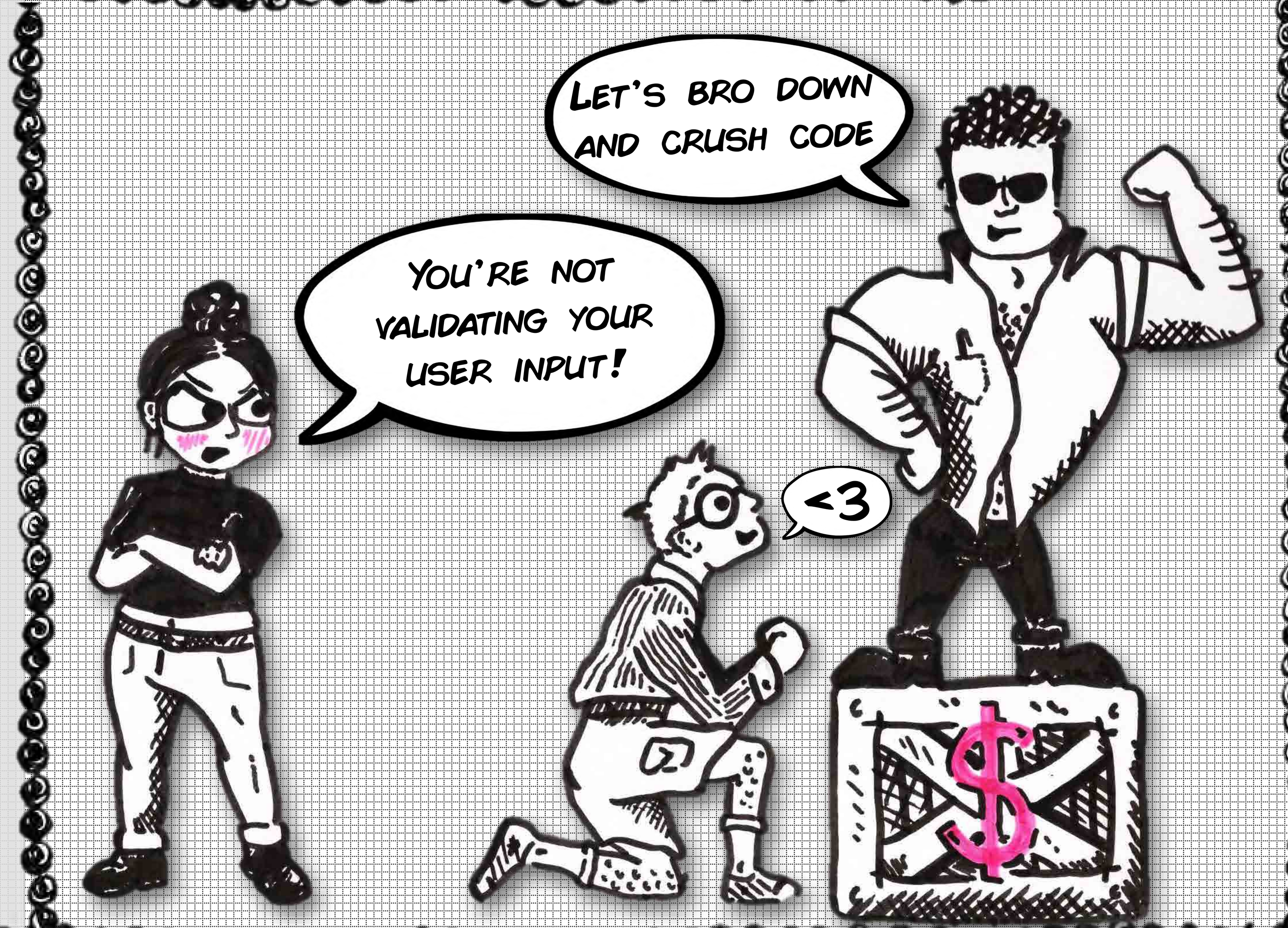

## Brogrammers *noun/US/informal:*

Overt sexism and hyper-masculinity has emerged in hackathons in recent years, in the form of the **BROGRAMMER**. The term is often used pejoratively but some male programmers use it positively to reflect how social and outgoing they are.

It gained popularity following a Mother Jones article entitled **"Gangbang interviews and Bikini Shots"** which saw the recasting geek identity with a frat house swagger as a dangerous game.

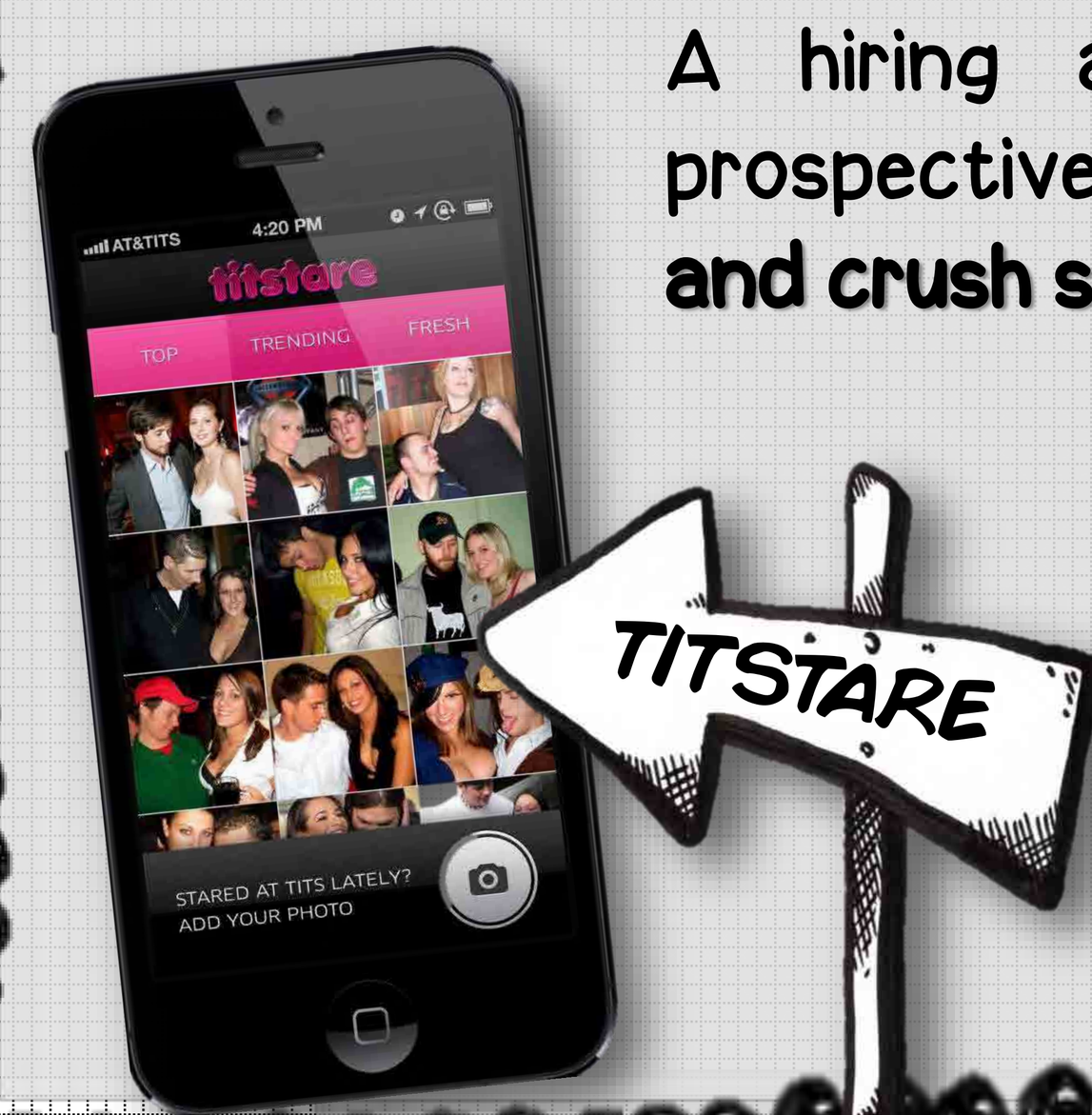

A hiring advert at Stanford University asked prospective employees if they wanted to **"Bro down and crush some code"**.

At the Disrupt 2013 hackathon the fictional mobile app **TitStare** was pitched.

Such discourses are typical of the programming subculture and the evolution of the uncool **GEEK**.

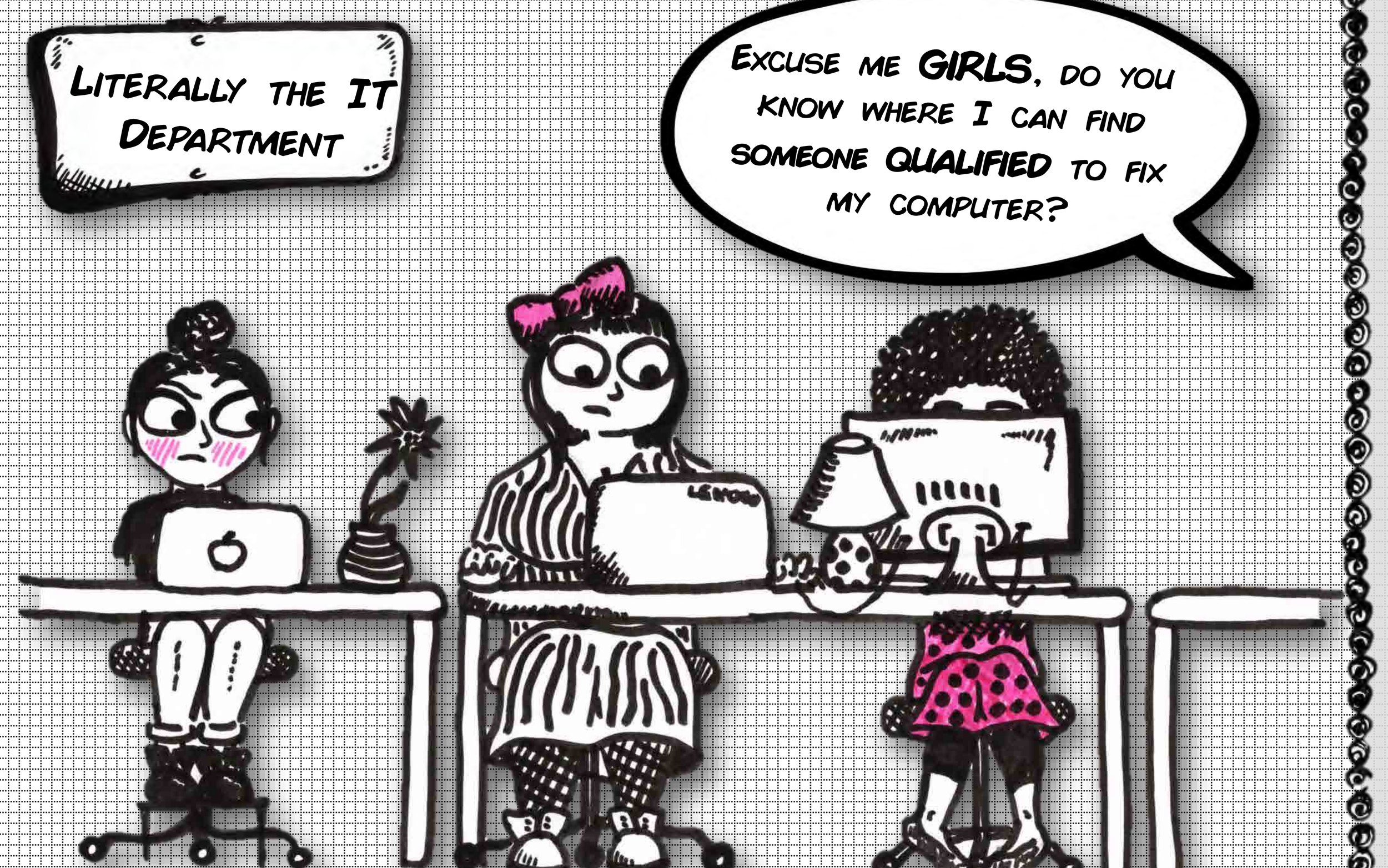

## The Oblivious Hacker

Much scholarship points to the dearth of women in creative coding but provides little explanation for this absence. Tanczer (2015) shows that women do participate in the social spaces of hacking, but online they are masked by the obliviousness of male discourse.

## Social Justice Warriors (SJWs)

ESR (2015) dismisses those who critique the male dominance of hacking as SJWs and the enemy of the hackers "cult of meritocracy". A SJW is seen to be motivated by personal validation rather than political conviction. Those who speak out are **FEMINIST KILLJOYS** (Ahmed, 2010) disrupting a collective delusion of happiness and equality.

## Hacker's Wives

Levy's (2010) work *Hackers: Heroes of the Computer Revolution* begins with a "Who's Who of Hacker Culture" with 52 men, 10 computers, and three women.

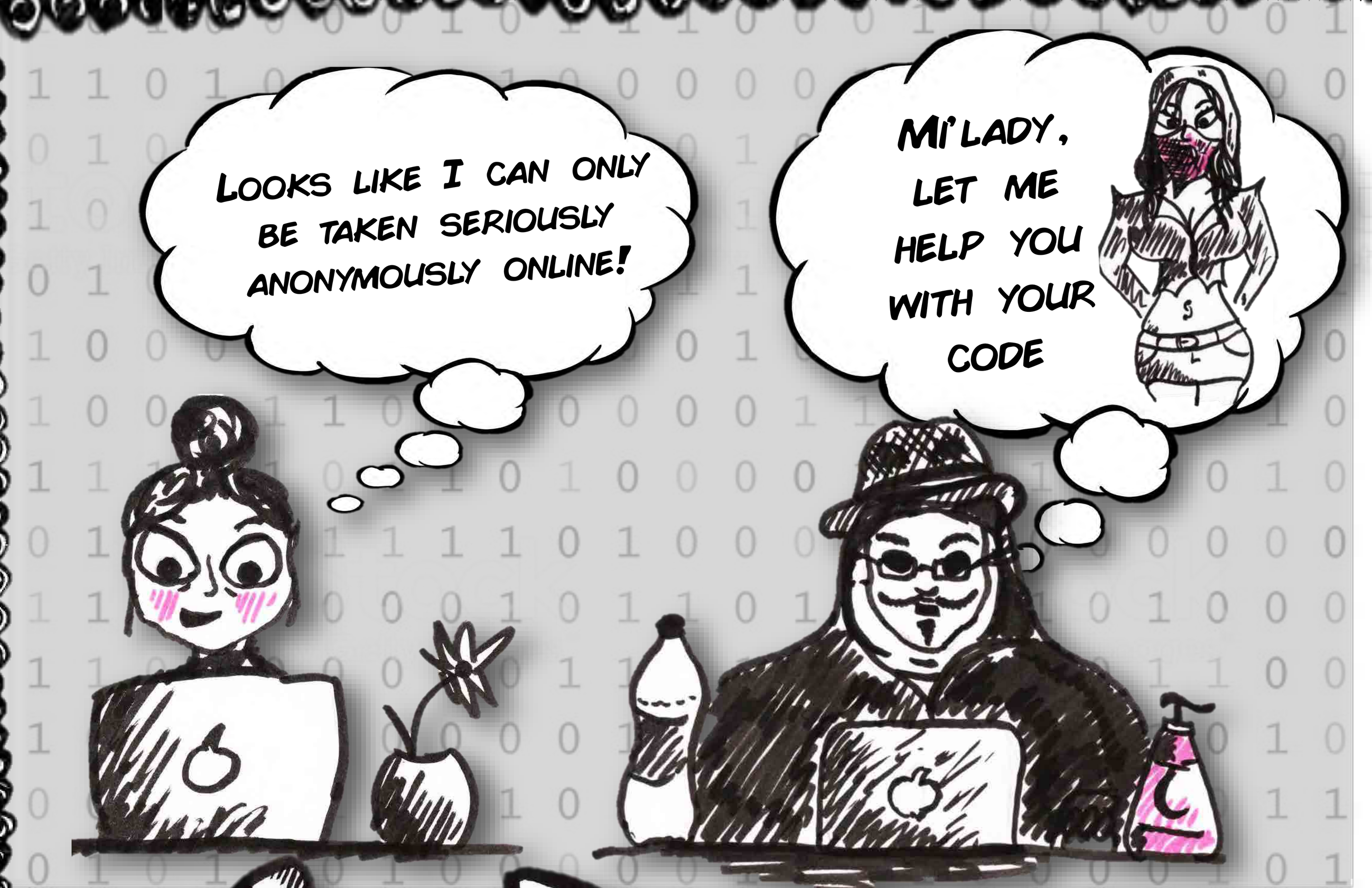

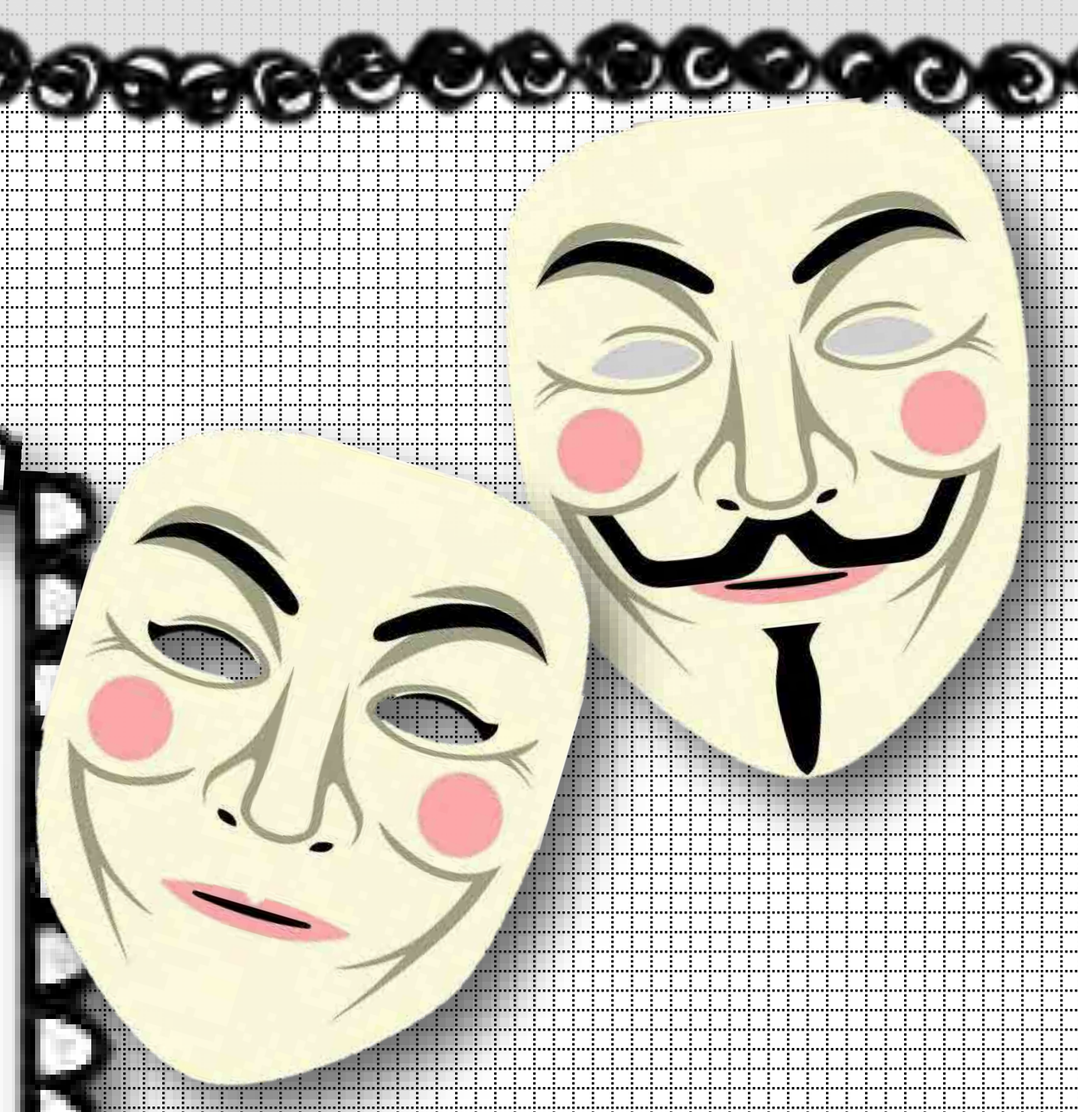

## Breaking Gender Code

Women have developed hackerspaces and places in which femininity in hacking is normative and necessary. The physical spaces are formed on the basis of interest and openness, rather than the pre-proven ability of the Hacker Ethics.

Hackathons are fundamental to inclusivity in hacking, in which the physical and digital spaces of hacking **"powerfully reinforce each other"** (Coleman, 2010, p.49). Organised spaces include **HACKERMOMs**, a women's only space with a DIY ethic and onsite childcare bringing domesticity into hacking.

## Deconstructing Resistance

The feminine physical spaces of women in hacking continue to define femininity corporeally; you may only be a woman hacker by presenting a female identifying body. Women-only spaces are still defined by the absence of masculinity.

**The cultural privilege and normativity of masculinity in relation to the feminine other is reinforced and the social hierarchy of hacking as a masculine space is re created.**

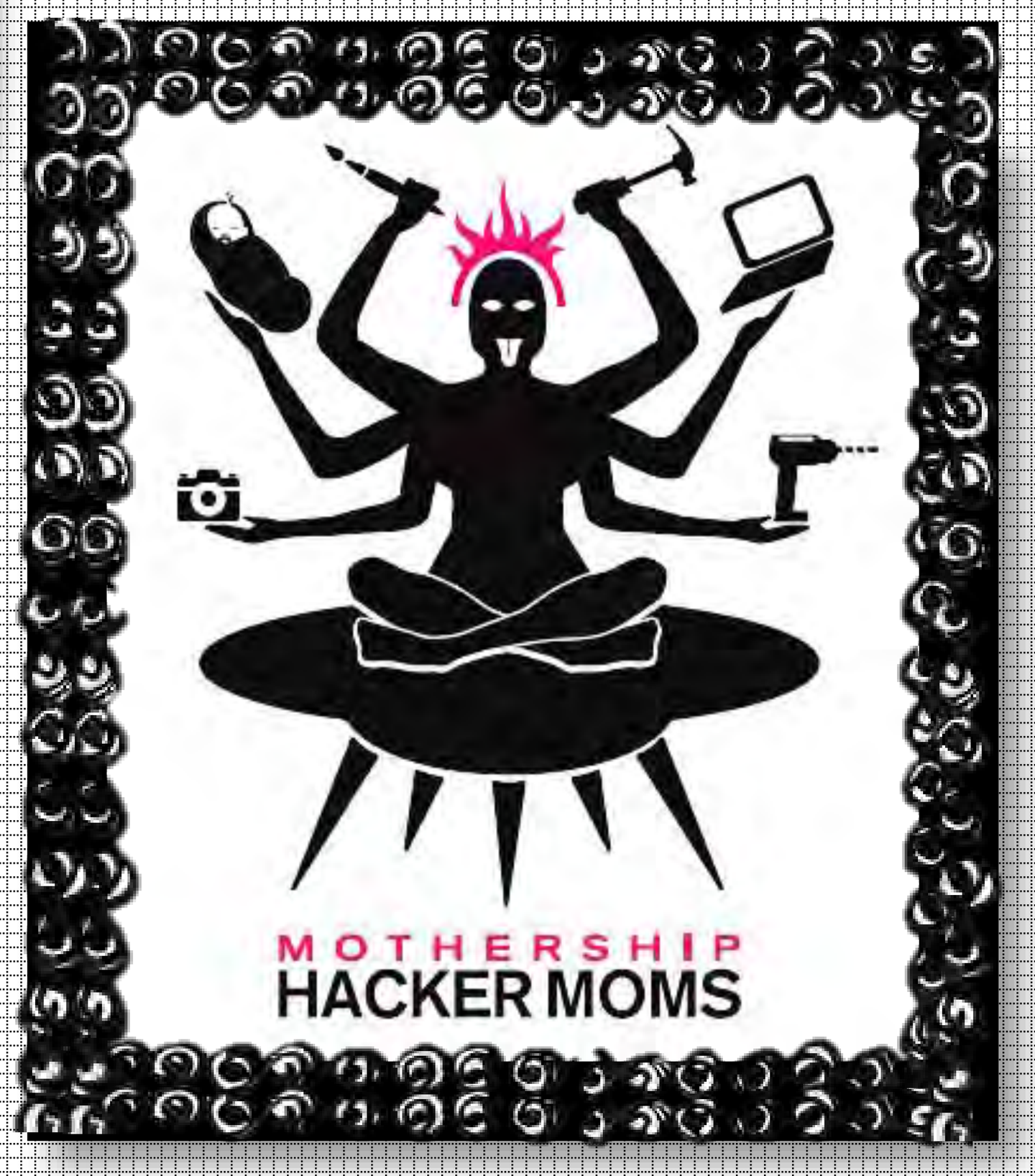

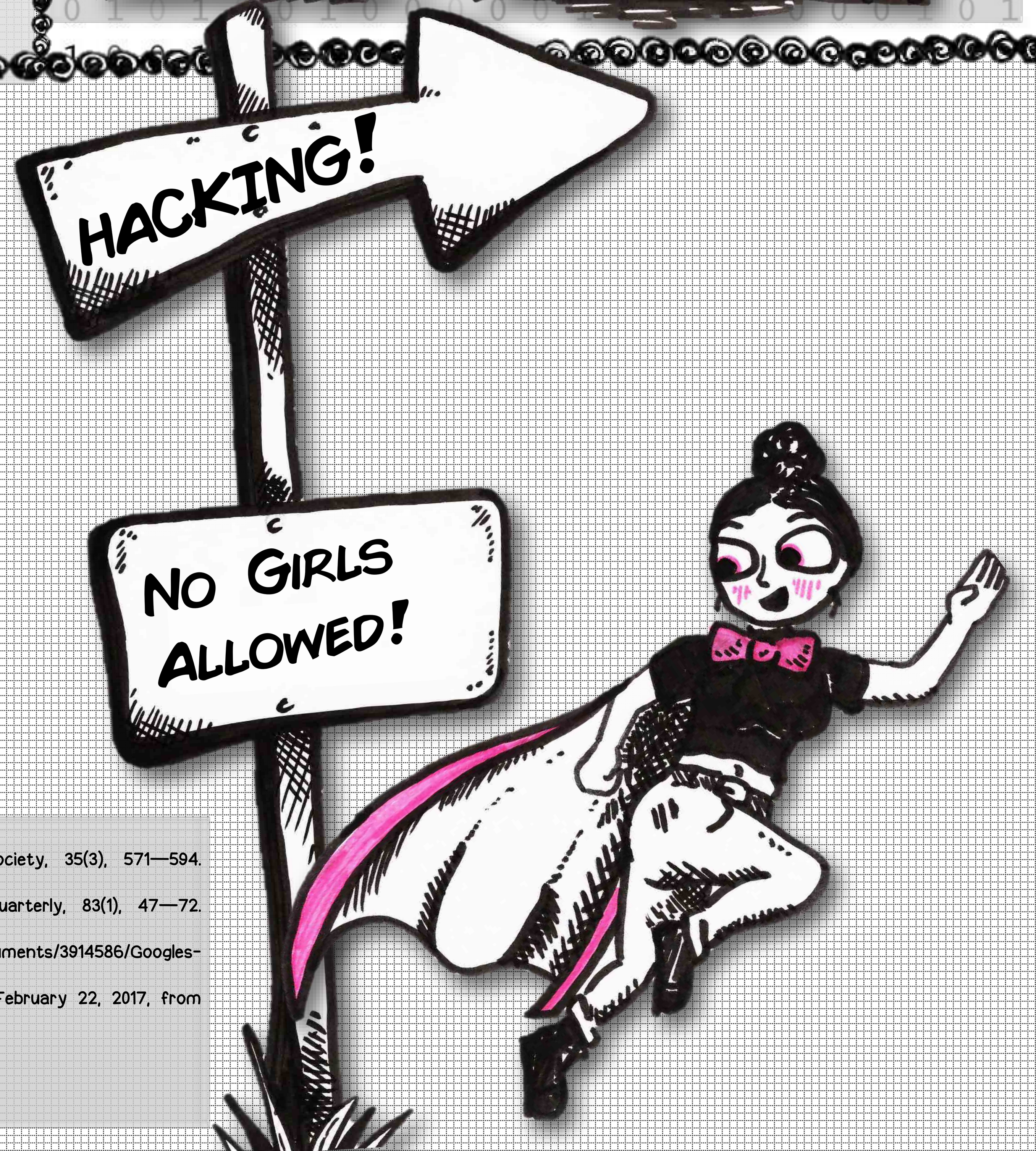

# ENGAGING TIME-POOR DESIGNERS IN PHILANTHROPIC ACTIVITIES

**OUR GOALS AND CONSTRAINTS**

1) Help young people do impactful “design for good” as a philanthropic activity.
2) Provide a meaningful experience for young people to understand UX design as a career choice.
3) Start young people on a path to life-long engagement in philanthropy.
4) Increase the reputation of Bloomberg among UX academics, professionals and students for the purpose of recruiting.

**OUR QUESTIONS**

1) How can a user-centered design process fit into a 1-day event?
2) How best to engage time-poor students in this event?
3) What is the role for UX professionals in the event?
4) What format of deliverable can best achieve the goals?
5) Should the procedures be piloted before the event?
6) How do we measure the success of the event?

**Bonnie E. John**
**Paddy O’Flaherty**
**Shahtab Wahid**
**Anthony Viviano**

Bloomberg

# Collaborative community events in open source biological research

Brad Chapman
Harvard Chan School Bioinformatics Core

HARVARD T.H. CHAN SCHOOL OF PUBLIC HEALTH

Open Bioinformatics Foundation OBF

## Open Bioinformatics Codefest

- Two day community working session
- Open Source
- Bioinformatics
  (Biology + Programming)
- Extension of yearly conference
- Build existing online relationships
- 9th year

## Successes to share

- Collaboration, not competition
- Self-organizing groups
- Standards development
- Maintenance and documentation

Another flaw in the human character is that everybody wants to build and nobody wants to do maintenance.
- Kurt Vonnegut

## Experiments we're trying

- Flexible conference format
- Training -> Conference -> Codefest
- CollaborationFest -- beyond code
- More days for working
- Additional events during the year

## Help please

- Improve diversity
- More welcoming to newcomers
- Better training and orientation
- Scaling to more attendees
- Build leadership and recruiting team

# Hacking Academia

## How Academic Communities are Evolving a New Breed of Hackathon

Margaret Drouhard, Brittany Fiore-Gartland, and Anissa Tanweer

## Hack Weeks

Hack weeks are designed to build out infrastructure, practices, and culture for members of an academic discipline or sub-field.

Our access to longitudinal data from multiple iterations of various hack weeks, as well as our ability to triangulate analysis using survey responses from participants, have allowed us to explore some of the motivations that draw participants to hack weeks and how these events are perceived to fit into the broader ecology of academia and scientific exploration.

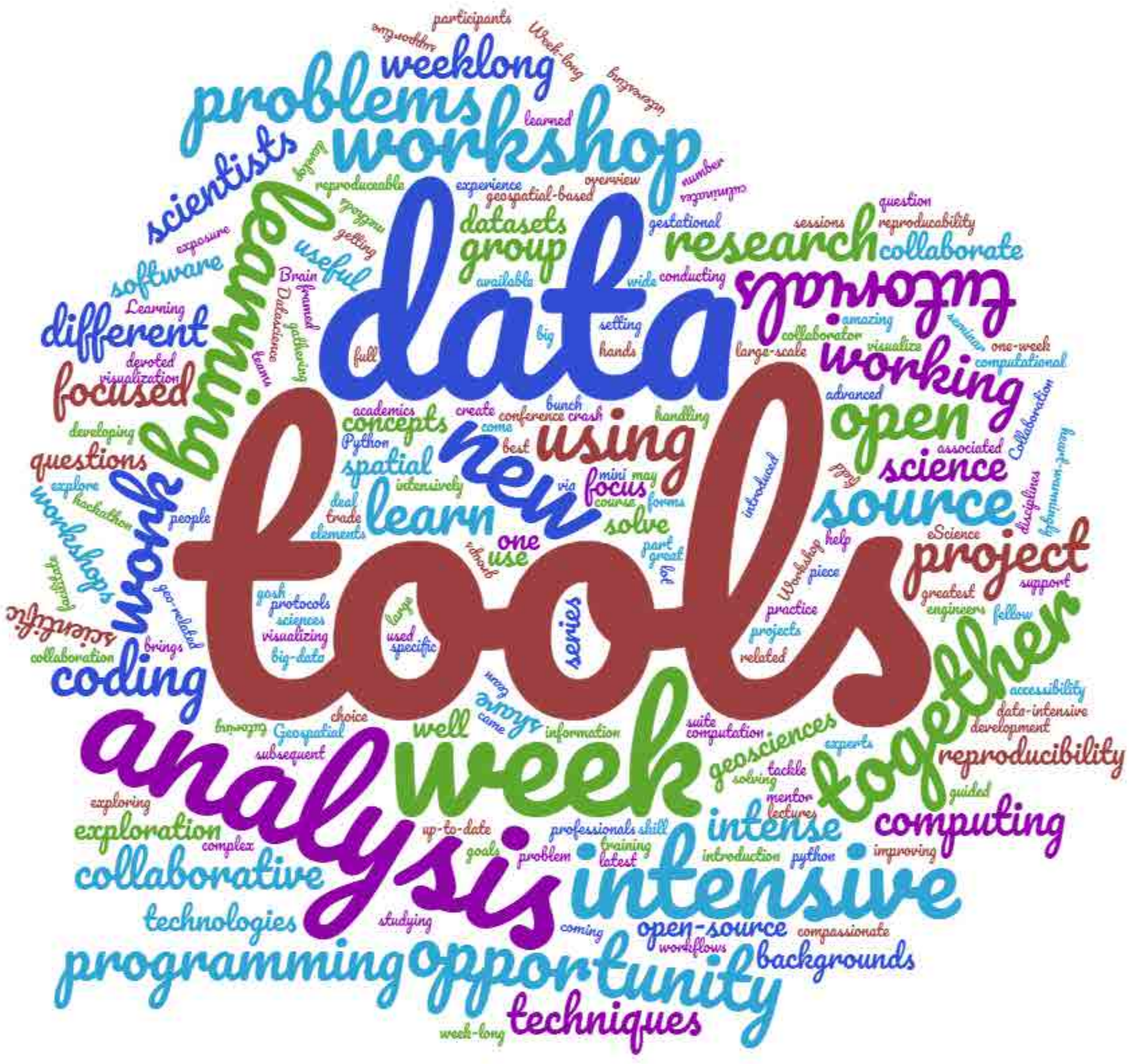

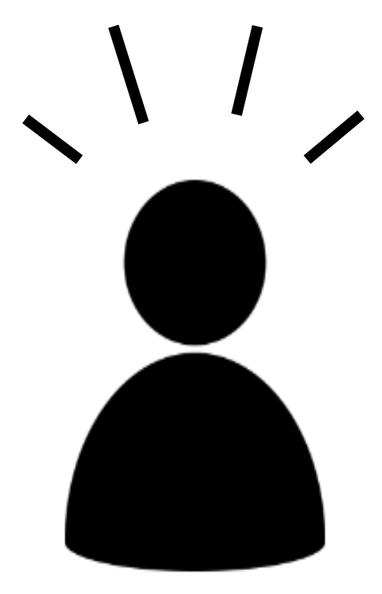

## Methods

- Ethnographers embedded within academic data science environments
- 175+ hours of participant observation
- Observations of multiple iterations of hack weeks and organizing work in advance

## How Participants describe Hack Weeks

"A week of intensive, but heart-warmingly supportive and compassionate, exposure to a wide suite of technologies for improving data science workflows, accessibility, and scientific reproducibility.."

## Findings

- Motivations for participation:
  - hack week community
  - space for a different kind of work
- Lenses for institutional change:
  - stopgap measures
  - response to material exigencies

HUMAN CENTERED DESIGN & ENGINEERING
UNIVERSITY of WASHINGTON

Contact: mdrouhard@acm.org

# Hackathon Team Leadership: Supporting Innovation Through Teaming at Time-Bounded Events

*Eureka Foong & Elizabeth Gerber, PhD | Northwestern University |*
*eureka@u.northwestern.edu, egerber@northwestern.edu*

Northwestern University
DELTA LAB

## How might hackathon event organizers plan for more effective individual teaming?

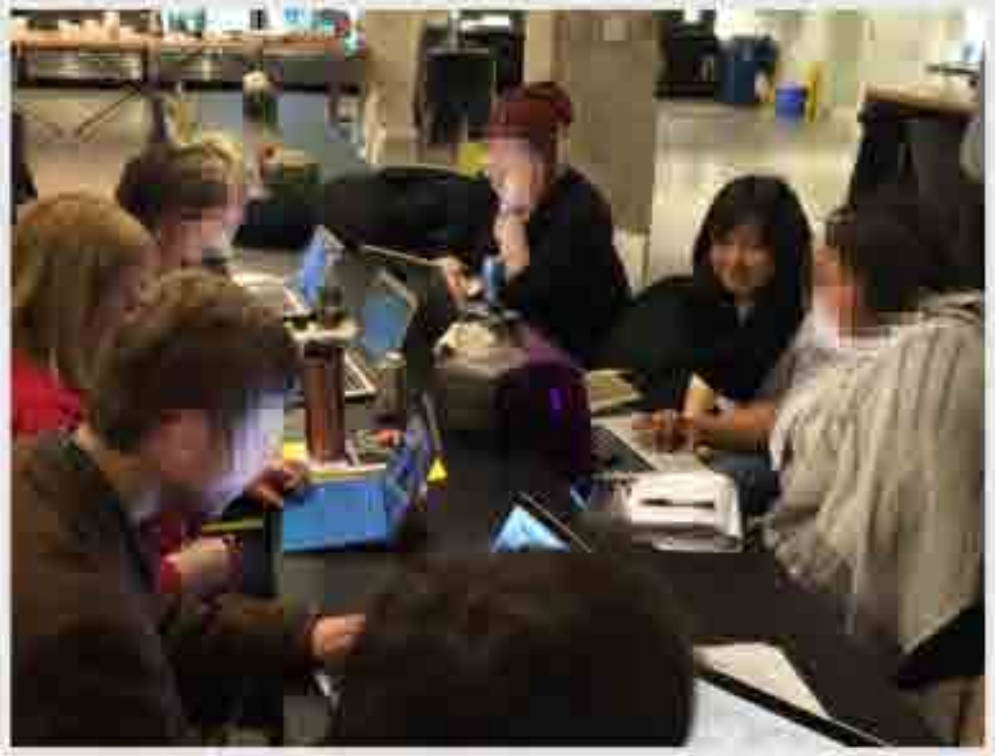

## Motivation

Although exposure to diverse ideas and skills is essential to individual team innovation, innovative teams must also prioritize teaming processes, such as setting meaningful goals and cultivating psychological safety. However, hackathon event organizers currently pay little attention to these processes, instead attending to event logistics.

## Method

The first author conducted a 6-week participant observation of a weekly civic hackathon in Chicago (Jan-Apr 2016). We analyzed memos and an interview with an event organizer using Edmonson's (2013) Framework for Effective Teaming.

## CHALLENGES AT HACKATHONS

**Aiming High** — 1. Setting inspiring goals is dependent on the presentation skills of individual team leaders

**Teaming Up** — 2. Newer group members lack familiarity with team history to be able to contribute to discussions

**Failing Well** — 3. Meetings with community members help individual teams test ideas, but are coordinated by individual team leaders

**Learning Fast** — 4. Time spent need-finding takes time away from individual teams to prototype solutions

**Repeating the Process** — 5. Individual team leaders spend valuable time outside of meetings to document individual team progress

*Edmonson's (2013) Framework for Effective Teaming*

## Recommendations for Supporting Individual Team Leaders at Hackathons

1. Encourage psychological safety in teams, especially for new members

2. Help team leaders form and maintain connections with community partners

3. Provide alternatives for documenting team progress beyond code repositories

## Future Work

We will test the effectiveness of these recommendations by conducting additional interviews with event organizers and participant observations. We will assess how these recommendations influence the frequency of participation of new members, the involvement of community partners, a team's ability to iterate, and the effectiveness of solutions produced by teams.

HARVARD | BUSINESS | SCHOOL

# Iterative Coordination in Organizational Search

Sourobh Ghosh, Andy Wu
Harvard Business School

## Abstract

Firms use iterative coordination, or periodic coordination meetings, in their technology development on a presumed link to both exploratory innovation and exploitative performance. We critically evaluate this practice and identify boundary conditions to its effectiveness. With a leading technology firm, we embed a field experiment within a software development competition to measure iterative coordination's effect on firm outcomes and search process. We find that iteratively coordinating firms conduct more productive overall search, but face trade-offs between exploration and exploration: while sampling more distant neighborhoods on their landscapes, these firms ultimately exploit at the expense of exploring. Our findings contribute to literatures on organizational search and strategy formation in entrepreneurial settings. Methodologically, we introduce a novel experimental data collection methodology enabling granular minute-level search measures.

## Firm Outcomes

Third party panel of experienced judges evaluated day-end firm apps for *Appeal*, *Creativity*, and *Completion*

*Appeal* as measured by attractiveness to customers as outcome of exploitation, whereas *Creativity* is that of exploration

*Completion* compared to rule out potential confound of judge perceptions

**TABLE 3: Firm Judge Scores.** Models are OLS, with robust standard errors clustered at the firm level. $p$-values are displayed in brackets. The *Forfeit* variable, which takes a value of 1 for firms that forfeited and did not undergo the judging process, is included in odd numbered models below

| | (3-1) Appeal | (3-2) Appeal | (3-3) Appeal | (3-4) Appeal | (3-5) Creativity | (3-6) Creativity | (3-7) Creativity | (3-8) Creativity | (3-9) Completion | (3-10) Completion | (3-11) Completion | (3-12) Completion |
|---|---|---|---|---|---|---|---|---|---|---|---|---|
| Treatment Group | 0.614 | 0.846 | 0.546 | 0.661 | -0.499 | -0.687 | -0.692 | -0.960 | 0.279 | 0.385 | 0.251 | 0.336 |
| | [0.009] | [0.007] | [0.014] | [0.042] | [0.082] | [0.079] | [0.046] | [0.018] | [0.435] | [0.434] | [0.562] | [0.578] |
| Current Student | | | 0.725 | 0.991 | | | 0.145 | 0.638 | | | 0.336 | 0.917 |
| | | | [0.050] | [0.145] | | | [0.714] | [0.431] | | | [0.563] | [0.338] |
| Graduate Degree | | | 0.430 | 0.365 | | | -0.458 | -0.877 | | | -1.065 | -1.640 |
| | | | [0.273] | [0.544] | | | [0.386] | [0.224] | | | [0.112] | [0.078] |
| GitHub | | | 0.120 | 0.106 | | | 0.893 | 0.907 | | | -0.223 | 0.013 |
| | | | [0.881] | [0.914] | | | [0.378] | [0.423] | | | [0.855] | [0.992] |
| Google Development | | | -0.207 | -0.310 | | | 0.898 | 1.006 | | | -0.127 | -0.352 |
| | | | [0.618] | [0.707] | | | [0.082] | [0.165] | | | [0.838] | [0.707] |
| Software Development | | | -0.050 | -0.039 | | | 0.007 | -0.003 | | | -0.042 | -0.043 |
| | | | [0.080] | [0.354] | | | [0.872] | [0.957] | | | [0.480] | [0.578] |
| Prior Hackathons | | | -0.144 | -0.182 | | | -0.096 | -0.046 | | | -0.094 | 0.021 |
| | | | [0.108] | [0.106] | | | [0.444] | [0.782] | | | [0.529] | [0.944] |
| Firm Size | | | -0.113 | -0.096 | | | -0.222 | -0.421 | | | -0.169 | -0.195 |
| | | | [0.274] | [0.527] | | | [0.280] | [0.075] | | | [0.424] | [0.524] |
| Forfeit | -3.497 | | -3.702 | | -3.336 | | -3.331 | | -2.772 | | -2.929 | |
| | [0.000] | | [0.000] | | [0.000] | | [0.000] | | [0.000] | | [0.000] | |
| Constant | 3.274 | 3.154 | 3.590 | 3.456 | 3.518 | 3.615 | 3.284 | 3.782 | 2.670 | 2.615 | 4.012 | 3.640 |
| | [0.000] | [0.000] | [0.000] | [0.009] | [0.000] | [0.000] | [0.008] | [0.017] | [0.000] | [0.000] | [0.009] | [0.114] |
| Observations | 38 | 27 | 38 | 27 | 38 | 27 | 38 | 27 | 38 | 27 | 38 | 27 |

## Introduction/Motivation

Iterative coordination (e.g., Agile stand-up meetings) used by over 70% of organizations to manage software and non-software projects

Practitioner expectations of simultaneous exploration (e.g. creativity) and exploitation (e.g. product quality) are in conflict with conventional wisdom in academic literature

Non-random selection into Agile makes studying this phenomenon within a company difficult

Approach: embed field experiment in one-day app development hackathon tracking early-stage software projects

## Search in Firm Software Code

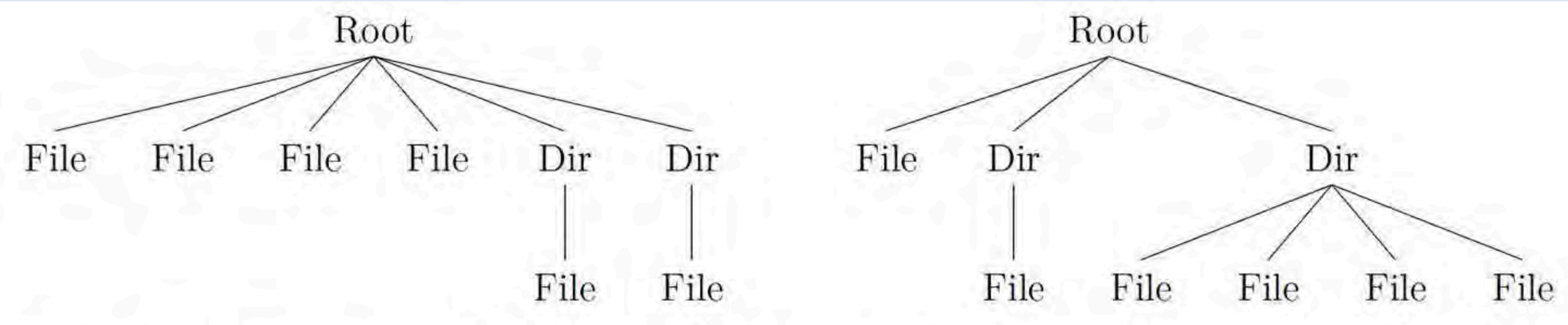

**FIGURE 1: File Hierarchy Branching Factors.** The two file hierarchies above share the same number of files and directories and a depth of 2 levels but demonstrate different patterns of exploratory and exploitative search.

| Variable | Definition | Search Interpretation |
|---|---|---|
| *Files* | Number of files across the file hierarchy | Overall search |
| *Directories* | Number of directories across the file hierarchy | Overall search |
| *Lower Descendants* | Total nodes (files + directories) below root/first level of file hierarhcy | |
| *Overall Branching* | Average branching factor for the entire file hierarchy. Equal to $\frac{files+directories}{directories}$ | |
| *Upper Branching* | Branching factor at the root/first of level of file tree. Equal to files + directories directly below the root | Exploitative search |
| *Lower Branching* | Average branching factor for directories below root/first level of a given file hierarhcy | Exploratory search |

**TABLE 6: Overall Search Productivity.** Models are OLS, with robust standard errors clustered at the firm level. $p$-values are displayed in brackets.

| | *Files* | | *Directories* | | *Lower Descendants* | |
|---|---|---|---|---|---|---|
| | (6-1) | (6-2) | (6-3) | (6-4) | (6-5) | (6-6) |
| Standups × Post | 0.373 | 0.567 | 0.443 | 0.613 | 0.426 | 0.652 |
| | [0.460] | [0.087] | [0.363] | [0.077] | [0.481] | [0.103] |
| ln(Lines + 1) | | 0.452 | | 0.396 | | 0.528 |
| | | [0.000] | | [0.000] | | [0.000] |
| Firm FE | Yes | Yes | Yes | Yes | Yes | Yes |
| Minute FE | Yes | Yes | Yes | Yes | Yes | Yes |
| Observations | 20520 | 20520 | 20520 | 20520 | 20520 | 20520 |

**TABLE 7: Comparing Search Process Across File Hierarchies.** Models are OLS, with robust standard errors clustered at the firm level.

| | *Overall Branching* | | *Upper Branching* | | *Lower Branching* | |
|---|---|---|---|---|---|---|
| | (7-1) | (7-2) | (7-3) | (7-4) | (7-5) | (7-6) |
| Standups × Post | −0.007 | 0.036 | 0.185 | 0.258 | −0.200 | −0.169 |
| | [0.941] | [0.499] | [0.291] | [0.047] | [0.139] | [0.054] |
| ln(Lines + 1) | | 0.100 | | 0.170 | | 0.127 |
| | | [0.000] | | [0.000] | | [0.000] |
| Firm FE | Yes | Yes | Yes | Yes | Yes | Yes |
| Minute FE | Yes | Yes | Yes | Yes | Yes | Yes |
| Observations | 20520 | 20520 | 20520 | 20520 | 12454 | 12454 |

## Conclusion

**We find** that iterative coordination increases search productivity, while favoring exploitative search over exploratory search

We contribute to/empirically validate the **theoretical search literature**, introducing 1) **novel data collection methodology** via Git and 2) **the use of hackathons** as empirical setting for strategy and entrepreneurship research

Study limitations include group size, inability to capture communication mechanisms, junior engineers

Future research should focus on boundaries to our findings (organization size, problem-solving context), other outcomes (e.g. emergence of shadow hierarchy)

## Method

Population (112): sophomore+/post-grad CS majors, professional developers, etc.

Participants registered as firms of 2-4

Hackathon as entrepreneurial setting, providing market competition with judges representing consumer choice

### *Task*

Competing firms developed apps that achieved some prosocial goal

Firms required to 1) develop off of toolkits provided by sponsor, and 2) use Git for version control

### *Experiment*

Mentors facilitate (but do not run) stand-ups

Treatment: every two hours, three questions:

*"What have you accomplished since your last check-in?"*

*"What are your goals until the next check-in 2 hours from now?"*

*"What are your goals for the end of the day (and have they changed)?"*

Control: every two hours, null interaction

Difference-in-differences design with 2.5 hour pre-treatment period (for search panel data):

$$ln(Y_{it}+1) = \beta(Standups_i \times Post_{it}) + ln(Lines_{it}) + \alpha_i + \delta_t + \epsilon_{it}$$

## Discussion

How does an organization solve universal problems of task division and allocation when system-level goals are in flux?

Before task division and task allocation can occur, agents in organization must reach a working agreement over their system-level goal

We define iterative coordination in the context of organizational search as an explicit update to the organization's goal definition

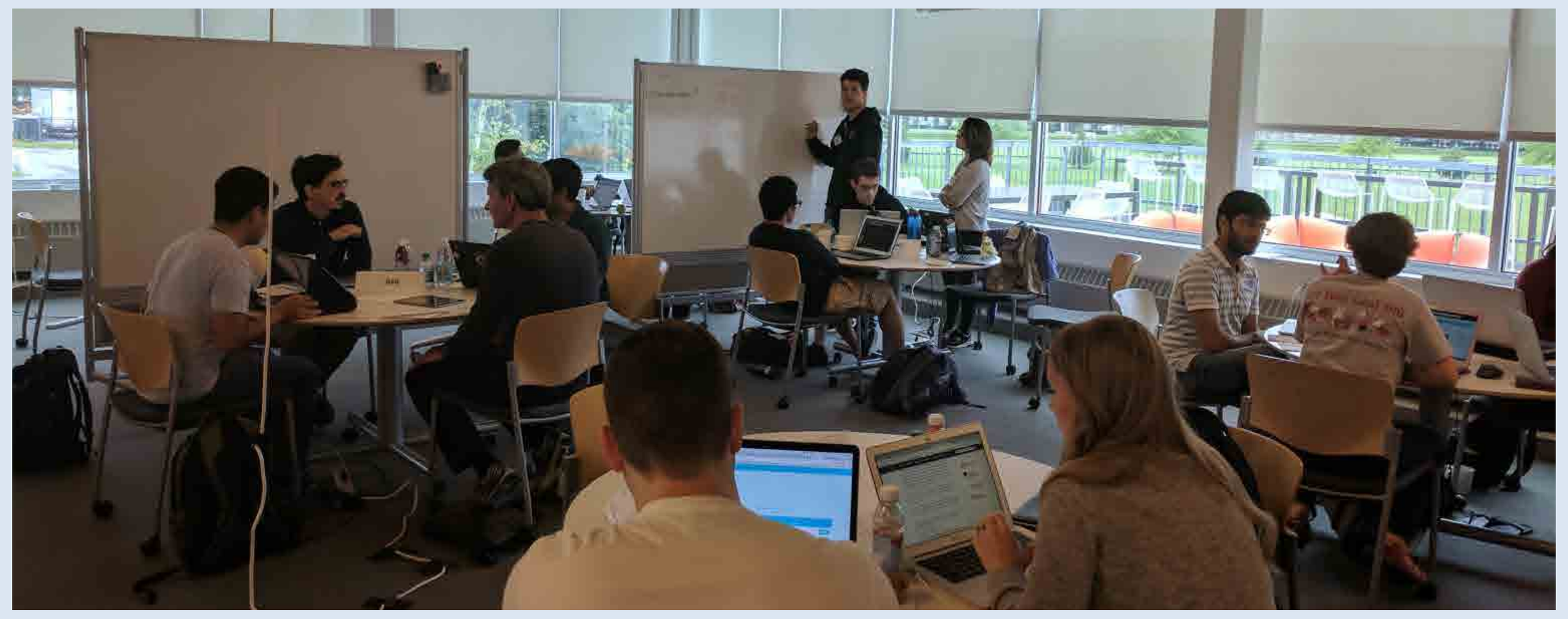

### *Interpretations*

Increased iteration may lead to imperfect shared mental representations across organization (Knudsen & Srikanth 2014)

Simplifications in mental representations can help reduce search effort (Csaszar & Levinthal 2016), potentially increasing search productivity (Ethiraj & Levinthal 2009)

Inaccurate representations preferable to counterfactual of one that doesn't distinguish between alternatives (Puranam & Swamy 2016)

MCTS
Munich Center for Technology in Society
Technische Universität München

# Hacking Creativity

## Key Concepts

**Openness**
- *Preliminary Anonymity*
- *Funnyfication*

**Rhythm**
- *Tacit Schedule*
- *Time Boundedness*

**Diversity**
- *of Prospective*
- *of Knowledge*
- *of Perspective*

## Problem Conceptualization

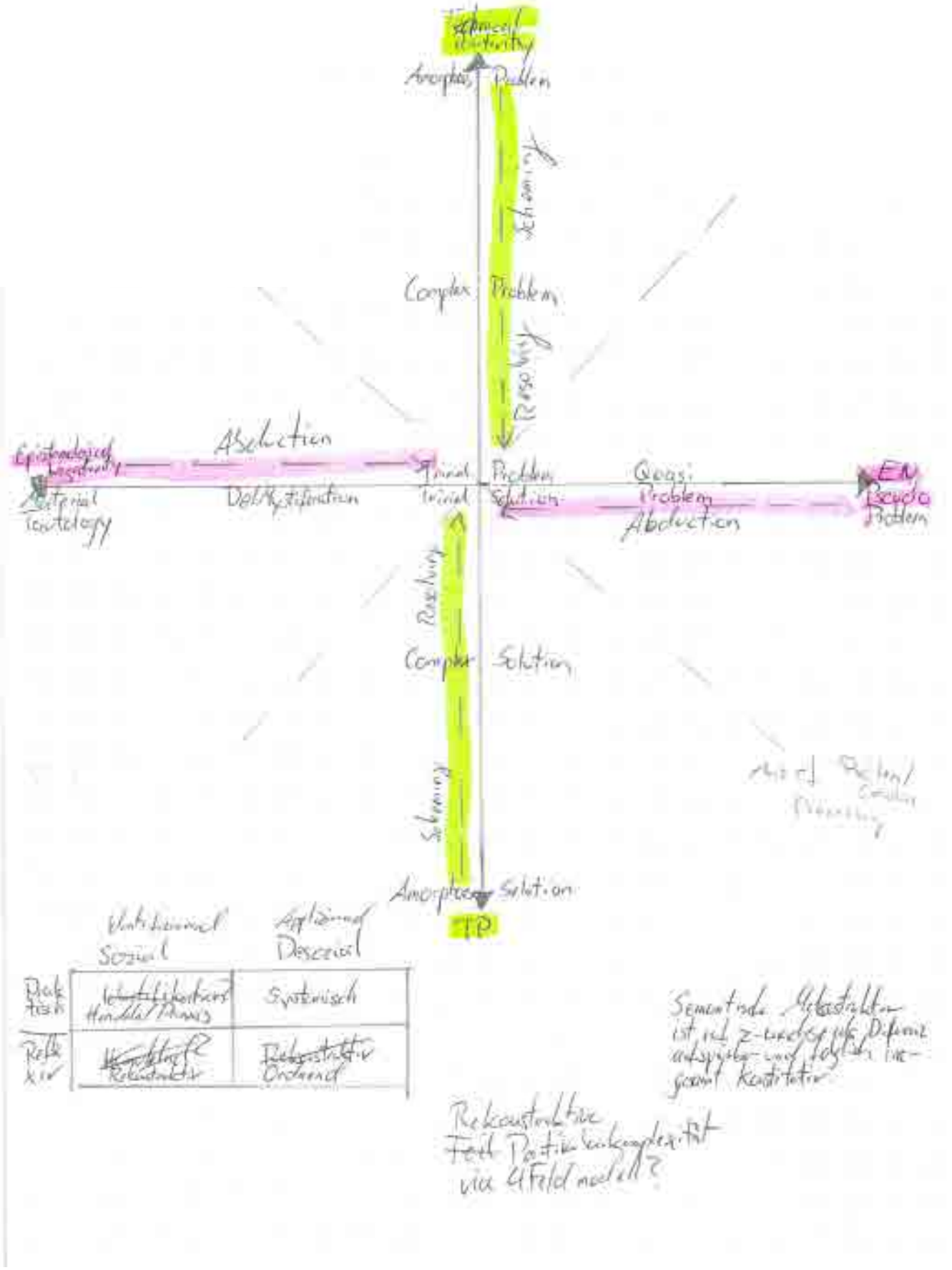

## Need support!

Are there greater institutional shifts concerning „hackafication" of academia, economy, etc? Where can they be seen/studied?

How to understand (and resolve) problems of transferring hackathon outcomes into *established* organization contexts?

What are your experiences with „hacking problems"?
How to define or even operationalize invention/creativity?

Can we make „social creativity" a ordinary, non-event bounded feature of everyday life work?

Carnegie Mellon University

isr institute for SOFTWARE RESEARCH

# Project Selection, Goal Setting, and Coordination of Teams in Short-term, Intensive Collocation

Ei Pa Pa Pe Than[1], Alexander Nolte[2], Anna Filippova[3], Christian Bird[4], Steve Scallen[5], James D. Herbsleb[1]

[1]Institute for Software Research, School of Computer Science, Carnegie Mellon University, Pittsburgh, PA, USA

[2]Institute of Computer Science, University of Tartu, Tartu, Estonia

[3]GitHub Inc., San Francisco, CA, USA, [4]Microsoft Research, Redmond, WA, USA, [5]Microsoft Garage, Redmond, WA, USA

## BACKGROUND

- Hackathons or short-term, intense collaboration
  - ❖ People come together for a few days, assemble into small teams and create artifacts - most commonly software prototypes
  - ❖ Variously known as data dives, codefests, hack days, sprints, edit-a-thons, etc.
- Hackathon designs vary along multiple dimensions
  - ❖ Collaborative vs. Competitive
  - ❖ Innovation vs. Community Building vs. Learning vs. Just having fun
  - ❖ Diversity in skills, expertise, familiarity, etc.

## RESEARCH GAPS AND QUESTIONS

- Collocation for extended period facilitates coordination and productivity
  - ❖ Does the hackathon yield the same benefits? What coordination activities they emphasize in the hackathon? What kinds of difficulties do they run into?

**RQ: In hackathons, how do pre-existing teams (PET) and flash teams (FT):**

1) **select their projects?**
2) **set their goals?**
3) **coordinate their work?**

## METHODS: THE SETTING

- The 2017 Microsoft OneWeek Hackathon
- At Redmond site - 6,700 participants and 1,800 registered projects
- "HackBox" was used - A tool for project creation, team building, and skill matching
- Team selection criteria: Team size, diversity on roles, org units, prior experience of working together, and code or non-code projects
- Empirical grounded theory procedures were used to analyze observation and post-interview data

## METHODS: DATA COLLECTION PROCEDURE

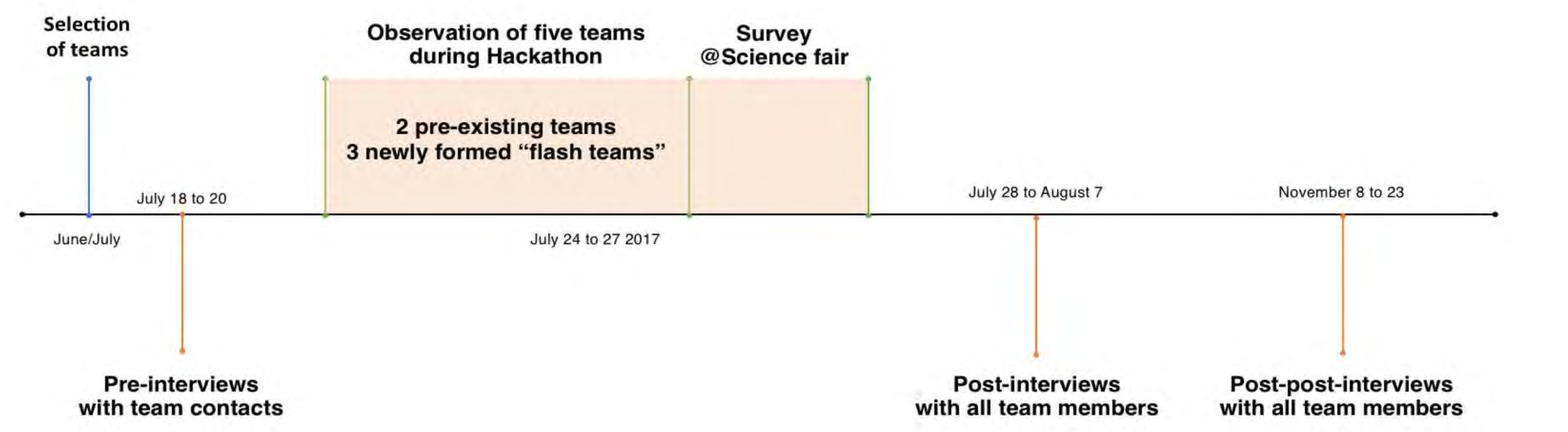

## FINDINGS AND IMPLICATIONS

### Project Selection

- PET used the hackathon for regular project-related work
- FT used the hackathon to address a personally experienced need or a passion, collaborating with unfamiliar people
- FT's joiner's project selection criteria used - Technology and topic of interest, an opportunity for new connections, apply existing skills in a new context
- PET recruited joiners mainly via proposers' existing social network while FT leveraged both proposers' network and HackBox

### Goal Setting

- PET strove to have a product complete enough to serve the team's needs
- FT focused on official hackathon outcomes of demos and videos and hoped their ideas would find home somewhere in the company

### Coordination

- PET fell back to regular work process with only minor modification, just being another day at the office
- FT used a form of role-based coordination where members coordinated based on roles signed up on HackBox

### Implications

- Tools like HackBox that match up projects with potential participants are useful
- Role-based coordination is beneficial to FT
- Needs to manage mismatched expectation, especially with first-time participants
  - ❖ Explicit mentoring opportunities by having two roles - "Expert" and "Apprentice"
  - ❖ A carefully curated selection of prior projects to help set realistic expectations
- Extends the theory of **radical collocation** by demonstrating how PET and FT responses to time pressure imposed by the hackathon

**Contact Info: Ei Pa, Post-Doctoral Reseacher (eipapapt@cs.cmu.edu)**

# Appendix C

## Online Guides on Organizing Hackathons

Major league hacking (MLH) – an organization that primarily organizes student hackathons affiliated with universities

- https://mlh.io/event-membership
- https://guide.mlh.io/

Gartener – the planning kit derived based on NASA's international annual hackathons

- https://www.gartner.com/smarterwithgartner/plan-a-successful-hackathon/

Socrata

- https://socrata.com/open-data-field-guide/how-to-run-a-hackathon/

Rally team

- https://rallyteam.com/blog/

Hackathon: Your guide to running a hackathon by J Mac

- https://www.amazon.ca/Hackathon-Your-guide-running-hackathon-ebook/dp/B00JLT24BY

Guide to civic hackathons from DC – based on five successful years of open data day D

- https://hackathon.guide/

CHI Hacknight

- https://chihacknight.org/blog/2015/11/23/10-lessons-from-organizing-the-chi-hack-night.html

## Online Listing of Hackathons

Devpost – https://devpost.com/hackathons

- LinkedIn competitor for Hackathon → organizers are free to post hackathons, participants can use it to submit their projects to hackathons

Eventbrite - https://www.eventbrite.com

- Generic event website, you can keyword search for "hackathon"

Hackathon.com – https://www.hackathon.com/

- Just found this with a Google search, but it seems to have event listings

MLH – https://mlh.io/seasons/na-2018/events

- Major League Hacking list of events for the 2017-2018 school year